%% file: main.tex
\documentclass[journal,table]{IEEEtran}

\usepackage[prependcaption,colorinlistoftodos]{todonotes}

\usepackage{cite}
\usepackage{amssymb,amsfonts}
\usepackage{graphicx}
\usepackage{textcomp}
\usepackage{booktabs}
\usepackage{multirow}
\usepackage{tabularx}
\usepackage{siunitx}
\DeclareSIUnit \voltampere { VA }
\usepackage{lipsum}

\usepackage{enumitem}

\usepackage{amsthm}
\usepackage[cmex10]{amsmath}

\usepackage{tikz}
\usetikzlibrary{positioning}
\usetikzlibrary{arrows}

\usepackage{algorithm}
\usepackage{algpseudocode}

\newcommand{\bbR}{\mathbb{R}}

\newcommand{\calY}{\mathcal{Y}}

\newcommand{\calC}{\mathcal{C}}

\newcommand{\calH}{\mathcal{H}}

\newcommand{\calB}{\mathcal{B}}
\newcommand{\calA}{\mathcal{A}}

\newcommand{\calQ}{\mathcal{Q}}
\newcommand{\calE}{\mathcal{E}}

\definecolor{backgroundblue}{rgb}{0, 0.4470, 0.7410}
\definecolor{hydro}{rgb}{0.101,0.639,1}
\definecolor{backgroundgreen}{rgb}{0.4660, 0.6740, 0.1880}
\definecolor{steam}{rgb}{0.9290, 0.6940, 0.1250}
\definecolor{PV}{rgb}{1, 0.752, 0}
\definecolor{battery}{rgb}{1, 0.4, 1}
\definecolor{wind}{rgb}{0.4, 0.8, 1}
\definecolor{SC}{rgb}{0.572, 0.816, 0.313}
\definecolor{STATCOM}{rgb}{0, 0.8, 0.6}

\usepackage[capitalize,nameinlink,compress]{cleveref}

\crefname{figure}{Fig.}{Figures} % Use Figure instead of Fig.
\crefname{line}{line}{lines} % Make sure line is not capitalized
\crefname{claim}{Claim}{Claims} % Make sure line is not capitalized
\crefname{equation}{}{} % No Eq. for equations
\crefname{problem}{Problem}{Problems}
\crefname{assumption}{Assumption}{Assumptions}

\usepackage{cuted}
\usepackage{flushend}
\usepackage{stfloats}

\usepackage[font=footnotesize]{caption}
\usepackage[font=footnotesize]{subcaption}
\usepackage[bottom]{footmisc}

\usepackage{accents}
\newcommand{\ubar}[1]{\underaccent{\bar}{#1}}

\usepackage{xcolor,soul}
\usepackage{framed}
\colorlet{shadecolor}{yellow}
\soulregister\cite7
\soulregister\cref7
\soulregister\ref7
\soulregister\pageref7
\soulregister\eqref7

\renewcommand{\baselinestretch}{0.95}

\begin{document}

\title{Control Design of Dynamic Virtual Power Plants: An Adaptive Divide-and-Conquer Approach}

\author{Verena Häberle, Michael W. Fisher, Eduardo Prieto-Araujo and~Florian Dörfler% <-this % stops a space
\thanks{\hspace{-4mm}This paper is based upon work supported by the King Abdullah University of Science and Technology (KAUST) Office of Sponsored Research (award No. OSR-2019-CoE-NEOM-4178.11) and by the European Union's Horizon 2020 research and innovation program (grant agreement No. 883985).}
\thanks{\hspace{-4mm}V. Häberle, M. W. Fisher and F. Dörfler are with the Automatic Control Laboratory, ETH Zurich, 8092 Zurich, Switzerland.}% <-this % stops a space
\thanks{\hspace{-4mm}E. Prieto-Araujo is a Serra Húnter Lecturer with the Centre d’Innovació Tecnològica en Convertidors Estàtics i Accionamients, Department d’Enginyeria Elèctrica, Universitat Politècnica de Catalunya, 08028 Barcelona, Spain.}
\thanks{\hspace{-4mm}Email:\{verenhae,mfisher,dorfler\}@ethz.ch; eduardo.prieto-araujo@upc.edu}% <-this % stops a space
}

%\markboth{IEEE Transactions on Power Systems,~Vol.~xx, No.~x, month~year}%

\maketitle

\begin{abstract}
In this paper, we present a novel control approach for dynamic virtual power plants (DVPPs). In particular, we consider a group of heterogeneous distributed energy resources (DERs) which collectively provide desired dynamic ancillary services such as fast frequency and voltage control. Our control approach relies on an adaptive divide-and-conquer strategy: first, we disaggregate the desired frequency and voltage control specifications of the aggregate DVPP via adaptive dynamic participation matrices (ADPMs) to obtain the desired local behavior for each device. Second, we design local linear parameter-varying (LPV) $\mathcal{H}_\infty$ controllers to optimally match this local behaviors. In the process, the control design also incorporates the physical and engineered limits of each DVPP device. Furthermore, our adaptive control design can properly respond to fluctuating device capacities, and thus include weather-driven DERs into the DVPP setup. Finally, we demonstrate the effectiveness of our control strategy in a case study based on the IEEE nine-bus system. 
\end{abstract}


\IEEEpeerreviewmaketitle

\setlength{\textfloatsep}{6pt}% Remove \textfloatsep
\section{Introduction}
\IEEEPARstart{F}{uture} power systems will contain an increasing penetration of non-synchronous distributed energy resources (DERs). In this regard, reliable ancillary services provision, as currently ensured by conventional generators, has to be shouldered by DERs. This imposes great challenges to cope with the fluctuating nature of renewable energy sources\cite{milano2018foundations}, as well as their device-specific limitations.

As early as 1997, the concept of \textit{virtual power plants (VPPs)} has been proposed to pave the way for future ancillary services by DERs\cite{awerbuch2012virtual}. VPPs are collections of distributed generators (all with individual device limitations), aggregated to have the same visibility, controllability and market functionality as a unique power plant\cite{saboori2011virtual,caldon2004optimal,pudjianto2007virtual}. Today, most commercial implementations as well as the scientific landscape are restricted to VPPs providing static ancillary services in the form of tracking power and voltage set points, see, e.g.,\cite{dall2017optimal}.

In this work, we are interested in the vastly underexplored concept of a \textit{dynamic virtual power plant (DVPP)} consisting of \textit{heterogeneous} DERs, which all-together can provide desired \textit{dynamic ancillary services} beyond mere set point tracking\cite{posytyf}. In particular, we are interested in dynamic ancillary services on faster time scales, such as fast frequency and voltage control, which cannot be provided by existing VPP setups restricted to tracking set points. The key to success is heterogeneity: Only a sufficiently heterogeneous group of devices (complementing each other in terms of energy/power availability, response times, and weather dependency) can reliably provide dynamic ancillary services across all power and energy levels and time scales, while none of the individual devices is able to do so. 

Motivating examples of collections of heterogeneous energy sources for dynamic ancillary services provision include hydro-power with initially inverse
response dynamics compensated by batteries on short time scales\cite{ghasemi2019investigation}, synchronous condensers
(with rotational energy) paired with converter-based generation\cite{kenyon2020grid}, or hybrid storage
pairing batteries with supercapacitor providing regulation on different frequency ranges\cite{li2008power}. However, the coordination of all these collections is highly customized, and not (even conceptually) extendable to other device aggregations. Further, none of these collections are controlled to match a desired aggregate dynamic behavior, therefore lacking optimal performance and reliability during ancillary services provision. In contrast, other works in\cite{joak,zhong2021coordinated} propose more versatile DVPP approaches to achieve a desired short-term frequency response on an aggregate level. In particular,\cite{zhong2021coordinated} relies on static participation factors and a coordinated control signal which is communicated to each device, but therefore subject to communication delays and single point of failure risk. As opposed to this,\cite{joak} presents a fully decentralized control strategy based on dynamic participation factors, which can be used to take local device dynamics into account. However, both\cite{zhong2021coordinated} and\cite{joak} are restricted to provide frequency control, do not consider device-level constraints, and are non-adaptive, therefore prone to failure during temporal variability of weather-driven DERs.

In this work, we present a novel multivariable control approach for DVPPs, capable of providing multiple desired dynamic ancillary services at once. We particularly focus on fast frequency and voltage control objectives, specifying them as a \textit{desired dynamic multi-input multi-output (MIMO) behavior} of the aggregate DVPP, given in terms of a desired transfer matrix from frequency and voltage to active and reactive power. In addition to the desired aggregate output, our DVPP control strategy also incorporates the \textit{DVPP internal
constraints} of the devices (e.g. speed limitations, capacities, current constraints, etc.), to ensure they are not exceeded during normal operating conditions. We pursue a local control strategy and design individual feedback controllers for each DVPP device, subject to its own limitations, but so that the aggregate behavior meets the desired MIMO specification. More specifically, our control approach relies on an \textit{adaptive divide-and-conquer} strategy composed of two steps: first, we disaggregate the MIMO specification among the devices using \textit{adaptive dynamic participation matrices (ADPMs)} which take the form of MIMO transfer matrices, and basically represent a multidimensional and adaptable version of the dynamic participation factors in\cite{joak}. Second, we employ local linear parameter-varying (LPV) $\mathcal{H}_\infty$ methods\cite{montagner2005lmi} to optimally match the obtained local desired behavior of each device, while satisfying transient device-level constraints. Further, we propose centralized and distributed update strategies to \textit{adapt the ADPMs online} towards capacity fluctuations, so that our control design can properly respond to temporal variability of DERs. This allows for a DVPP setup including weather-driven DERs, which are typically treated as non-dispatchable, and not employed for fast ancillary services provision\cite{morales2013integrating}. 

The remainder of the paper is structured as follows: in~\cref{sec:control_setup}, we introduce the novel DVPP control setup for fast frequency and voltage control. We provide a simplified setup using the formalism of linear systems, which makes it convenient to develop our control design. \cref{sec:divide_conquer} presents the divide-and-conquer strategy involving the disaggregation via ADPMs and the local $\mathcal{H}_\infty$ control. In~\cref{sec:test_case}, we demonstrate the performance of our control design on a test case of the IEEE nine-bus system using detailed and nonlinear system and device models. \cref{sec:conclusion} concludes the paper.

\begin{figure}[b!]
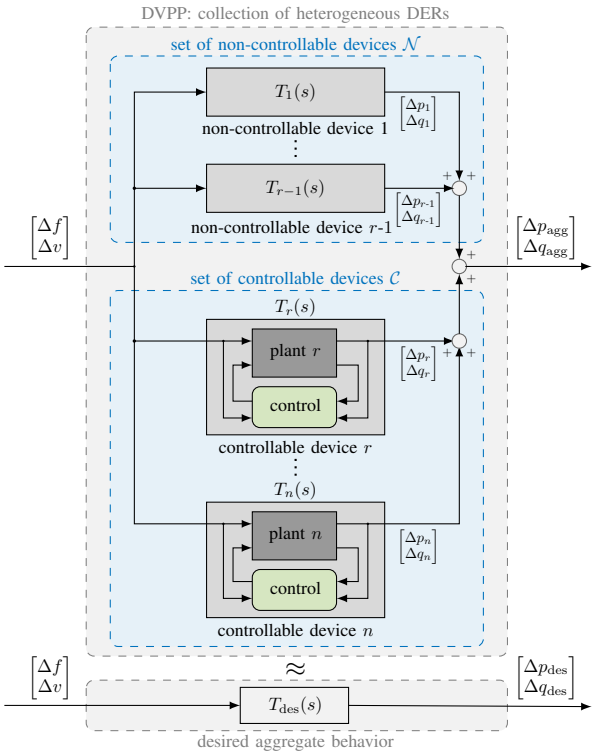

	\centering
	\include{Figures/DVPP_ctrl_setup}
	\vspace{-8mm}
	\caption{Schematic of the DVPP control setup.}
	\label{fig:DVPP_setup}
\end{figure}

\section{DVPP Control Setup}\label{sec:control_setup}
We consider a DVPP control setup for a group of heterogeneous DERs (\cref{fig:DVPP_setup} and \cref{tab:secII_variables}), including both a collection~$\mathcal{N}$ of \textit{non-controllable} devices (e.g., installed synchronous generators and condensers), as well as a collection~$\mathcal{C}$ of \textit{controllable} devices (e.g., converter-based generators). We assume that all devices of a DVPP are connected at the same bus of the transmission grid, where they receive an input signal in terms of the measured bus frequency $\Delta f$ and voltage magnitude deviation $\Delta v$ (\cref{fig:DVPP_setup}). The active and reactive power deviation output of each device $i$, namely $\Delta p_i$ and $\Delta q_i$, respectively (deviating from the respective power set point), sum up to the aggregate active and reactive power deviation output of the DVPP, namely $\Delta p_\mathrm{agg}$ and $\Delta q_\mathrm{agg}$, respectively, i.e.,
\begin{align} 
\begin{bmatrix}\Delta p_\mathrm{agg}\\\Delta q_\mathrm{agg}\end{bmatrix}=\textstyle\sum_{i\in\mathcal{N}\cup\mathcal{C}}\,\begin{bmatrix}\Delta p_i\\\Delta q_i\end{bmatrix}.
\end{align}

We assume that all non-controllable devices $i\in\mathcal{N}$ have a pre-existing frequency and voltage control (e.g., installed turbine and governor controls or automatic voltage regulators (AVRs)). Hence, their local closed-loop transfer $2\times2$ matrices $T_i(s)$ (from frequency and voltage magnitude to active and reactive power) are considered as fixed within the DVPP setup.

The local closed-loop transfer $2\times2$ matrices $T_i(s)$ of the controllable devices $i\in\mathcal{C}$, in turn, can be shaped freely by employing appropriate feedback controls (cf.~\cref{sec:local_model_match}). 

Considering the local closed-loop transfer matrices $T_i$ of both the non-controllable and controllable devices $i\in\mathcal{N}\cup\mathcal{C}$, the aggregate DVPP behavior is given by
\begin{align}
\begin{bmatrix}\Delta p_\mathrm{agg}(s)\\\Delta q_\mathrm{agg}(s)\end{bmatrix}=\textstyle\sum_{i\in\mathcal{N}\cup\mathcal{C}}\,T_i(s)\begin{bmatrix}\Delta f(s)\\ \Delta v(s)\end{bmatrix}.
\end{align}

To compensate for ancillary services conventionally provided by synchronous generators in transmission networks, a decoupled $\mathrm{f}$-$\mathrm{p}$ and $\mathrm{v}$-$\mathrm{q}$ behavior is specified for the aggregate DVPP as a desired diagonal\footnote{Our formalism directly extends to full MIMO specifications, potentially relevant for future ancillary services or in other types of networks\cite{de2007voltage}.} MIMO transfer matrix as
\begin{align}\label{eq:MIMO_spec}
\begin{bmatrix}\Delta p_\mathrm{des}(s)\\\Delta q_\mathrm{des}(s)\end{bmatrix}=\underset{=:T_\mathrm{des}(s)}{\underbrace{\begin{bmatrix}T_\mathrm{des}^\mathrm{fp}(s)&0\\0&T_\mathrm{des}^\mathrm{vq}(s)\end{bmatrix}}}\begin{bmatrix}\Delta f(s)\\ \Delta v(s)\end{bmatrix}.
\end{align}

\renewcommand{\arraystretch}{1.2}
\begin{table}[t!]\scriptsize
    \centering
     \caption{List of notation for the DVPP control setup.}
    \vspace{-1mm}
    \begin{tabular}{c||c}
     \toprule
         Description & Symbol  \\ \hline
         Set of non-controllable DVPP devices & $\mathcal{N}$ \\
         Set of controllable DVPP devices & $\mathcal{C}$ \\ 
         DVPP device index & $i$ \\\hline
         Measured bus frequency deviation& $\Delta f$ \\
         Measured bus voltage magnitude deviation& $\Delta v$ \\
          Active power deviation output of device $i$& $\Delta p_i$ \\
          Reactive power deviation output of device $i$& $\Delta q_i$ \\
          Aggregate active power deviation output of the DVPP & $\Delta p_\mathrm{agg}$ \\
           Aggregate reactive power deviation output of the DVPP & $\Delta q_\mathrm{agg}$ \\
           Desired active power deviation output of the DVPP & $\Delta p_\mathrm{des}$ \\
           Desired reactive power deviation output of the DVPP & $\Delta q_\mathrm{des}$ \\\hline
            Local closed-loop transfer matrix of device $i$ & $T_i(s)$ \\
            Desired MIMO transfer matrix of the DVPP & $T_\mathrm{des}(s)$ \\
            Desired DVPP transfer function for the $\mathrm{f}$-$\mathrm{p}$ channel & $T_\mathrm{des}^\mathrm{fp}(s)$ \\
            Desired DVPP transfer function for the $\mathrm{v}$-$\mathrm{q}$ channel & $T_\mathrm{des}^\mathrm{vq}(s)$ \\
    \bottomrule
    \end{tabular}
    	
    \label{tab:secII_variables}
\end{table}
\renewcommand{\arraystretch}{1} \normalsize

Since the transfer matrices $T_i$ of the non-controllable devices $\mathcal{N}$ are fixed, the DVPP control design problem is to find local controllers for the controllable devices $\mathcal{C}$, such that the following \textit{aggregation condition} holds\footnote{The sign $``\stackrel{!}{=}"$ indicates that the terms on the left hand side of the equality must be designed in such a way that the equality is satisfied.}:
\begin{align}\label{eq:DVPP_specification}
\textstyle\sum_{i\in\mathcal{N}\cup\mathcal{C}}\,T_i(s)\stackrel{!}{=}T_\mathrm{des}(s).
\end{align} 

Furthermore, during the control design it is important to ensure that physical and engineered \textit{device limitations}, including response time constraints as well as (potentially time-varying) limits on power availability and current capacity, are not exceeded during normal operating conditions. 

Of course, to meet the aggregation condition~\cref{eq:DVPP_specification}, the power park comprising the DVPP has to be sufficiently diverse covering all time scales and energy/power levels. Further, it is assumed that the desired behavior $T_\mathrm{des}$ is provided by the power system operator, therefore reasonably specified so that it is collectively achievable by the devices, while rendering the closed-loop power system stable, and being robust to model uncertainties and parameter variations in the rest of the grid.

\section{Adaptive Divide-and-Conquer Strategy}\label{sec:divide_conquer}
Our approach to solve the previous DVPP control design problem is based on an adaptive divide-and-conquer strategy, composed of two steps:
\begin{enumerate}
\item Disaggregate the desired DVPP behavior by dividing the MIMO transfer matrix $T_\mathrm{des}$ among the DVPP devices using adaptive dynamic participation matrices (ADPMs) to obtain local desired behaviors. The latter are defined by the product of each ADPM and $T_\mathrm{des}$, respectively.
\item Design a local feedback control for each device to optimally match the local desired behavior. We will resort to a linear parameter-varying (LPV) $\mathcal{H}_\infty$ method.
\end{enumerate}

\subsection{Disaggregation via ADPMs}\label{sec:disaggregation}
We disaggregate the desired MIMO transfer matrix to the individual devices by imposing the \textit{local matching condition}
\begin{align}\label{eq:control_problem}
T_i(s)&\stackrel{!}{=}M_{i}(s)\cdot T_\mathrm{des}(s),\quad\quad \forall i\in\mathcal{N}\cup\mathcal{C},
\end{align} 
where the $2\times 2$ transfer matrices $M_{i}$ are adaptive dynamic participation matrices (ADPMs) of the form
\begin{align}
 	M_i(s) = \begin{bmatrix}
 	m_{i}^\mathrm{fp}(s)&0\\0&m_{i}^\mathrm{vq}(s)
 	\end{bmatrix},
 	\quad \forall i\in\mathcal{N}\cup\mathcal{C},
\end{align}
with the diagonal elements $m_{i}^\mathrm{fp},\,m_{i}^\mathrm{vq}$ being adaptive dynamic participation factors (ADPFs) for the $\mathrm{f}$-$\mathrm{p}$ and $\mathrm{v}$-$\mathrm{q}$ channel, respectively (see below). Using the matching condition~\eqref{eq:control_problem}, the aggregation condition~\eqref{eq:DVPP_specification} can be disaggregated as
\begin{align}
\textstyle\sum_{i\in\mathcal{N}\cup\mathcal{C}}\,T_i(s)\stackrel{!}{=}\sum_{i\in\mathcal{N}\cup\mathcal{C}}\,M_{i}(s)\cdot T_\mathrm{des}(s) = T_\mathrm{des}(s), 
\end{align}
where $\textstyle\sum_{i\in\mathcal{N}\cup\mathcal{C}}\,M_{i}(s)\stackrel{!}{=}I_{2}$, and $I_2$ is the identity matrix. This results in the \textit{participation condition}
\begin{align}\label{eq:sum_part_fact_one}
\textstyle\sum_{i\in\mathcal{N}\cup\mathcal{C}}\,m_{i}^\mathrm{fp}(s)\stackrel{!}{=}1,\quad\quad \textstyle\sum_{i\in\mathcal{N}\cup\mathcal{C}}\,m_{i}^\mathrm{vq}(s)\stackrel{!}{=}1.
\end{align}

Considering the solvability of the local matching condition in \cref{eq:control_problem}, we require that each reference model $M_i\cdot T_\mathrm{des}$ is selected carefully in such a way that it can be matched by the associated device dynamics during normal operating conditions. On the one hand, as mentioned before, we therefore require that $T_\mathrm{des}$ is reasonably specified, and on the other hand, as outlined in the following, we need to carefully select the ADPMs $M_i$ according to the individual device limitations.

\subsubsection*{ADPF Selection}\label{sec:dyn_chan_part_fact} The ADPFs for the $\mathrm{f}$-$\mathrm{p}$ and $\mathrm{v}$-$\mathrm{q}$ channel are selected independently, but according to the same principle. In the following, we hence address both channels simultaneously using the variable $k\in\{\mathrm{fp},\mathrm{vq}\}$. A list of notation is provided in\cref{tab:secIIIA_variables}. 

For the non-controllable devices with fixed $T_{i}^k, i\in\mathcal{N}$, for each channel $k$, the ADPFs are obtained as 
\begin{align}\label{eq:non_ctrl_part_fact}
	\begin{split}
m_{i}^k(s)&:=T_{i}^k(s)(T_{\mathrm{des}}^k(s))^{-1}, \quad\forall i\in\mathcal{N},\, k\in\{\mathrm{fp},\mathrm{vq} \},
\end{split}
\end{align}
so that the matching condition~\cref{eq:control_problem} holds trivially. Given the fixed ADPFs in~\eqref{eq:non_ctrl_part_fact}, the ADPFs of the controllable devices are selected such that the participation condition in~\eqref{eq:sum_part_fact_one} is satisfied, while simultaneously respecting the heterogeneous time scales of local device dynamics along with steady-state power capacity limits. Hence, for each ADPF, we envision (see case studies in~\cref{sec:case1,sec:case2} for examples)
\begin{itemize} 
	\item a \textit{low-pass filter (LPF)} participation factor for devices that can provide regulation on longer time scales on channel $k$ including steady-state contributions,
	\item a \textit{high-pass filter (HPF)} participation factor for devices able to provide regulation on very short time scales on channel $k$, and
	\item a \textit{band-pass filter (BPF)} participation factor for devices able to cover the intermediate regime.
\end{itemize}
To accomplish this, we specify the ADPFs by two parameters: a channel-specific time constant $\tau_{i}^k$ for the roll-off frequency to account for different time scales of local device dynamics on channel $k\in\{\mathrm{fp},\mathrm{vq} \}$, and a DC gain $m_{i}^k(s=0):=~\theta_{i}^k$ to account for device power capacity limits. In particular, the ADPFs with a BPF or HPF behavior will always have a constant zero DC gain by definition, i.e., $m_{i}^k(s=0)=0$. In contrast, for all devices $\mathcal{C}_\mathrm{lpf}^k$ participating as a LPF on channel $k$, the LPF DC gains $\theta_{i}^k, i \in\mathcal{C}_\mathrm{lpf}^k$ have to satisfy
\begin{align}
	\textstyle\sum_{i\in\mathcal{N}\cup\mathcal{C}}\,m_{i}^k(s=0)=1, \quad k\in\{\mathrm{fp},\mathrm{vq} \}
\end{align}
to meet the participation condition in~\eqref{eq:sum_part_fact_one}.

Finally, for each channel $k\in\{\mathrm{fp},\mathrm{vq} \}$ separately, we sort the devices in descending order w.r.t. their channel-specific time constant, and apply Algorithm 1 to compute the respective ADPFs as LPFs, BPFs and HPF according to the devices' response time and capacity limitations. 
\begin{algorithm}[t!]
	\caption{Sort Algorithm for Channel $k\in\{\mathrm{fp}, \mathrm{vq}\}$}
	\begin{algorithmic}[1]
		\State // Fix ADPFs of non-controllable devices $1,...,r-1$ via~\eqref{eq:non_ctrl_part_fact}
		\State // Define steady-state ADPFs as LPFs
		\State$m_{i}^k(s)\leftarrow \tfrac{\theta_{i}^k}{\tau_{i}^ks+1},\,\forall i\in\mathcal{C}_\mathrm{lpf}^k$ 
		\State // Fix intermediate ADPFs as BPFs
		\For {$i = r+|\mathcal{C}_\mathrm{lpf}^k|:n-1$}
		\State \hspace{-3mm}$m_{i}^k(s)\leftarrow \tfrac{1}{(\tau_i^ks+1)^{d_i}}\left( \tfrac{1}{\tau_{i}^ks+1}-\textstyle\sum_{l=1}^{i-1}m_{l}^k(s) \right),\,d_i\in \mathbb{N}_0$
		\EndFor
		\State // Fix fastest device's ADPF as HPF
		\State $m_{n}^k(s)\leftarrow  \left(1-\textstyle\sum_{i=1}^{n-1}m_{i}^k(s) \right)$
	\end{algorithmic}
\end{algorithm}

\subsection{Online Adaptation of LPF DC gains}\label{sec:online_adaption} We specify the LPF DC gains $\theta_{i}^k, i \in\mathcal{C}_\mathrm{lpf}^k$ in such a way that they can be adapted online, proportionately to the time-varying power capacity limits of the devices. For each channel $k\in\{\mathrm{fp},\mathrm{vq} \}$, we consider the optimal quadratic allocation
\begin{align}\label{eq:resource_allocation}
\begin{split}
	\underset{\theta_{i}^k(t), \forall i\in\mathcal{C}_\mathrm{lpf}^k}{\text{minimize}}\quad&\textstyle\sum_{i\in\mathcal{C}_\mathrm{lpf}^k}\tfrac{1}{{y}_{i}^{\mathrm{max},k}(t)}(\theta_{i}^k(t))^2\\
	\text{subject to}\quad&\textstyle\sum_{i\in\mathcal{C}_\mathrm{lpf}^k}\theta_{i}^k(t)+\textstyle\sum_{j\in\mathcal{N}}m_{j}^k(s=0)=1\\
	&\theta_{i}^k(t)\geq 0,\,\raggedleft\forall i \in \mathcal{C}_\mathrm{lpf}^k,
\end{split}
\end{align} 
where the equality constraint assures the participation condition~\cref{eq:sum_part_fact_one}, and ${y}_{i}^{\mathrm{max,fp}}(t)\equiv{p}_{i}^\mathrm{max}(t)\in[\ubar{p}_{i}^\mathrm{max},\bar{p}_{i}^\mathrm{\,max}], \,\forall i\in\mathcal{C}_\mathrm{lpf}^\mathrm{fp}$ (or ${y}_{i}^{\mathrm{max,vq}}(t)\equiv{q}_{i}^\mathrm{max}(t)\in[ \ubar{q}_{i}^\mathrm{max},\bar{q}_{i}^\mathrm{\,max}],\, \forall i\in\mathcal{C}_\mathrm{lpf}^\mathrm{vq}$) represents the time-varying active (or reactive) power capacity limit of device~$i$. The LPF DC gains are given by the optimal solution of~\eqref{eq:resource_allocation}, i.e., for all $i \in \mathcal{C}_\mathrm{lpf}^k$ and $k\in\{\mathrm{fp},\mathrm{vq}\}$, we get \begin{align}\label{eq:time_var_DC_gain}
 \theta_{i}^k(t)=\left(1-\textstyle\sum_{j\in\mathcal{N}}m_{j}^k(s=0) \right) \tfrac{{y}_{i}^{\mathrm{max},k}(t)}{\textstyle\sum_{l\in\mathcal{C}_\mathrm{lpf}^k} {y}_{l}^{\mathrm{max},k}(t)},
 \end{align}
where the quantity in the parentheses in~\eqref{eq:time_var_DC_gain} is the contribution to the DC gain coming from the non-controllable devices. Obviously, the LPF DC gains in~\eqref{eq:time_var_DC_gain} are bounded on an interval $\theta_i^k(t)\in [ \ubar{\theta}_{i}^k,\bar{\theta}_{i}^k]$, where the lower and upper bounds depend on the lower and upper power capacity limits.
\renewcommand{\arraystretch}{1.2}
\begin{table}[t!]\scriptsize
    \centering
    \caption{List of notation for the disaggregation via ADPMs.}
            \vspace{-1mm}
    \begin{tabular}{c||c}
     \toprule
         Description & Symbol  \\ \hline
         Control channel index &\hspace{-1mm} $k\in\{\mathrm{fp,vq}\}$\hspace{-1mm} \\\hline
         ADPM of device $i$ & $M_i(s)$ \\
         ADPF of device $i$ for channel $k$ & $m_i^k(s)$ \\\hline
         Closed-loop transfer function of device $i$ for channel $k$& $T_i^k(s)$\\
        Desired DVPP transfer function for channel $k$& $T_\mathrm{des}^k(s)$\\\hline
        Time constant of device $i$ for channel $k$& $\tau_i^k$\\
        Time varying DC gain of device $i$ for channel $k$& $\theta_i^k(t)$\\
        Upper/lower DC gain of device $i$ for channel $k$& $\ubar{\theta}_i^k,\,\bar{\theta}_i^k$\\\hline
        \hspace{-2mm}Set of contrl. devices with LPF participation on channel $k$& $\mathcal{C}^k_\mathrm{lpf}$\\
        \hspace{-2mm}Set of communication partners of device $i$ for channel $k$& $\mathcal{I}^k_i$\\\hline
        Time-varying active power capacity limit of device $i$& \hspace{-1mm}${y}_{i}^{\mathrm{max,fp}}\hspace{-0.75mm}=\hspace{-0.5mm}{p}_{i}^\mathrm{max}(t)\hspace{-2mm}$\\
        Lower/Upper active power capacity limit of device $i$& $\ubar{p}_{i}^\mathrm{max},\,\bar{p}_{i}^\mathrm{max}$\\
        Time-varying reactive power capacity limit of device $i$& \hspace{-1mm}${y}_{i}^{\mathrm{max,vq}}\hspace{-0.75mm}=\hspace{-0.5mm}{q}_{i}^\mathrm{max}(t)\hspace{-1.5mm}$\\
        Lower/Upper reactive power capacity limit of device $i$& $\ubar{q}_{i}^\mathrm{max},\,\bar{q}_{i}^\mathrm{max}$\\
        \hspace{-2mm}Time-varying power capacity limit of device $i$ for channel $k\hspace{-1.5mm}$& ${y}_{i}^{\mathrm{max},k}(t)\hspace{-1.5mm}$\\
    \bottomrule
    \end{tabular}
    \label{tab:secIIIA_variables}
\end{table}
\renewcommand{\arraystretch}{1} \normalsize

\begin{figure}[t!]
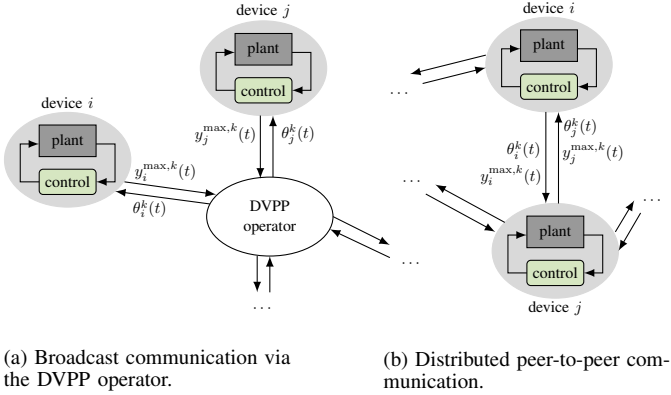

	\begin{subfigure}[h!]{0.21\textwidth}
		\centering
		\include{Figures/communication_strucutre_centralized}
		\vspace{-8mm}
		\caption{Broadcast communication via the DVPP operator.}
		\label{fig:centralized_communication}
	\end{subfigure}
	\hspace{10.5mm}
	\begin{subfigure}[h!]{0.21\textwidth}
			\centering
\include{Figures/communication_structure_distributed}
	\vspace{-8.1mm}
	\caption{Distributed peer-to-peer communication.}
	\label{fig:peer_to_peer_communication}
	\end{subfigure}
	\vspace{-1mm}
	\caption{Different options of communication structures to update the ADPFs.}
\end{figure}

During power system operation, the DC gains in~\eqref{eq:time_var_DC_gain} can be updated in a centralized (broadcast) fashion, where the DVPP operator collects all ${p}_{i}^\mathrm{max}(t), \forall i\in\mathcal{C}_\mathrm{lpf}^\mathrm{fp}$ and ${q}_{i}^\mathrm{max}(t), \forall i\in\mathcal{C}_\mathrm{lpf}^\mathrm{vq}$, and communicates back the solution~\eqref{eq:time_var_DC_gain} for $k\in\{\mathrm{fp},\mathrm{vq}\}$ to all devices (\cref{fig:centralized_communication}). The updates can be communicated either continuously or in an event-triggered fashion. If a distributed implementation is more desired, one could alternatively use the consensus-filters\cite{simpson2015secondary}
\begin{align}\label{eq:consensus_filter}
	\begin{split}
	&\tfrac{d}{dt}\theta_{i}^k(t)=\textstyle\sum_{l\in\mathcal{I}_i^k}\left( \tfrac{\theta_{l}^k(t)}{{y}_{l}^{\mathrm{max},k}(t)}-\tfrac{\theta_{i}^k(t)}{{y}_{i}^{\mathrm{max},k}(t)} \right)\text{ with }\\
	&\textstyle\sum_{i\in\calC_\mathrm{lpf}^k}\theta_{i}^k(0)=\left( 1-\textstyle\sum_{j\in\mathcal{N}}m_{j}^k(0) \right)
		\end{split}
\end{align}
where $\mathcal{I}^k_i$ is the set of communication partners for all $i\in\mathcal{C}_\mathrm{lpf}^k,\,k\in\{\mathrm{fp},\mathrm{vq}\}$ to track the allocation rule~\eqref{eq:time_var_DC_gain} via peer-to-peer communication of $\theta_{i}^k(t)$ and $y_i^{\mathrm{max},k}(t)$ (\cref{fig:peer_to_peer_communication}). Other possible implementations with a reduced amount of communication could be based on adaptive-increase/multiplicative-decrease (AIMD) methods\cite{fan2020optimized,ferraro2018stochastic}, which however, require a time-scale separation and are thus rather slow\cite{aastrom2013adaptive}. A more careful investigation of different update and communication strategies, along with the impact of communication delays (which are not explicitly modelled in this paper) is subject to future work.

Similar to the LPF DC gains, online adaptation could also be applied to all other ADPF parameters.

\subsection{Design of Local Matching Control}\label{sec:local_model_match}
Next, we need to find local feedback controls for the controllable devices $\mathcal{C}$ to ensure their closed-loop transfer matrix $T_i$ satisfies the matching condition~\eqref{eq:control_problem}. Inspired by existing methods on multivariable\cite{huang2020h,kammer2018convex,yang2010robust,chen2021generalized} and adaptive\cite{erfanmanesh2015performance},\cite{muhando11} $\mathcal{H}_\infty$ control of power converters, we address these local matching control designs with a robust and optimal LPV $\mathcal{H}_\infty$ control, which is well-suited to deal with the ADPMs, being parameter-varying with respect to the time-varying LPF DC gains. More specifically, the LPV $\mathcal{H}_\infty$ control is computationally tractable and provides a suboptimality certificate for
the entire parameter space of the LPF DC gains, thereby limiting the performance degradation over all possible operating conditions. Further, we ensure transient time-domain limits of the devices are not violated during normal operating conditions. 

\subsubsection*{Tolerating Mismatch}\label{sec:model_mismatch_high_freq}
We first relax the restrictions on the matching control and therefore modify the participation condition in~\eqref{eq:sum_part_fact_one} by allowing a mismatch in the high frequency range of the Bode plot. In particular, since the measurement unit (e.g. a phase-locked loop (PLL)) for the bus frequency and voltage measurement is limited by some bandwidth $1/\tau_\mathrm{c}$, it suffices if the participation condition in~\eqref{eq:sum_part_fact_one} only holds for the frequency range below, i.e., 
\begin{align}\label{eq:sum_part_one_relaxed}
	\begin{split}
		\textstyle\sum_{i\in\mathcal{N}\cup\mathcal{C}}m_{i}^k(s)\stackrel{!}{=}\tfrac{1}{\tau_\mathrm{c}s+1},\quad k\in\{\mathrm{fp},\mathrm{vq} \}.
	\end{split}
\end{align}
Consequently, line 9 in Algorithm 1 is adjusted as 
\begin{align}\label{eq:line9relaxed}
m_{n}^{k}\leftarrow \left( \tfrac{1}{\tau_{\mathrm{c}}s+1}-\textstyle\sum_{i=1}^{n-1}m_{i}^{k}(s) \right),\quad k\in\{\mathrm{fp},\mathrm{vq}\}.
\end{align}

To simplify notation, we will drop the device index $i\in\calC$ in the following, as the remainder of this subsection refers to the local matching control of one controllable device $i$.

\subsubsection*{LPV $\calH_\infty$ matching control} 
\begin{figure}[b!]
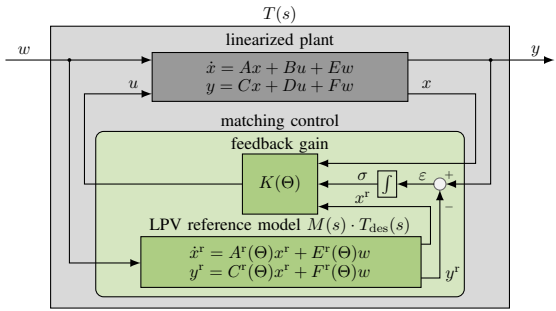

	\centering
	\include{Figures/Local_matching_ctrl}
	\vspace{-8mm}
	\caption{Setup for matching control design of device $i\in\mathcal{C}$.}
	\label{fig:local_matching_ctrl}
\end{figure}

For each controllable device, we attempt to find a matching controller where the specification in~\eqref{eq:control_problem} serves as a local reference model to be matched. Let $y:=[\Delta p,\,\Delta q]'$ and $w=:[\Delta f,\, \Delta v]'$. The control design setup is shown in~\cref{fig:local_matching_ctrl}, where 
\begin{align}\label{eq:ss_plant}
	\begin{split}
\Dot{x}&=Ax+Bu+Ew\\
y&=Cx+Du+Fw
\end{split}
\end{align}
is the linearized reduced-order system of the plant (e.g., representing the primary source technology and/or the associated grid-side converter interface, see~\cref{sec:test_case}), and
\begin{align}\label{eq:LPV_ss_reference}
	\begin{split}
\Dot{x}^{\mathrm{r}}&=A^{\mathrm{r}}(\Theta(t))x^{\mathrm{r}}+E^{\mathrm{r}}(\Theta(t))w\\
y^\mathrm{r}&=C^{\mathrm{r}}(\Theta(t))x^{\mathrm{r}}+F^{\mathrm{r}}(\Theta(t))w,
\end{split}
\end{align}
is the LPV system representation of the local reference model $M\cdot T_\mathrm{des}$, which is included into the control setup, but not a physical part of the system itself. The state-space matrices in~\eqref{eq:LPV_ss_reference} depend affinely on the time-varying vector of parameters $\Theta(t)\in\mathbb{R}^{\mathrm{log}_2(q)}$, which, according to the selected ADPM of the device, is composed of the different DC gain parameters $\theta_{i}^k(t),\,i\in\calC_\mathrm{lpf}^k, k\in\{\mathrm{fp},\mathrm{vq}\}$. Since the latter are varying on the interval $[ \ubar{\theta}_{i}^k,\bar{\theta}_{i}^k]$ for all $t\geq 0$ (cf.~\eqref{eq:time_var_DC_gain}), the vector $\Theta(t)$ ranges over a fixed polytope of vertices $\hat{\Theta}^{(1)},...,\hat{\Theta}^{(q)}$.

\begin{table*}
	\begin{subequations}
		\label{eq:R}
		\begin{align}\label{eq:R_l}
			\begin{bmatrix}
				\calA^{(l)}\calQ\hspace{-0.5mm}+\hspace{-0.5mm}\calQ{\calA^{(l)}}'\hspace{-0.5mm}+\hspace{-0.5mm}\calB^{(l)}\calY^{(l)}\hspace{-0.5mm}+\hspace{-0.5mm}{\calY^{(l)}}'\hspace{-0.5mm}{\calB^{(l)}}'\hspace{-0.5mm}&\hspace{-0.5mm}\star\hspace{-0.5mm}&\hspace{-0.5mm}\star \\
				{{\mathcal{E}}^{(l)'}}\hspace{-0.5mm}&\hspace{-0.5mm}{-}\gamma I\hspace{-0.5mm}&\hspace{-0.5mm}\star \\
				\calC^{(l)}\calQ\hspace{-0.5mm}+\hspace{-0.5mm}\mathcal{D}^{(l)}\calY^{(l)}&{\mathcal{F}}^{(l)}&{-}\gamma I
			\end{bmatrix}\hspace{-1mm}&\prec 0,\quad\,\,\, l=1,...,q\\ \label{eq:R_lt}
			\begin{bmatrix}
				(\calA^{(l)}\hspace{-0.5mm}+\hspace{-0.5mm}\calA^{(t)})\calQ\hspace{-0.5mm}+\hspace{-0.5mm}\calQ(\calA^{(l)}\hspace{-0.5mm}+\hspace{-0.5mm}\calA^{(t)})'\hspace{-0.5mm}+\hspace{-0.5mm}\calB^{(l)}\calY^{(t)}\hspace{-0.5mm}+\hspace{-0.5mm}\calB^{(t)}\calY^{(l)}\hspace{-0.5mm}+\hspace{-0.5mm}{\calY^{(l)}}'\hspace{-0.5mm}{\calB^{(t)}}'\hspace{-0.5mm}+\hspace{-0.5mm}{\calY^{(t)}}'\hspace{-0.5mm}{\calB^{(l)}}'\hspace{-0.5mm}&\hspace{-0.5mm}\star\hspace{-0.5mm}&\hspace{-0.5mm}\star \\
				{{\mathcal{E}}^{(l)'}}\hspace{-0.5mm}+\hspace{-0.5mm}{{\calE}^{(t)'}}\hspace{-0.5mm}&\hspace{-0.5mm}{-}2\gamma I\hspace{-0.5mm}&\hspace{-0.5mm}\star \\
				(\calC^{(l)}\hspace{-0.5mm}+\hspace{-0.5mm}\calC^{(t)})\calQ\hspace{-0.5mm}+\hspace{-0.5mm}\mathcal{D}^{(l)}\calY^{(t)}\hspace{-0.5mm}+\hspace{-0.5mm}\mathcal{D}^{(t)}\calY^{(l)}\hspace{-0.5mm}&\hspace{-0.5mm}{\mathcal{F}}^{(l)}\hspace{-0.5mm}+\hspace{-0.5mm}{\mathcal{F}}^{(t)}\hspace{-0.5mm}&\hspace{-0.5mm}{-}2\gamma I
			\end{bmatrix}\hspace{-1mm}&\prec 0,\quad \begin{array}{c} l=1,...,q\hspace{-0.5mm}-\hspace{-0.5mm}1\\t=l\hspace{-0.5mm}+\hspace{-0.5mm}1,...,q\end{array}
		\end{align}
	\end{subequations}
	\hrulefill
\vspace{-5mm}
\end{table*}

We combine~\eqref{eq:ss_plant} and~\eqref{eq:LPV_ss_reference}, and add an integral state~$\dot{\sigma}=~\varepsilon$ of the matching error $\varepsilon:=y-y^\mathrm{r}$ as a feedback signal to robustify the matching design against stationary model uncertainties and nonlinearities of the actual device. This yields the augmented LPV system used for control design
\begin{align}\label{eq:augmented_LPV}
\begin{split}
	\underset{\dot{z}}{\underbrace{\begin{bmatrix}\Dot{x}\\\Dot{x}^{\mathrm{r}}\\\Dot{\sigma}\end{bmatrix}}}
	\hspace{-1.5mm}&=\hspace{-1.5mm}\underset{\mathcal{A}(\Theta)}{\underbrace{\begin{bmatrix}A\hspace{-1mm}&\hspace{-1mm}\displaystyle{O}\hspace{-1mm}&\hspace{-1mm}\displaystyle{O}\\\displaystyle{O}\hspace{-1mm}&\hspace{-1mm}A^{\mathrm{r}}(\hspace{-0.1mm}\Theta\hspace{-0.1mm})\hspace{-1mm}&\hspace{-1mm}\displaystyle{O}\\C\hspace{-1mm}&\hspace{-1mm}-C^{\mathrm{r}}(\hspace{-0.1mm}\Theta\hspace{-0.1mm})\hspace{-1mm}&\hspace{-1mm}\displaystyle{O}\end{bmatrix}}}
	\underset{z}{\underbrace{\begin{bmatrix}{x}\\{x}^{\mathrm{r}}\\\sigma\end{bmatrix}}}
	\hspace{-1.5mm}+\hspace{-1.5mm}\underset{\mathcal{B}(\Theta)}{\underbrace{\begin{bmatrix}B\\\displaystyle{O}\\D\end{bmatrix}}}\hspace{-0.7mm}u\hspace{-0.5mm} +\hspace{-1.4mm}\underset{{\mathcal{E}}(\Theta)}{\underbrace{\begin{bmatrix}E\\E^{\mathrm{r}}(\hspace{-0.1mm}\Theta\hspace{-0.1mm})\\F\hspace{-0.7mm}-\hspace{-0.7mm}F^{\mathrm{r}}(\hspace{-0.1mm}\Theta\hspace{-0.1mm})\end{bmatrix}}}\hspace{-0.7mm}w\\
	\varepsilon&=\hspace{-1mm}\underset{\mathcal{C}(\Theta)}{\underbrace{\begin{bmatrix}C\hspace{-1mm}&\hspace{-1mm} -C^{\mathrm{r}}(\hspace{-0.1mm}\Theta\hspace{-0.1mm})\hspace{-1mm}&\hspace{-1mm}\displaystyle{O}\end{bmatrix}}}	\underset{z}{\underbrace{\begin{bmatrix}{x}\\{x}^{\mathrm{r}}\\\sigma\end{bmatrix}}}\hspace{-1.5mm}+\hspace{-1.5mm}\underset{\mathcal{D}(\Theta)}{\underbrace{\begin{bmatrix}D\end{bmatrix}}}\hspace{-0.5mm}u\hspace{-0.5mm}+\hspace{-1mm}\underset{\mathcal{F}(\Theta)}{\underbrace{\begin{bmatrix}F\hspace{-0.7mm}-\hspace{-0.7mm}F^{\mathrm{r}}(\hspace{-0.1mm}\Theta\hspace{-0.1mm})\end{bmatrix}}}w,
	\end{split}
\end{align}
where $\displaystyle{O}$ is the zero matrix with appropriate dimension. The associated static control law is given by $u(t)=K(\Theta(t))z(t)$, where the state-feedback gain $K(\Theta)$ is parameter-dependent, and the resulting (open-circuit) closed-loop system is given as
\begin{align}\label{eq:aug_LPV_cl}
\begin{split}
	\dot{z}&=\underset{\mathcal{A}_{\mathrm{cl}}(\Theta)}{\underbrace{(\mathcal{A}(\Theta)+\mathcal{B}(\Theta)K(\Theta))}}z+{\mathcal{E}}(\Theta)w\\
	\varepsilon&=\underset{\mathcal{C}_{\mathrm{cl}}(\Theta)}{\underbrace{(\mathcal{C}(\Theta)+\mathcal{D}(\Theta)K(\Theta))}}z+{\mathcal{F}}(\Theta)w.
	\end{split}
\end{align}

\begin{figure}[b!]
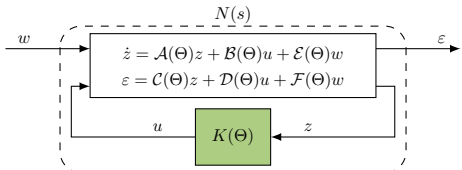

	\vspace{-3mm}
	\centering
	\include{Figures/H_inf_matching_ctrl}
	\vspace{-8mm}
	\caption{$\mathcal{H}_\infty$ control design configuration for model-matching.}
	\label{fig:H_inf_matching_ctrl}
\end{figure}
Under the configuration in~\cref{fig:H_inf_matching_ctrl}, the parameter-dependent state-feedback gain $K(\Theta(t))$ is obtained by minimizing the matching error $\varepsilon(s)=N(s)w(s)$ in the $\mathcal{H}_\infty$-norm as
\begin{subequations}
\label{eq:H_inf_opt_problem_unconstr}
\begin{align}
\underset{K(\Theta)}{\text{minimize}}\quad&\gamma\\\label{eq:H_inf_opt_problem_unconstr2}
\text{subject to}\quad&||N(s)||_\infty<\gamma,
\end{align}
\end{subequations}
where $N(s)$ is the transfer function obtained from~\cref{eq:aug_LPV_cl}. As a common practice in $\mathcal{H}_\infty$ control, the $\mathcal{H}_\infty$-bound in~\eqref{eq:H_inf_opt_problem_unconstr2} can be reformulated in terms of a linear matrix inequality (LMI) to obtain a tractable convex optimization problem that can be solved efficiently. Namely, by virtue of the Bounded Real Lemma (BRL)\cite{scherer1990riccati}, the $\mathcal{H}_\infty$ bound in~\eqref{eq:H_inf_opt_problem_unconstr2} is equivalent to the LMI ($\star$ indicates symmetric blocks)
\begin{align}\label{eq:BRL}
	\begin{split}
	\begin{bmatrix}
\calQ\mathcal{A}_{\mathrm{cl}}(\Theta)'+\mathcal{A}_{\mathrm{cl}}(\Theta)\calQ&\star&\star\\
{\mathcal{E}}(\Theta)'&-\gamma I& \star\\
\mathcal{C}_{\mathrm{cl}}(\Theta)\calQ&{\mathcal{F}}(\Theta)&-\gamma I
\end{bmatrix}\prec 0,
\end{split}
\end{align}
ensuring (open-circuit) closed-loop stability of the system in~\eqref{eq:aug_LPV_cl}, if and only if there exists a symmetric positive definite solution $\calQ\hspace{-0.64mm}=\hspace{-0.64mm}\calQ'\hspace{-0.68mm}\succ\hspace{-0.68mm}0$ for all admissible values of $\Theta(t)$\cite{montagner2005lmi,apkarian1995self}.

\subsubsection*{Parametric State-Feedback Gain} The key challenge of~\eqref{eq:BRL} is the infinite number of constraints it imposes, as we consider the entire polytope of possible parameter values of $\Theta(t)$. For polytopic LPV systems, however,~\eqref{eq:BRL} can be reduced to a finite set of constraints\cite{apkarian1995self,montagner2005lmi}. In particular, since the state-space matrices in~\eqref{eq:augmented_LPV} depend affinely on $\Theta(t)$, and $\Theta(t)$ varies in a polytope of vertices $\hat{\Theta}^{(1)},...,\hat{\Theta}^{(q)}$, the state-space matrices are constrained on the polytope with vertices
\begin{align}\label{eq:matrix_vertices}
	\begin{split}
\left(\hspace{-0.5mm}\mathcal{A}^{(l)}\hspace{-1.2mm}, \mathcal{B}^{(l)}\hspace{-1.2mm},{\mathcal{E}}^{(l)}\hspace{-1.2mm},\mathcal{C}^{(l)}\hspace{-1.2mm},\mathcal{D}^{(l)}\hspace{-1.2mm},{\mathcal{F}}^{(l)}\hspace{-0.65mm}\right)\hspace{-0.5mm}:=\hspace{-0.5mm}\left(\hspace{-0.1mm}\mathcal{A},\mathcal{B},{\mathcal{E}},\mathcal{C},\mathcal{D},{\mathcal{F}}\hspace{-0.1mm}\right)\hspace{-0.5mm}\left(\hspace{-0.5mm}\hat{\Theta}^{(l)}\hspace{-0.65mm}\right)
\end{split}
\end{align}
for $l=1,...,q$. Given the latter, it can be shown that~\eqref{eq:BRL} will hold for all $\Theta(t)$ if there exist a symmetric positive definite matrix $\calQ$ and matrices $\calY^{(l)}$ for $l=1,...,q$, such that the vertex LMIs in~\eqref{eq:R} are satisfied\cite{montagner2005lmi}. In this regard, we can transform the infinite-dimensional optimization problem in~\eqref{eq:H_inf_opt_problem_unconstr} into a convex finite-dimensional optimization problem
\begin{subequations}
\label{eq:LMI_H_inf_opt_problem_unconstr_vertices}
\begin{align}
\underset{\calQ,\calY^{(1)},...\calY^{(q)}}{\text{minimize}}\quad&\gamma\\
\text{subject to}\quad&\calQ=\calQ'\succ0\\
&\eqref{eq:R_l}\,\&\, \eqref{eq:R_lt},
\end{align}
\end{subequations}
to compute the vertex controllers $K^{(l)}:=\mathcal{Y}^{(l)}\mathcal{Q}^{-1}$. The associated state-feedback control gain is then obtained as a convex combination of the vertex controllers, with convex coefficients corresponding to the instantaneous $\Theta(t)$. I.e.,
\begin{align}\label{eq:LPV_ctrl}
	K(\Theta(t))\hspace{-0.75mm}=\hspace{-0.75mm}\textstyle\sum_{l=1}^{q}\hspace{-0.5mm}\lambda^{(l)}(\Theta)K^{(l)}\hspace{-1mm},\,\,\,\lambda^{(l)}\hspace{-0.75mm}\geq\hspace{-0.5mm} 0,\,\,\,\textstyle\sum_{l=1}^{q}\hspace{-0.5mm}\lambda^{(l)}\hspace{-0.75mm}=\hspace{-0.75mm}1,
\end{align}
where, for a given $\Theta(t)$ during power system operation, the coefficients $\lambda^{(l)}(\Theta)$ of the convex combination of the vertices are computed such that\cite{montagner2005lmi}
\begin{align*}
\textstyle\sum_{l=1}^{q}\lambda^{(l)}(\Theta)\hat{\Theta}^{(l)}=\Theta(t),\,\,\, \lambda^{(l)}(\Theta)\geq 0,\,\,\,\textstyle\sum_{l=1}^{q}\lambda^{(l)}(\Theta)=1. 
\end{align*}
For the parameter polytopes we are considering in our applications, i.e., simplices and parallelotopes, there exist analytical closed form-expressions of the coefficients $\lambda^{(l)}(\Theta)$\cite{schurmann2016closed}.

\subsubsection*{Controller Tuning} To ensure accurate model matching, the $\mathcal{H}_\infty$ design in~\cref{eq:LMI_H_inf_opt_problem_unconstr_vertices} generally selects large vertex control gains $K^{(l)}$. This results in an overly aggressive controller which is not favorable in the presence of nonlinearities, state and input constraints (e.g. transient current limits of a converter.), or during off-steady-state conditions.

We therefore regularize the problem in~\eqref{eq:LMI_H_inf_opt_problem_unconstr_vertices} by incorporating state and input constraints of the system in~\cref{eq:augmented_LPV} into control design. To do so, we employ an idealized method for design purposes, i.e., we do not aim to represent actual state and input constraints, but rather provide a tool, which, after tuning parameters, can lead to these constraints being satisfied. 

The tuning method for state and input constraint satisfaction is based on the reformulation of these constraints in terms of LMIs, which can immediately be included into the problem in~\eqref{eq:LMI_H_inf_opt_problem_unconstr_vertices}. Namely, as derived in\cite{boyd1994linear,chen2002constrained}, for $w$ limited in energy and $z(0)=0$, the input and state constraints $\max_{t\geq 0}||u(t)||\leq\mu$ and $\max_{t\geq 0}|z_{j}(t)|\leq\zeta_{j}$ for some states $j\in\{1,...,v\}$, are enforced for all $t\geq 0$, if $\calQ$ and $\calY^{(l)}$ satisfy
\begin{subequations}\label{eq:ellipsoidal}
\vspace{-3.5mm}
	\begin{align}\label{eq:ellipsoidal_u}
	\begin{bmatrix}\mathcal{Q}&{\calY^{(l)}}'\\\calY^{(l)}&\tfrac{\mu^2}{\alpha}I\end{bmatrix}&\succeq 0,\quad l=1,...,q, \\ \label{eq:ellipsoidal_x}
	\begin{bmatrix}\tfrac{1}{\alpha}\mathrm{diag}(\zeta_{j}^2)&Z\calQ\\(Z\calQ)'&\calQ\end{bmatrix}&\succeq 0,\quad j\in\{1,...,v\},
	\end{align}
\end{subequations}
where $Z=[Z_1' \cdots Z_v']'$ and $Z_j$ is a row vector of zeros with a '1' at position $j$, $||\cdot||$ is the Euclidean norm, and $\mu,\zeta_{j},\alpha\in\bbR$ are tuning parameters to handle several performance requirements (see\cite{boyd1994linear,chen2002constrained} for details). In this regard, we also implicitly tune the integral gain of the controller by adjusting the bound $\zeta_{j}$ for the state $\sigma$.

We include the LMIs in~\eqref{eq:ellipsoidal} into the problem in~\eqref{eq:LMI_H_inf_opt_problem_unconstr_vertices} and arrive at the final $\calH_\infty$ model-matching problem
\begin{subequations}
\label{eq:LMI_H_inf_opt_problem_vertices_final}
	\begin{align}
		\underset{\calQ,\calY^{(1)},...\calY^{(q)}}{\text{minimize}}\quad&\gamma\\
		\text{subject to}\quad&\calQ=\calQ'\succ0\\
		&\eqref{eq:R_l}\,\&\, \eqref{eq:R_lt},\\
		&\eqref{eq:ellipsoidal_u}\,\&\, \eqref{eq:ellipsoidal_x},
			\end{align}
\end{subequations}
where the resulting vertex controllers $K^{(l)}\hspace{-0.5mm},l\hspace{-0.5mm}=\hspace{-0.5mm}1,...,q$, are used to compute the LPV feedback gain $K(\Theta(t))$ online via~\eqref{eq:LPV_ctrl}.

\section{Test Case}\label{sec:test_case}
To verify our DVPP controls, we use Simscape Electrical to perform an electromagnetic transients (EMT) simulation based on the IEEE nine-bus system\footnote{The MATLAB/Simulink implementation is available online\cite{simulink_model}.}. In particular, to demonstrate the basic idea of our DVPP control strategy in an instructive way, we consider a deliberately simple test system assembled with DVPPs containing only a few devices. Our proposed method, however, can be easily extended to larger power systems comprising DVPPs with a larger number of devices, especially since our local matching control is independent of the size of the power system and the number of DVPP devices. An investigation of such scenarios will be part of future work.

In a first case study, we start with a tutorial example of a DVPP composed of only weather-independent DERs, specified to improve the fast frequency response of the initial system in\cite{anderson} in a non-adaptive fashion. In a second case study, we investigate a multivariable DVPP setup replacing the fast frequency and voltage control of a thermal-based generator, while additionally including online-adaptation of the ADPMs to handle temporal variability of weather-dependent DERs. Finally, in a third case study, we conceptually demonstrate the benefits of our DVPP control strategy based on ADPFs over competing approaches to DVPP control in\cite{joak,zhong2021coordinated}, relying on dynamic, or even only static participation factors.

\renewcommand{\arraystretch}{1.2}
\begin{table}[t!]\scriptsize
    \centering
     \caption{IEEE nine-bus system parameters.}
     \vspace{-1mm}
    \begin{tabular}{c||c}
     \toprule
         Parameter & Value  \\ \hline
         System base power &  $100\,\text{MVA}$ \\
         System base voltage (ph-ph, rms)  & $230\,\text{kV}$\\
         System base frequency & $50\,\text{Hz}$ \\ \hline
         Power rating, SG1 & $250\,\text{MVA}$ \\
         Power rating, SG2 & $96\,\text{MVA}$\\
         Power rating, SG3 & $64\,\text{MVA}$\\ \hline
         Voltage rating, SG1 (ph-ph, rms) & $16.5\,\text{kV}$ \\
         Voltage rating, SG2 (ph-ph, rms) & $18\,\text{kV}$\\
         Voltage rating, SG3 (ph-ph, rms) & $13.8\,\text{kV}$\\
    \bottomrule
    \end{tabular}
    \label{tab:9bus_parameters}
\end{table}
\renewcommand{\arraystretch}{1} \normalsize

\begin{figure}[b!]
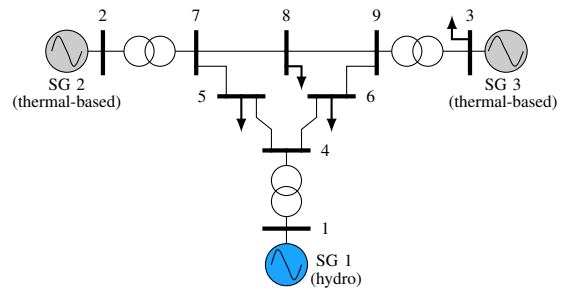

	\centering
	\include{Figures/9bus_system}
	\vspace{-8mm}
	\caption{IEEE nine-bus system with two thermal-based power plants and one hydro power plant\cite{anderson} (SG = synchronous generator). }
	\label{fig:9bus}
\end{figure}

\subsection{System Model}
As in the nine-bus system in\cite{anderson}, we consider two conventional thermal-based power plants and one hydro power plant for our initial test case (\cref{fig:9bus} and \cref{tab:9bus_parameters}). The implementation is based on the system model in\cite{tayyebi2020frequency}, where the transmission lines are modelled via nominal $\pi$ sections, and the transformers via three-phase linear transformer models. The loads are modelled as constant impedance loads. We adopt an 8th-order model for the synchronous machines equipped with a ST1A excitation system with built-in automatic voltage regulator (AVR) and a power system stabilizer (PSS)\cite{tayyebi2020frequency}. The governors are modelled as a proportional speed-droop control with first-order delay, and the hydro and steam turbine parameters are taken from\cite{kundur2007power}.

\subsection{Grid-Side Converter Model and Control Architecture}
\begin{figure}[t!]
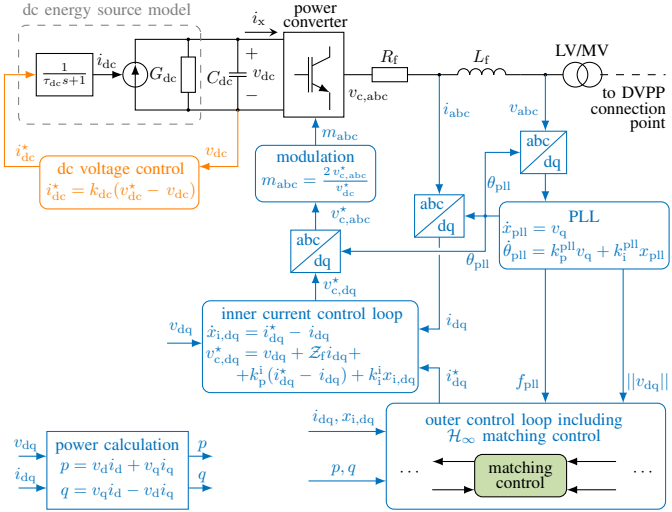

	\centering
	\include{Figures/Conv_model_extended}
	\vspace{-8mm}
	\caption{Converter model in per unit, where $\mathcal{Z}_\mathrm{f}=(\tfrac{L_\mathrm{f}\omega_\mathrm{pll}}{\omega_\mathrm{b}}\mathcal{J}_2+R_\mathrm{f}\mathcal{I}_2)$ with $\mathcal{J}_2=[0\,\text{-}1;\,1\,0]$ and $\mathcal{I}_2=[1\,0;\,0\,1]$.}
	\label{fig:conv_model}
\end{figure}

All DVPP case studies include (among others) converter-based generators which are interfaced to the grid via power converters, and thus considered as controllable ($\in\calC$) within their respective DVPP control setup. The proposed grid-side converter model used for dynamic simulation represents an aggregation of multiple commercial converter modules, and is based on a state-of-the-art converter control scheme\cite{yazdani2010voltage}, into which we have incorporated the $\calH_\infty$ matching control (\cref{fig:conv_model}). Namely, we employ a grid-supporting control scheme that is synchronized with the grid voltage and contributes to the regulation of the grid frequency and voltage according to the local desired DVPP specifications $M_i\cdot T_\mathrm{des}$, respectively. 

Similar to\cite{tayyebi2020frequency}, we assume that the dc current $i_\mathrm{dc}$ is supplied by a controllable dc current source (\cref{fig:conv_model}), e.g. representing the machine-side converter of a wind power plant, a PV system, or an energy storage. In particular, we use a coarse-grain model of the underlying primary source technology and model its response time by a first-order delay with time constant $\tau_\mathrm{dc}$\cite{tayyebi2020frequency}, e.g. representing the resource associated dynamics, communication delays and/or actuation delays.

The grid-side converter control is separated into two control loops for the dc and the ac side (\cref{fig:conv_model}). The dc-side control regulates the dc voltage through the dc current source and a proportional controller\cite{tayyebi2020frequency}. The ac-side control is used to control the network current magnitudes. It is implemented in a $\mathrm{dq}$-coordinate frame oriented via a phase-locked loop (PLL) which tracks the system frequency after the $RL$-filter, keeping the converter synchronized with the grid voltage\cite{yazdani2010voltage}. The ac-side control is composed of a cascaded control loop, where the outer loop, containing the $\calH_\infty$ matching control, provides the ac current reference $i^\star_\mathrm{dq}$ to the inner current control loop\footnote{The proposed $\mathcal{H}_\infty$ matching control scheme can also be adapted to other type of converter controls, e.g. more classic versions where the grid-side converter is regulating the dc voltage.}. The converter and control parameters are given in~\cref{tab:converter_parameters}.

\renewcommand{\arraystretch}{1.2}
\begin{table}[t!]\scriptsize
    \centering
    \caption{Converter model and control parameters.}
        \vspace{-1mm}
    \begin{tabular}{c||c|c}
     \toprule
         Parameter & Symbol & Value  \\ \hline
         dc link capacitor & $C_\mathrm{dc}$ & $0.096\,\text{pu}$ \\
         dc link conductance & $G_\mathrm{dc}$ & $0.05\,\text{pu}$ \\ \hline
         $RL$-filter resistance & $R_\mathrm{f}$ & $0.01\,\text{pu}$ \\
         $RL$-filter inductance & $L_\mathrm{f}$ & $0.0942\,\text{pu}$ \\\hline
         dc voltage control gain & $k_\mathrm{dc}$ & $100$ \\
         PLL control gains & $k_\mathrm{p}^\mathrm{pll},\,k_\mathrm{i}^\mathrm{pll}$ & $0.4775,\,5.4113$ \\
         Current control gains & $k_\mathrm{p}^\mathrm{i},\,k_\mathrm{i}^\mathrm{i}$ & $0.73,\,1.19$ \\
         Reactive power control gains & $k_\mathrm{p}^\mathrm{q},\,k_\mathrm{i}^\mathrm{q}$ & $0.005,\,0.0005$\\
    \bottomrule
    \end{tabular}
        \caption*{\scriptsize Converter rated at $S_\mathrm{r}$, $v_\mathrm{r}\hspace{-0.5mm}=\hspace{-0.5mm}\sqrt{2/3}\,\text{kV}$ (ph-n, peak), and $v_\mathrm{dc}^\star=3 v_\mathrm{r}$. \normalsize}
        \vspace{-3mm}
        \caption*{\scriptsize Parameters in per unit, normalized w.r.t. to the dc- and ac-side ratings, respectively.\normalsize}
    \label{tab:converter_parameters}
\end{table}
\renewcommand{\arraystretch}{1} \normalsize

\subsection{Case Study I: Supplementing Hydro in Frequency Response}\label{sec:case1}
\begin{figure}[b!]
	\begin{subfigure}[h!]{0.2\textwidth}
		\centering
		\include{Figures/9bus_system_C1}
		\vspace{-5mm}
		\caption{\textit{Case study I:} IEEE nine-bus system with a DVPP at bus~1.}
		\label{fig:9bus_DVPP_hydro}
	\end{subfigure}
	\hspace{3.5mm}
	\begin{subfigure}[h!]{0.21\textwidth}
			\centering
	\scalebox{0.465}{\includegraphics[]{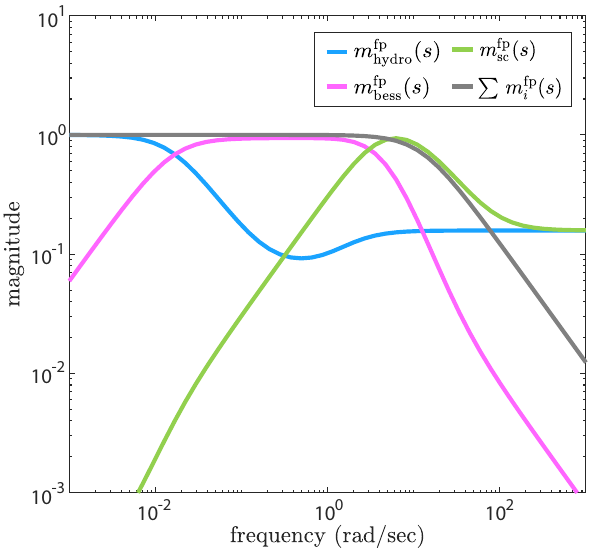}}
	\vspace{-4mm}
	\caption{Magnitude Bode plots of the ADPFs of the DVPP~1 devices.}
	\label{fig:bode_DVPP1}
	\end{subfigure}
	\vspace{-1mm}
	\caption{System model and ADPFs of case study I.}
\end{figure}

The initial system (\cref{fig:9bus}) is characterized by a poor short-term frequency response, caused by the transient droop compensation and the non-minimum phase zero of the hydro turbine\cite{kundur2007power}. As proposed in\cite{ghasemi2019investigation,saarinen2016full}, this poor response behavior can be compensated by a battery energy storage system (BESS) connected to the same bus. In addition, we complement the hydro turbine and BESS by supercapacitor ($\mathrm{sc}$) for fast frequency response, as in hybrid energy storage systems\cite{li2008power}. This gives us a DVPP at bus~1 (\cref{fig:9bus_DVPP_hydro}), for which we specify an aggregate frequency response behavior identical to a proportional $\mathrm{f}$-$\mathrm{p}$ droop control, i.e.,
\begin{align}\label{eq:T_des_DVPP1}
	\Delta p(s) = T_\mathrm{des}(s) \Delta f(s),\quad T_\mathrm{des}(s)\hspace{-0.5mm}:=\tfrac{-D}{\tau s+1},
\end{align}
where $D$ is the desired droop coefficient, and the denominator with $\tau$ is included to filter out high frequency dynamics. The parameter values are provided in~\cref{tab:DVPP1_parameters}. The voltage control at bus~1 is fully provided by the installed AVR of the hydro plant and not part of the DVPP control. Thus,~\cref{eq:T_des_DVPP1} represents a one-dimensional version of the aggregate specification in~\cref{eq:MIMO_spec}.

To provide each DVPP device with the DVPP input signal, given by the bus frequency deviation at bus 1, we use individual bus measurements of the devices (instead of one common bus measurement), assuming that independent measurements are sufficiently similar.

\subsubsection*{ADPF Selection}\label{sec:DVPP1}
Since the aggregate specification of DVPP~1 in~\eqref{eq:T_des_DVPP1} is a one-dimensional transfer function, the ADPMs of the devices reduce to the ADPFs for $\mathrm{f}$-$\mathrm{p}$ control, respectively, i.e., $M_i(s)=m_i^\mathrm{fp}(s),\,i\in\{\mathrm{hydro,bess,sc}\}$.

A small-signal model of the hydro governor and the turbine dynamics with input being the rotor frequency deviation and output the mechanical active power deviation is given by\footnote{Since the hydro power plant is naturally grid-forming, we only consider the ``grid-following'' part of the hydro power plant for the DVPP participation, given by the hydro governor and turbine dynamics.}\cite{kundur2007power}
\begin{align}\label{eq:hydro_IO}
	T_\mathrm{hydro}(s)\hspace{-0.5mm}=\hspace{-0.5mm}\underset{\text{speed droop}}{\underbrace{\tfrac{-1/R_\mathrm{g}}{\tau_\mathrm{g}s+1}}}\underset{\text{transient droop}}{\underbrace{\tfrac{\tau_\mathrm{r}s+1}{(R_\mathrm{t}/R_\mathrm{g})\tau_\mathrm{r}s+1}}}\,\,\underset{\text{turbine}}{\underbrace{\tfrac{1-\tau_\mathrm{w}s}{1+0.5\tau_\mathrm{w}s}}},
\end{align}
where the associated parameter values are given in~\cref{tab:DVPP1_parameters}. Since~\eqref{eq:hydro_IO} describes a pre-installed setup of the hydro unit at bus~1, it is considered as fixed ($\in\mathcal{N}$) within the DVPP setup for $\mathrm{f}$-$\mathrm{p}$ control. In particular, for a choice of $1/D=R_\mathrm{g}$ and $\tau=\tau_\mathrm{g}$ in~\eqref{eq:T_des_DVPP1}, the hydro ADPF is given as
\begin{align}\label{eq:hydro_ADPF}
	m^\mathrm{fp}_\mathrm{hydro}(\hspace{-0.2mm}s\hspace{-0.2mm})\hspace{-0.5mm}=\hspace{-0.5mm}T_\mathrm{des}(\hspace{-0.2mm}s\hspace{-0.2mm})^{\text{-}1}T_\mathrm{hydro}(\hspace{-0.2mm}s\hspace{-0.2mm})\hspace{-0.5mm}=\hspace{-0.5mm}\tfrac{\tau_\mathrm{r}s+1}{(R_\mathrm{t}/R_\mathrm{g})\tau_\mathrm{r}s+1}\tfrac{1-\tau_\mathrm{w}s}{1+0.5\tau_\mathrm{w}s}
\end{align}
i.e., by the design choice of~\eqref{eq:T_des_DVPP1}, the hydro unit establishes the full DC gain of DVPP~1, so that $T_\mathrm{hydro}(0)=T_\mathrm{des}(0)$.

The converter-based BESS and supercapacitor are controllable ($\in\mathcal{C}$) and used to complement the hydro response on faster time scales, therefore participating in the specified $\mathrm{f}$-$\mathrm{p}$ control in~\eqref{eq:T_des_DVPP1} as BPF and HPF, respectively. Their ADPFs are selected by Algorithm 1, i.e.,
\begin{align}\label{eq:ADPM_BESS}
	m^\mathrm{fp}_\mathrm{bess}(s) &= \tfrac{1}{\tau_\mathrm{bess}s+1}\left(\tfrac{1}{\tau_\mathrm{bess}s+1}-m^\mathrm{fp}_\mathrm{hydro}(s) \right)\\\label{eq:ADPM_SC}
	m^\mathrm{fp}_\mathrm{sc}(s) &= \tfrac{1}{\tau_\mathrm{c}s+1}-m^\mathrm{fp}_\mathrm{bess}(s)-m^\mathrm{fp}_\mathrm{hydro}(s),
\end{align}
where the magnitude Bode plots are shown in~\cref{fig:bode_DVPP1} and the parameter values are provided in~\cref{tab:DVPP1_parameters}. The time constant $\tau_\mathrm{bess}$ corresponds to the dc time constant in the associated converter model in~\cref{fig:conv_model}, representing actuation delays of the BESS technology. In contrast to the hydro unit, the frequency measurements for the BESS and the supercapacitor are given by their respective PLL. Therefore, the HPF ADPF of the supercapacitor is cut at the PLL-bandwidth $1/\tau_\mathrm{c} < 1/\tau_\mathrm{sc}$ according to the relaxation in~\eqref{eq:line9relaxed}. 

Since $m^\mathrm{fp}_\mathrm{hydro}$ is non-adaptive by definition and establishes the full DC gain of DVPP~1, all ADPFs in~\cref{eq:hydro_ADPF,eq:ADPM_BESS,eq:ADPM_SC} are free of any adaptive DC gain parameters, which is in accordance with the weather-independence of all three devices. 

\subsubsection*{Local matching control}
\begin{figure}[t!]
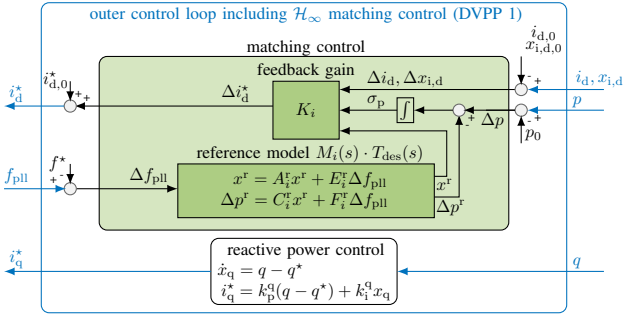

	\centering
	\include{Figures/Conv_matching_DVPP1}
	\vspace{-8mm}
	\caption{Outer control loop with matching control for DVPP~1 converters, where $K_i,\,i\in\{\mathrm{bess,sc}\}$ is provided with the states of the nonlinear converter model in~\cref{fig:conv_model}. The reactive power set point is regulated with a PI controller.}
	\label{fig:outer_ctrl_DVPP1}
\end{figure}

To match the local closed-loop dynamics of the converter-based BESS and supercapacitor with their local reference model $M_i\cdot T_\mathrm{des},\,i\in\{\mathrm{bess,sc}\}$, respectively, we employ the previously introduced $\mathcal{H}_\infty$ matching control in the outer control loop of their respective grid-side converter (\cref{fig:conv_model}). As shown in~\cref{fig:outer_ctrl_DVPP1}, the matching control is defined by the local reference model $M_i\cdot~T_\mathrm{des},\,i\in\{\mathrm{bess,sc}\}$ and the associated state-feedback gain $K_i$. In particular, since the ADPFs of the BESS and the supercapacitor are fixed transfer functions (and with that their reference models $M_i\cdot T_\mathrm{des}$), the matching control design simplifies in such a way, that the respective optimization problem in~\eqref{eq:LMI_H_inf_opt_problem_vertices_final} has to be solved for a parameter polytope consisting of one vertex only. Consequently, each matching controller $K_\mathrm{bess}(\Theta_\mathrm{bess}(t))=~\hspace{-1.5mm}K_\mathrm{bess}$ and $K_\mathrm{sc}(\Theta_\mathrm{sc}(t))=K_\mathrm{sc}$ is parameter-independent, and directly obtained as the solution of~\eqref{eq:LMI_H_inf_opt_problem_vertices_final}, respectively. 

Beyond that, we consider a one-dimensional DVPP specification for the $\mathrm{f}$-$\mathrm{p}$ control of DVPP~1 (cf.~\eqref{eq:T_des_DVPP1}), and therefore employ the matching control only in the converter's ``active power path'' to control the $\mathrm{d}$-component $i^\star_\mathrm{d}$ (\cref{fig:outer_ctrl_DVPP1}).

\subsubsection*{Simplified Converter Model for Control Design}According to~\cref{sec:local_model_match}, a linearized representation of the converter model in~\cref{fig:conv_model} is required to design the feedback gain $K_i$ in~\cref{fig:outer_ctrl_DVPP1}. While linearizing the converter model, we proceed by making a reasonable model reduction, which supports the optimization problem in~\cref{eq:LMI_H_inf_opt_problem_vertices_final} in computation and scope.

Firstly, we disregard the dc side of the converter by assuming $v_\mathrm{dc}\approx v_\mathrm{dc}^\star$ and only consider the ac-side dynamics for control design. In particular, since the ADPFs $m_i^\mathrm{fp},\,i\in\{\mathrm{bess,sc}\}$ include the time constants $\tau_\mathrm{bess}$ and $\tau_\mathrm{c}>\tau_\mathrm{sc}$ of the converter's primary source (cf.~\cref{eq:ADPM_BESS,eq:ADPM_SC}), the regulation of active power on the ac side, imposed by the local reference model $M_i\cdot T_\mathrm{des}$, will be slower or equal than the dc-side control. 

Considering the ac-side dynamics, we extract the PLL from the model, as it provides the DVPP input signal in terms of the frequency deviation $\Delta f_\mathrm{pll}$. Therefore, the remaining system for control design consists of the $RL$-filter dynamics, the current loop, and the power computation, given in the $\mathrm{dq}$-coordinate frame of the PLL as
\begin{subequations}\label{eq:conv_model_ctrl_design_nonlinear}
\begin{align}\label{eq:conv_model_ctrl_design_nonlinear_RL}
	\tfrac{L_\mathrm{f}}{\omega_\mathrm{b}}\dot{i}_\mathrm{dq}&=-\mathcal{Z}_\mathrm{f}i_\mathrm{dq}+v_\mathrm{c,dq}-v_\mathrm{dq}\\
	\dot{x}_\mathrm{i,dq}&=i^\star_\mathrm{dq}-i_\mathrm{dq}\\\label{eq:conv_model_ctrl_design_nonlinear_curr_ctrl}
	v^\star_\mathrm{c,dq}&=v_\mathrm{dq}+\mathcal{Z}_\mathrm{f}i_\mathrm{dq}+k_\mathrm{p,i}(i^\star_\mathrm{dq}-i_\mathrm{dq})+k_\mathrm{i,i}x_\mathrm{i,dq}\\
	p &= v_\mathrm{d}i_\mathrm{d}+v_\mathrm{q}i_\mathrm{q},\quad\quad
	q = v_\mathrm{q}i_\mathrm{d}-v_\mathrm{d}i_\mathrm{q}.
\end{align}
\end{subequations} 

We proceed by approximating $v_\mathrm{d}\approx v^\star$ and $v_\mathrm{q}\approx 0$ as constants, such that the active and reactive power expressions become decoupled, i.e., $p\approx v^\star i_\mathrm{d}$ and $q\approx -v^\star i_\mathrm{q}$, assuming to stay close to the nominal operating point. In particular, since we include the reference model $M_i\cdot T_\mathrm{des}$ into the matching control, we can compensate for the mismatch between the desired active power injection and the latter approximation.
\renewcommand{\arraystretch}{1.2}
\begin{table}[b!]\scriptsize
    \centering
        \caption{DVPP 1 parameters.}
        \vspace{-1mm}
    \begin{tabular}{c||c|c}
     \toprule
         Parameter & Symbol & Value  \\ \hline
         Power rating, DVPP 1 & $S_\mathrm{dvpp1,r}$ & $250\,\text{MVA}$ \\
         Power rating, hydro & $S_\mathrm{hydro,r}$ & $250\,\text{MVA}$ \\
         Power rating, BESS & $S_\mathrm{bess,r}$ & $50\,\text{MVA}$ \\
         Power rating, supercapacitor & $S_\mathrm{sc,r}$ & $25\,\text{MVA}$ \\ \hline
         Hydro droop control gains & $R_\mathrm{g},\, R_\mathrm{t}$ & $0.03,\,0.38$ \\
         Hydro time constants & $\tau_\mathrm{g},\,\tau_\mathrm{r},\,\tau_\mathrm{w}$ & $0.2\,\text{s},\,5\,\text{s},\,1\,\text{s}$ \\
    Device time constants & $\tau_\mathrm{bess},\,\tau_\mathrm{sc},\,\tau_\mathrm{c}$ & $0.2\,\text{s},\,0.01\,\text{s},\,0.081\,\text{s}$ \\\hline
    $T_\mathrm{des}$ parameters & $D,\,\tau$ & $1/0.03,\,0.2\,\text{s}$ \\\hline
    & $\alpha_\mathrm{bess}\hspace{-0.5mm}=\hspace{-0.5mm}\alpha_\mathrm{sc}$ & $5\cdot10^{\text{-}5}$ \\
    $\mathcal{H}_\infty$ tuning parameters & $\mu_\mathrm{bess}\hspace{-0.5mm}=\hspace{-0.5mm}\mu_\mathrm{sc}$ & 1\\
    & $\zeta_{\mathrm{bess}_{\sigma_p}}\hspace{-0.5mm}=\hspace{-0.5mm}\zeta_{\mathrm{sc}_{\sigma_p}}$ & $2.5\cdot10^{\text{-}4}$\\
    \bottomrule
    \end{tabular}
    \label{tab:DVPP1_parameters}
\end{table}
\renewcommand{\arraystretch}{1} \normalsize

Moreover, since $v_\mathrm{dc}\approx v_\mathrm{dc}^\star$, we can assume $v_\mathrm{c,dq}\approx v_\mathrm{c,dq}^\star$ and therefore reduce~\eqref{eq:conv_model_ctrl_design_nonlinear_curr_ctrl} and~\eqref{eq:conv_model_ctrl_design_nonlinear_RL} to 
\begin{align}
		\tfrac{L_\mathrm{f}}{\omega_\mathrm{b}}\dot{i}_\mathrm{dq}&=k_\mathrm{p,i}(i^\star_\mathrm{dq}-i_\mathrm{dq})+k_\mathrm{i,i}x_\mathrm{i,dq}.
\end{align}
By linearizing around the nominal operating point, we obtain
\begin{subequations}\label{eq:conv_model_ctrl_design_small_signal}
	\begin{align}
		\tfrac{L_\mathrm{f}}{\omega_\mathrm{b}}\Delta\dot{i}_\mathrm{dq}&=k_\mathrm{p,i}(\Delta i^\star_\mathrm{dq}-\Delta i_\mathrm{dq})+k_\mathrm{i,i}\Delta x_\mathrm{i,dq}\\
		\Delta\dot{x}_\mathrm{i,dq}&=\Delta i^\star_\mathrm{dq}-\Delta i_\mathrm{dq}\\
		\Delta p &= v^\star\Delta i_\mathrm{d},\quad\quad \Delta q = -v^\star \Delta i_\mathrm{q},
	\end{align}
\end{subequations}
where all variables represent deviations from their respective equilibrium point, i.e., $\Delta i_\mathrm{dq}=i_\mathrm{dq}-i_\mathrm{dq,0}$, $\Delta i_\mathrm{dq}^\star=i_\mathrm{dq}^\star-i_\mathrm{dq,0}^\star$, $\Delta x_\mathrm{i,dq}=x_\mathrm{i,dq}-x_\mathrm{i,dq,0}$, $\Delta p = p-p_0$ and $\Delta q = q-q_0$.

Since we are only interested in controlling $i_\mathrm{d}^\star$ to regulate the active power, the decoupled system in~\eqref{eq:conv_model_ctrl_design_small_signal} can be reduced to only considering the $\mathrm{d}$-states and the active power output, i.e.,
\begin{subequations}\label{eq:conv_model_ctrl_design_small_signal_d}
\vspace{-4mm}
	\begin{align}
	\tfrac{L_\mathrm{f}}{\omega_\mathrm{b}}\Delta\dot{i}_\mathrm{d}&=k_\mathrm{p,i}(\Delta i^\star_\mathrm{d}-\Delta i_\mathrm{d})+k_\mathrm{i,i}\Delta x_\mathrm{i,d}\\
		\Delta\dot{x}_\mathrm{i,d}&=\Delta i^\star_\mathrm{d}-\Delta i_\mathrm{d}\\
		\Delta p &= v^\star\Delta i_\mathrm{d}.
	\end{align}
\end{subequations}

Finally, the system in~\cref{eq:conv_model_ctrl_design_small_signal_d} serves as a linearized plant model for the design of the BESS and supercapacitor matching controllers (cf.~\cref{fig:local_matching_ctrl}), where the PLL frequency deviation $\Delta f_\mathrm{pll}$ is an input to only the device's local reference model $M_i\cdot T_\mathrm{des}$ (i.e., $E=0$ in~\cref{eq:ss_plant}). In this regard, the augmented system in~\eqref{eq:augmented_LPV}, required to solve the optimization problem in~\cref{eq:LMI_H_inf_opt_problem_vertices_final}, is established by~\cref{eq:conv_model_ctrl_design_small_signal_d} and the individual reference model $M_i\cdot T_\mathrm{des},\,i\in\{\mathrm{bess,sc}\}$, respectively, where $T_\mathrm{des}$ is given in~\cref{eq:T_des_DVPP1}. The tuning parameters in~\eqref{eq:ellipsoidal} to limit transients of the converter currents as $||\Delta i^\star_\mathrm{d}||\leq \mu_\mathrm{bess}$ and $||\Delta i^\star_\mathrm{d}||\leq \mu_\mathrm{sc}$, as well as to shape the integral gain of the controllers as $|\sigma_\mathrm{p}|\leq \zeta_{\mathrm{bess}_{\sigma_\mathrm{p}}}$ and $|\sigma_\mathrm{p}|\leq \zeta_{\mathrm{sc}_{\sigma_\mathrm{p}}}$, are given in~\cref{tab:DVPP1_parameters}.

\subsubsection*{Simulation Results}
\begin{figure}[t!]
	\centering
	\scalebox{0.55}{\includegraphics[]{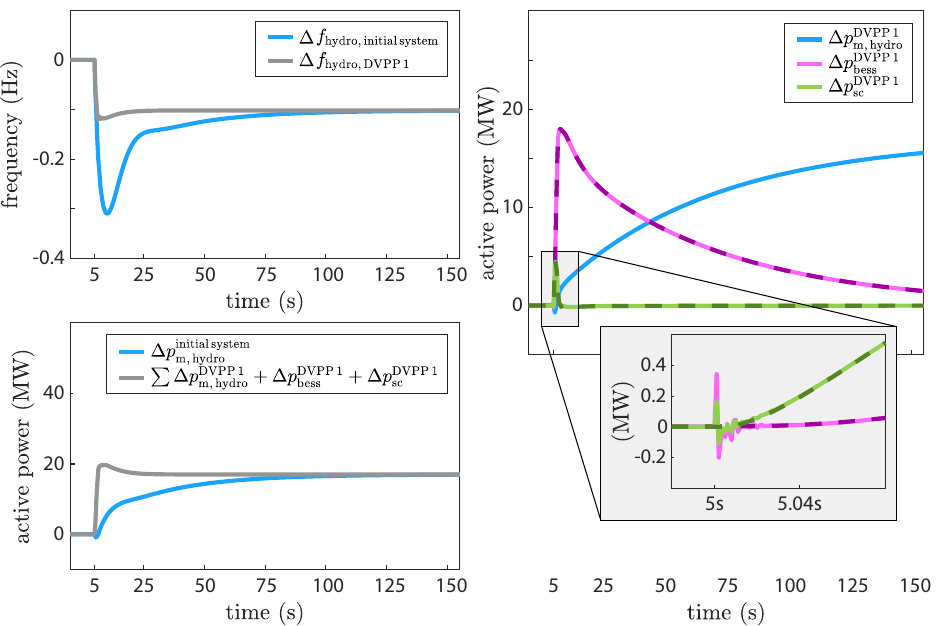}}
	\vspace{-1.5mm}
	\caption{Frequency and active power deviation of the hydro unit at bus 1 in the initial system (\cref{fig:9bus}) compared with DVPP~1 in case study I (\cref{fig:9bus_DVPP_hydro}). The dashed lines indicate the desired active power injection of the DVPP devices.}
	\label{fig:response_DVPP1}
\end{figure}

We compare the frequency response of the hydro unit at bus~1 in the initial system in~\cref{fig:9bus} with the response of DVPP~1 in~\cref{fig:9bus_DVPP_hydro} in detailed simulations based on nonlinear system and device models. In both scenarios, the hydro unit is assigned $60\%$ of the baseload, while the thermal-based power plants at buses 2 and 3 provide the remaining $40\%$. Since we focus on time scales of minutes to seconds or faster, the impact of the state of charge of the BESS and the supercapacitor are neglected. From the simulation results of the nine-bus system in~\cref{fig:response_DVPP1}, we can observe that the frequency response of the stand-alone hydro unit to a $30$~MW load step at bus 6 is quite poor in terms of both settling time and frequency nadir. In contrast, the DVPP highly improves the frequency response of the hydro plant, and the BESS and the supercapacitor match their desired power injection (dashed lines) very well. We also observed that the transient actuation constraints, i.e., converter current constraints, are not encountered during the load step (not shown).

\subsection{Case Study II: SG Replacement by Adaptive DVPP Control}\label{sec:case2}
\begin{figure}[b!]
	\begin{subfigure}[h!]{0.23\textwidth}
		\centering
		\hspace{7mm}
		\include{Figures/9bus_system_C2}
		\vspace{-4mm}
		\caption{\textit{Case study II:} IEEE nine-bus system with DVPPs at buses~1 and~3.}
		\label{fig:9bus_DVPPs}
	\end{subfigure}
	\hspace{10mm}
	\begin{subfigure}[h!]{0.21\textwidth}
			\centering
	\scalebox{0.62}{\includegraphics[]{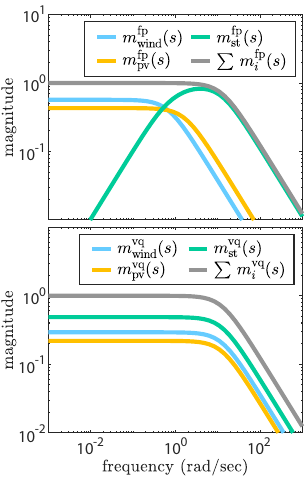}}
	\hspace{7mm}
	\vspace{-4mm}
	\caption{Magnitude Bode plots of the ADPFs of the DVPP~3 devices.}
	\label{fig:bode_ADPM_DVPP3}
	\end{subfigure}
	\vspace{-1mm}
	\caption{System model and ADPFs of case study II.}
\end{figure}

While keeping DVPP~1 at bus~1, we replace the thermal-based power plant at bus~3 by another DVPP, consisting of a wind power plant, a PV system and a STATCOM with battery (\cref{fig:9bus_DVPPs}). We want to substitute the services of the thermal-based power plant, and specify a $\mathrm{f}$-$\mathrm{p}$ and $\mathrm{v}$-$\mathrm{q}$ control as
\begin{align}\label{eq:T_des_DVPP3}
	\begin{bmatrix}\hspace{-0.5mm}\Delta p(\hspace{-0.1mm}s\hspace{-0.1mm})\hspace{-1.75mm}\\\hspace{-0.5mm}\Delta q(\hspace{-0.1mm}s\hspace{-0.1mm})\hspace{-1.75mm}\end{bmatrix}\hspace{-1.25mm}=\hspace{-0.5mm}T_\mathrm{des}(\hspace{-0.1mm}s\hspace{-0.1mm}) \hspace{-1mm}\begin{bmatrix}\hspace{-0.5mm}\Delta f(\hspace{-0.1mm}s\hspace{-0.1mm})\hspace{-0.75mm}\\ \hspace{-0.5mm}\Delta v(\hspace{-0.1mm}s\hspace{-0.1mm})\hspace{-0.75mm}\end{bmatrix}\hspace{-1.25mm},\,\,T_\mathrm{des}(\hspace{-0.1mm}s\hspace{-0.1mm})\hspace{-0.75mm}:=\hspace{-1.25mm}\begin{bmatrix}\hspace{-0.5mm}\tfrac{{-}D_\mathrm{p}{-}H_\mathrm{p}s}{\tau_\mathrm{p} s+1}\hspace{-4mm}&\hspace{-4mm}0\hspace{-1.75mm}\\\hspace{-0.5mm}0\hspace{-4mm}&\hspace{-4mm}\tfrac{{-}D_\mathrm{q}}{\tau_\mathrm{q} s+1}\hspace{-1.75mm} \end{bmatrix}\hspace{-1.5mm},
\end{align}
where $H_\mathrm{p}$ and $D_\mathrm{p}$ are the normalized virtual inertia and droop coefficients for the $\mathrm{f}$-$\mathrm{p}$ control, $D_\mathrm{q}$ is a high gain droop for the $\mathrm{v}$-$\mathrm{q}$ control, and the denominators with $\tau_\mathrm{p}$ and $\tau_\mathrm{q}$ are included to filter out high-frequency dynamics. The associated parameter values are given in~\cref{tab:DVPP3_parameters}.

Due to their volatile availability, neither PV nor wind are typically solely installed for frequency and voltage control. However, the complementary nature of wind and solar energy can be exploited to compensate each other's fluctuations\cite{nehrir2011review}. This motivates the combination of wind, PV and STATCOM for DVPP~3, which, thanks to our adaptive control approach (\cref{sec:divide_conquer}), is able to handle temporal variability of the DERs without affecting the overall DVPP response behavior.

\subsubsection*{ADPF Selection}\label{sec:DVPP3}
The ADPMs of wind, PV and STATCOM ($\mathrm{st}$) are composed of the ADPFs for $\mathrm{f}$-$\mathrm{p}$ and $\mathrm{v}$-$\mathrm{q}$ control, i.e.,
\begin{align}\label{eq:DVPP3_ADPMs}
	\begin{split}
M_i(s) = \begin{bmatrix}
m_i^\mathrm{fp}(s)&0\\0&m_i^\mathrm{vq}(s)
\end{bmatrix},\quad i\in\{\mathrm{wind,pv,st}\}.
\end{split}
\end{align}
Further, both the wind, PV and STATCOM are converter-based, thus considered as controllable ($\in\mathcal{C}$) within the DVPP setup for $\mathrm{f}$-$\mathrm{p}$ and $\mathrm{v}$-$\mathrm{q}$ control in~\cref{eq:T_des_DVPP3}. For the ADPFs of the $\mathrm{f}$-$\mathrm{p}$ channel, we select wind and PV to participate as LPFs:
\begin{align}\label{eq:ADPF_PF_wind_PV}
	m_\mathrm{wind}^\mathrm{fp}(s) = \tfrac{\theta_\mathrm{wind}^\mathrm{fp}}{\tau_\mathrm{wind}s+1},\quad
	m_\mathrm{pv}^\mathrm{fp}(s) = \tfrac{\theta_\mathrm{pv}^\mathrm{fp}}{\tau_\mathrm{pv}s+1},
\end{align}
where $\tau_\mathrm{wind}, \tau_\mathrm{pv}$ are the dc time constants employed in the associated converter model in~\cref{fig:conv_model}. The adaptive DC gains $\theta_\mathrm{wind}^\mathrm{fp}(t)$ and $\theta_\mathrm{pv}^\mathrm{fp}(t)$ are defined as in~\eqref{eq:time_var_DC_gain}, i.e., 
\begin{align}\label{eq:DC_gains_W_PV_Pf}
	\begin{split}
	\theta_\mathrm{wind}^\mathrm{fp}(t) &= \tfrac{{p}_\mathrm{wind}^\mathrm{max}(t)}{{p}_\mathrm{wind}^\mathrm{max}(t)+{p}_\mathrm{pv}^\mathrm{max}(t)}\quad\in[\ubar{\theta}_\mathrm{wind}^\mathrm{fp},\,\bar{\theta}_\mathrm{wind}^\mathrm{fp}],\\
	\theta_\mathrm{pv}^\mathrm{fp}(t)&= \tfrac{{p}_\mathrm{pv}^\mathrm{max}(t)}{{p}_\mathrm{wind}^\mathrm{max}(t)+{p}_\mathrm{pv}^\mathrm{max}(t)}\quad\in[\ubar{\theta}_\mathrm{pv}^\mathrm{fp},\,\bar{\theta}_\mathrm{pv}^\mathrm{fp}],
	\end{split}
\end{align}
where the active power capacity limits vary in the interval ${p}_\mathrm{wind}^\mathrm{max}(t)\in[\ubar{p}_\mathrm{wind}^\mathrm{max},\bar{p}_\mathrm{wind}^\mathrm{\,max}]$ and ${p}_\mathrm{pv}^\mathrm{max}(t)\in[\ubar{p}_\mathrm{pv}^\mathrm{max},\bar{p}_\mathrm{pv}^\mathrm{\,max}]$, respectively. The ADPF of the $\mathrm{f}$-$\mathrm{p}$ channel for the STATCOM with battery is obtained via Algorithm 1 as
\begin{align}
	m_\mathrm{st}^\mathrm{fp}(s) = \tfrac{1}{\tau_\mathrm{c}s+1}-m_\mathrm{wind}^\mathrm{fp}(s)-m_\mathrm{pv}^\mathrm{fp}(s),
\end{align}
where $1/\tau_\mathrm{c}$ is the PLL bandwidth. Note that $m_\mathrm{st}^\mathrm{fp}$ depends on the adaptive DC gains $\theta_\mathrm{wind}^\mathrm{fp}$ and $\theta_\mathrm{pv}^\mathrm{fp}$ of wind and PV.

Since the reactive power capability of the converters is independent of the dc source technology in our test case, we select the ADPFs of the $\mathrm{v}$-$\mathrm{q}$ channel for all three devices as
\begin{align}\label{eq:ADPFs_W_PV_ST_Qv}
	m_\mathrm{wind}^\mathrm{vq}\hspace{-0.25mm}(s)\hspace{-0.5mm} =\hspace{-0.5mm} \tfrac{\theta_\mathrm{wind}^\mathrm{vq}}{\tau_\mathrm{c}s+1},\,\,
	m_\mathrm{pv}^\mathrm{vq}\hspace{-0.25mm}(s) \hspace{-0.5mm}=\hspace{-0.5mm} \tfrac{\theta_\mathrm{pv}^\mathrm{vq}}{\tau_\mathrm{c}s+1}, \,\,
	m_\mathrm{st}^\mathrm{vq}\hspace{-0.25mm}(s) \hspace{-0.5mm}=\hspace{-0.5mm} \tfrac{\theta_\mathrm{st}^\mathrm{vq}}{\tau_\mathrm{c}s+1}.
\end{align} 
The adaptive DC gains are defined as in~\eqref{eq:time_var_DC_gain}, i.e.,
\begin{align}\label{eq:DC_gain_Qv}
	\begin{split}
	\theta_\mathrm{wind}^\mathrm{vq}(t) &= \tfrac{{q}_\mathrm{wind}^\mathrm{max}(t)}{{q}_\mathrm{wind}^\mathrm{max}(t)+{q}_\mathrm{pv}^\mathrm{max}(t)+{q}_\mathrm{st}^\mathrm{max}(t)}\quad\in[\ubar{\theta}_\mathrm{wind}^\mathrm{vq},\,\bar{\theta}_\mathrm{wind}^\mathrm{vq}],\\
	\theta_\mathrm{pv}^\mathrm{vq}(t)&= \tfrac{{q}_\mathrm{pv}^\mathrm{max}(t)}{{q}_\mathrm{wind}^\mathrm{max}(t)+{q}_\mathrm{pv}^\mathrm{max}(t)+{q}_\mathrm{st}^\mathrm{max}(t)}\quad\in[\ubar{\theta}_\mathrm{pv}^\mathrm{vq},\,\bar{\theta}_\mathrm{pv}^\mathrm{vq}],\\
\theta_\mathrm{st}^\mathrm{vq}(t)&= \tfrac{{q}_\mathrm{st}^\mathrm{max}(t)}{{q}_\mathrm{wind}^\mathrm{max}(t)+{q}_\mathrm{pv}^\mathrm{max}(t)+{q}_\mathrm{st}^\mathrm{max}(t)}\quad\in[\ubar{\theta}_\mathrm{st}^\mathrm{vq},\,\bar{\theta}_\mathrm{st}^\mathrm{vq}],
\end{split}
\end{align}
where the reactive power capacity limit ${q}_i^\mathrm{max}$ is related to the active power capacity limit ${p}_i^\mathrm{max},i\in\{ \mathrm{wind,pv,st}\}$ via the PQ-capability curve of the power converter\cite{johansson2004power}. Namely, for $v_\mathrm{dc}$ sufficiently large, the ac voltage converter limit can be disregarded, and, under the assumption of a constant voltage magnitude, the stationary current converter limit $||{i}_\mathrm{dq}||^\mathrm{max}=1$ pu relates the active and reactive power limits in per unit as
\begin{align}
	{q}_i^\mathrm{max}(t)\hspace{-0.5mm}=\hspace{-0.5mm}\sqrt{(||{i}_\mathrm{dq}||^\mathrm{max})^2\hspace{-0.5mm}-\hspace{-0.5mm}{p}_i^\mathrm{max}(t)^2}\hspace{-0.5mm}=\hspace{-0.5mm}\sqrt{1\hspace{-0.5mm}-\hspace{-0.5mm}{p}_i^\mathrm{max}(t)^2}.
\end{align}
Finally, the magnitude Bode plots of all ADPFs during nominal capacity conditions are shown in~\cref{fig:bode_ADPM_DVPP3}, and the associated parameter values are provided in~\cref{tab:DVPP3_parameters}.

\subsubsection*{Local matching control} 
\begin{figure}[t!]
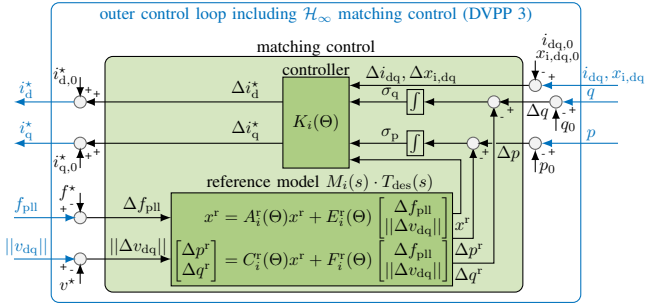

	\centering
	\include{Figures/Conv_matching_DVPP3}
	\vspace{-8mm}
	\caption{Outer control loop with matching control for DVPP~3 converters.}
	\label{fig:outer_ctrl_DVPP3}
\end{figure}

Similar as in case study I, we employ the $\mathcal{H}_\infty$ matching control in the outer control loop of each grid-side converter to match the local closed-loop dynamics of the DVPP 3 devices with their local reference model $M_i\cdot T_\mathrm{des},\,i\in\{\mathrm{wind,pv,st}\}$ (\cref{fig:outer_ctrl_DVPP3}). However, in contrast to the DVPP 1 converters, we now aim to regulate the entire reference current $i^\star_\mathrm{dq}$ to participate in the desired $\mathrm{f}$-$\mathrm{p}$ and $\mathrm{v}$-$\mathrm{q}$ control of DVPP 3 in~\cref{eq:T_des_DVPP3}. In this regard, we consider the previous linearized converter model in~\cref{eq:conv_model_ctrl_design_small_signal} with both the $\mathrm{d}$- and the $\mathrm{q}$-states, as well as the active and reactive power outputs as our plant model for control design. 

Furthermore, the ADPFs of the DVPP 3 devices (and with that their local reference models $M_i\cdot T_\mathrm{des},\,i\in\{\mathrm{wind,pv,st}\}$) rely on the adaptive DC gains introduced in~\cref{eq:DC_gains_W_PV_Pf,eq:DC_gain_Qv}, so that the associated state-feedback gains $K_i(\Theta)$ are parameter-dependent now. More precisely, the parameter vector $\Theta_i(t)$ of each device's augmented LPV system in~\eqref{eq:augmented_LPV} is given by the DC gains appearing in the individual ADPM in~\eqref{eq:DVPP3_ADPMs}, i.e.,
\begin{align}\label{eq:theta_DVPP3}
	\begin{split}
\Theta_\mathrm{wind}(\hspace{-0.25mm}t\hspace{-0.25mm})\hspace{-0.5mm}=\hspace{-0.5mm}[\theta_\mathrm{wind}^\mathrm{fp}&(\hspace{-0.25mm}t\hspace{-0.25mm}),\hspace{-0.25mm}\theta_\mathrm{wind}^\mathrm{vq}(\hspace{-0.25mm}t\hspace{-0.25mm})]'\hspace{-0.5mm},\, \Theta_\mathrm{pv}(\hspace{-0.25mm}t\hspace{-0.25mm})\hspace{-0.5mm}=\hspace{-0.5mm}[\theta_\mathrm{pv}^\mathrm{fp}(\hspace{-0.25mm}t\hspace{-0.25mm}),\hspace{-0.25mm} \theta_\mathrm{pv}^\mathrm{vq}(\hspace{-0.25mm}t\hspace{-0.25mm})]'\hspace{-0.5mm},\\ \Theta_\mathrm{st}(\hspace{-0.25mm}t\hspace{-0.25mm})&\hspace{-0.5mm}=\hspace{-0.5mm}[\theta_\mathrm{wind}^\mathrm{fp}(\hspace{-0.25mm}t\hspace{-0.25mm}),\hspace{-0.25mm} \theta_\mathrm{pv}^\mathrm{fp}(\hspace{-0.25mm}t\hspace{-0.25mm}),\hspace{-0.25mm}\theta_\mathrm{st}^\mathrm{vq}(\hspace{-0.5mm}t\hspace{-0.5mm})]'.
\end{split}
\end{align}

Hence, for each device, the optimization problem in~\eqref{eq:LMI_H_inf_opt_problem_vertices_final} is solved for the respective set of vertices $\hat{\Theta}_\mathrm{wind}^{(1)},...,\hat{\Theta}_\mathrm{wind}^{(4)}$, $\hat{\Theta}_\mathrm{pv}^{(1)},...,\hat{\Theta}_\mathrm{pv}^{(4)}$ and $\hat{\Theta}_\mathrm{st}^{(1)},...,\hat{\Theta}_\mathrm{st}^{(8)}$, to compute the associated vertex controllers, respectively. The tuning parameters in~\eqref{eq:ellipsoidal} to limit transients of the converter currents as $||\Delta i^\star_\mathrm{dq}||\hspace{-0.5mm}\leq\hspace{-0.5mm}\mu_i$, and to shape the integral gains of the controllers as $|\sigma_\mathrm{p}|\hspace{-0.5mm}\leq \hspace{-0.5mm}\zeta_{i_{\sigma_\mathrm{p}}}$ and $|\sigma_\mathrm{q}|\hspace{-0.5mm}\leq\hspace{-0.5mm}\zeta_{i_{\sigma_\mathrm{q}}}$, $\forall i\hspace{-0.5mm}\in\hspace{-0.5mm}\{\mathrm{wind,pv,st}\}$, are given in~\cref{tab:DVPP3_parameters}.

Finally, during online operation, we compute the state-feedback controllers $K_\mathrm{wind}(\Theta_\mathrm{wind}(t))$, $K_\mathrm{pv}(\Theta_\mathrm{pv}(t))$ and $K_\mathrm{st}(\Theta_\mathrm{st}(t))$ as the convex combination of the respective vertex controllers (cf.~\eqref{eq:LPV_ctrl}), where the convex combination coefficients are obtained via closed-form expressions derived in\cite{schurmann2016closed}, based on the instantaneous values of the vectors $\Theta_i(t),\,i\in\{\mathrm{wind,pv,st}\}$.

\begin{figure}[b!]
	\centering
	\scalebox{0.54}{\includegraphics[]{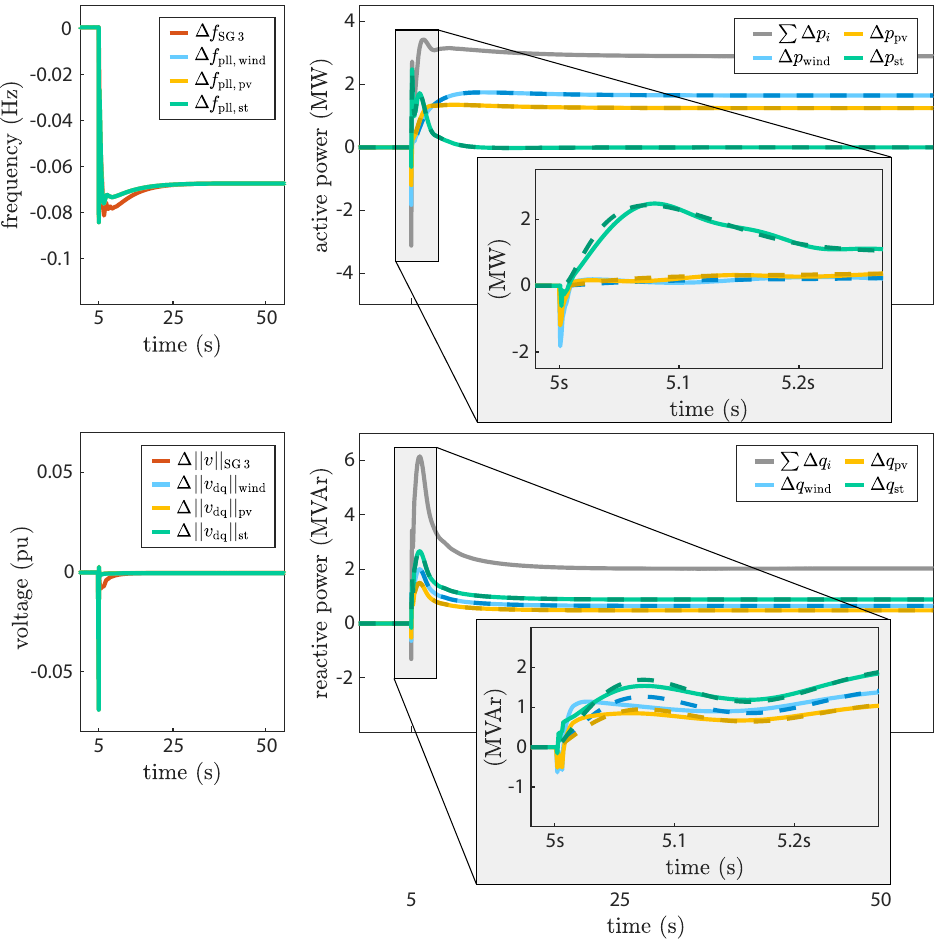}}
	\vspace{-1.5mm}
	\caption{System response in case study II after load step at bus~3. The dashed lines indicate the desired power injection of the DVPP devices.}
	\label{fig:response_DVPP3_load}
	\vspace{-2mm}
\end{figure}
\begin{figure}[b!]
	\centering
	\scalebox{0.54}{\includegraphics[]{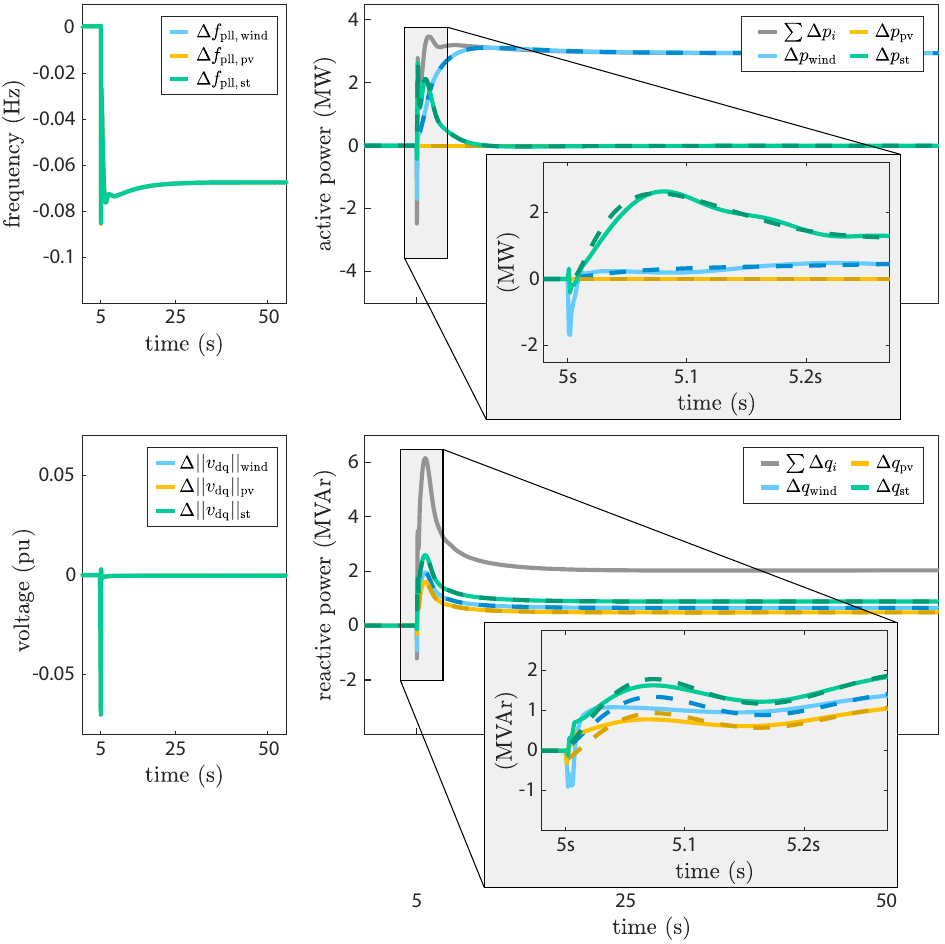}}
	\vspace{-1.5mm}
	\caption{System response in case study II after a PV generation loss at bus 3. The dashed lines indicate the desired power injection of the DVPP devices.}
	\label{fig:response_DVPP3_cloud}
\end{figure}
\subsubsection*{Simulation Results}
While keeping DVPP 1 at bus 1, we replace the synchronous generator (SG) at bus~3 by DVPP~3 as in~\cref{fig:9bus_DVPPs}, and investigate its frequency and voltage response behavior in detailed simulations using nonlinear system and device models. We adopt the same baseload supply for DVPP~3 as for the prior SG, provided by the wind power plant and the PV system. The latter (wind and PV) are operated under deloaded conditions with respect to their maximum power point tracking, allowing them to participate in frequency and voltage regulation\cite{dreidy2017inertia}. The DC gains of the ADPFs are adapted by the DVPP operator and communicated to all DVPP devices (the impact of communication delays is neglected).

We first consider nominal generation capacities of the DVPP devices, i.e., the ADPFs are given as in~\cref{fig:bode_ADPM_DVPP3}. During the simulation at $t=5\,$s, we impose a $18.7$ MW load step at bus 3. From the results in~\cref{fig:response_DVPP3_load}, it becomes apparent how all DVPP devices accomplish an accurate matching of their desired active and reactive power injection (dashed lines), such that their specified $\mathrm{f}$-$\mathrm{p}$ and $\mathrm{v}$-$\mathrm{q}$ controls establish an adequate replacement of the prior SG services. In particular, due to the design choice of $T_\mathrm{des}$ in~\cref{eq:T_des_DVPP3}, the aggregate DVPP behavior even outperforms the frequency and voltage responses of SG~3 (cf. red curves in~\cref{fig:response_DVPP3_load} from a separate simulation). Moreover, we also observed that the transient actuation constraints, i.e., converter reference current constraints, are not encountered during the load step (not shown).

\begin{figure}[t!]
	\centering
	\scalebox{0.77}{\includegraphics[]{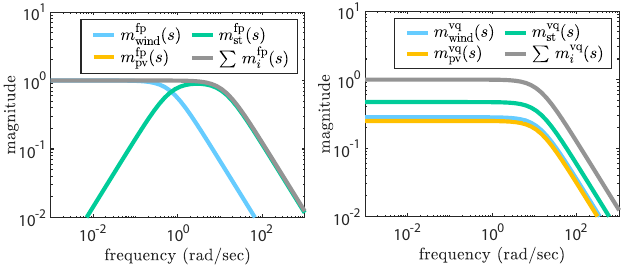}}
	\caption{Magnitude Bode plots of the ADPFs of the DVPP 3 devices during the PV generation loss.}
	\label{fig:bode_ADPMs_after_cloud}
\end{figure}

\renewcommand{\arraystretch}{1.2}
\begin{table}[t!]\scriptsize
    \centering
        \caption{DVPP 3 parameters.}
        \vspace{-1mm}
    \begin{tabular}{c||c|c}
     \toprule
         Parameter & Symbol & Value  \\ \hline
         Power rating, DVPP 3 & $S_\mathrm{dvpp3,r}$ & $64\,\text{MVA}$ \\
         Power rating, wind & $S_\mathrm{wind,r}$ & $70.5\,\text{MVA}$ \\
         Power rating, PV & $S_\mathrm{pv,r}$ & $53\,\text{MVA}$ \\
         Power rating, STATCOM & $S_\mathrm{st,r}$ & $80\,\text{MVA}$ \\ \hline
         \hspace{-1mm}Upper/lower capacity limit, wind\hspace{-1mm} & $\ubar{p}^\mathrm{max}_\mathrm{wind},\,\bar{p}^\mathrm{max}_\mathrm{wind}$ & $0\,\text{MW},\,37\,\text{MW}$ \\
         \hspace{-1mm}Upper/lower capacity limit, PV \hspace{-1mm}& $\ubar{p}^\mathrm{max}_\mathrm{pv},\,\bar{p}^\mathrm{max}_\mathrm{pv}$ & $0\,\text{MW},\,28\,\text{MW}$ \\ \hline
         Device time constants & $\tau_\mathrm{wind},\,\tau_\mathrm{pv}$& $1.5\,\text{s},\,0.6\,\text{s}$ \\
          Device time constants & $\tau_\mathrm{st},\,\tau_\mathrm{c}$ & $0.2\,\text{s},\,0.081\,\text{s}$ \\\hline
           $T_\mathrm{des}^\mathrm{fp}$ parameters & $D_\mathrm{p},\,H_\mathrm{p},\,\tau_\mathrm{p}$ & \hspace{-1mm}$1/0.03,13\,\text{s},0.2\,\text{s}$\hspace{-1mm} \\
           $T_\mathrm{des}^\mathrm{vq}$ parameters & $D_\mathrm{q},\,\tau_\mathrm{q}$ & $100,\,0.2\,\text{s}$ \\\hline
           & \hspace{-1mm}$\alpha_\mathrm{wind}\hspace{-0.5mm}=\hspace{-0.5mm}\alpha_\mathrm{pv}\hspace{-0.5mm}=\hspace{-0.5mm}\alpha_\mathrm{st}$ & $5\cdot10^{\text{-}4}$ \\
           & \hspace{-1mm}$\mu_\mathrm{wind}\hspace{-0.5mm}=\hspace{-0.5mm}\mu_\mathrm{pv}\hspace{-0.5mm}=\hspace{-0.5mm}\mu_\mathrm{st}$ & $0.25$ \\
           $\mathcal{H}_\infty$ tuning parameters & \hspace{-1mm}$\zeta_{\mathrm{wind}_{\sigma_p}},\,\zeta_{\mathrm{wind}_{\sigma_q}}$ & $0.011,\,0.0002$\\
            & $\zeta_{\mathrm{pv}_{\sigma_p}},\,\zeta_{\mathrm{pv}_{\sigma_q}}$ & $0.0047,\,0.0002$\\
             & $\zeta_{\mathrm{st}_{\sigma_p}},\,\zeta_{\mathrm{st}_{\sigma_q}}$ & \hspace{-1mm}$0.00175,\,0.000175$\hspace{-1mm}\\
    \bottomrule
    \end{tabular}
    \label{tab:DVPP3_parameters}
\end{table}
\renewcommand{\arraystretch}{1} \normalsize

To examine the DVPP control during time-varying limits on generation capacity, we induce a step decrease of the PV active power capacity limit ${p}_\mathrm{pv}^\mathrm{max}(t)$ at $t=5\,$s (e.g. caused by a cloud), resulting in an overall generation loss of $18.7$ MW at bus 3 (\cref{fig:response_DVPP3_cloud}). This causes the ADPMs of all DVPP devices to automatically adapt in response to the PV loss, such that the respective LPV controllers $K_\mathrm{wind}(\Theta_\mathrm{wind}(t)), K_\mathrm{pv}(\Theta_\mathrm{pv}(t))$ and $K_\mathrm{st}(\Theta_\mathrm{st}(t))$ are recomputed online. In particular, as we can see by comparing the aggregate DVPP response during the PV generation loss (\cref{fig:response_DVPP3_cloud}) and during the load step of equal size (\cref{fig:response_DVPP3_load}), the overall DVPP behavior in~\cref{eq:T_des_DVPP3} remains nearly unaffected, since the wind and STATCOM ADPMs are adapted to compensate for the missing DVPP control provided by the PV plant (see~\cref{fig:bode_ADPMs_after_cloud} in comparison to~\cref{fig:bode_ADPM_DVPP3}). Note that if we would stick to a non-adaptive controller without adapting the ADPMs online, one would observe a larger steady-state deviation in frequency due to the absent DC gain contribution of the PV plant.

\begin{figure}[b!]
	\centering
	\include{Figures/9bus_system_C3}
	\vspace{-8mm}
	\caption{\textit{Case study III:} IEEE nine-bus system with a DVPP at bus 1.}
	\label{fig:9bus_showcase_DVPP}
\end{figure}

\renewcommand{\arraystretch}{1.4}
\begin{table*}[]\scriptsize
    \centering
    \begin{tabular}{l||l|l|l}
     \toprule
         &\multicolumn{1}{c}{SPFs}&\multicolumn{1}{c}{DPFs}&\multicolumn{1}{c}{ADPFs} \\ \hline
         \hspace{-1.75mm}Hydro$\hspace{-1.5mm}$&\hspace{-2mm} $m_\mathrm{hydro}^\mathrm{fp}\hspace{-0.5mm}=\hspace{-0.5mm}\tfrac{1/R_\mathrm{g}}{D}\hspace{-1.5mm}$ &\hspace{-1.25mm}$m_\mathrm{hydro}^\mathrm{fp}(s)\hspace{-0.5mm}=\hspace{-0.5mm}\tfrac{1/R_\mathrm{g}}{D}\tfrac{\tau_\mathrm{r}s+1}{(R_\mathrm{t}/R_\mathrm{g})\tau_\mathrm{r}s+1}\tfrac{1-\tau_\mathrm{w}s}{1+0.5\tau_\mathrm{w}s}\hspace{-1.5mm}$&\hspace{-1.25mm}$m_\mathrm{hydro}^\mathrm{fp}(s)\hspace{-0.5mm}=\hspace{-0.5mm}\tfrac{1/R_\mathrm{g}}{D}\tfrac{\tau_\mathrm{r}s+1}{(R_\mathrm{t}/R_\mathrm{g})\tau_\mathrm{r}s+1}\tfrac{1-\tau_\mathrm{w}s}{1+0.5\tau_\mathrm{w}s}\hspace{-3mm}$ \\
         \hspace{-1.75mm}Wind$\hspace{-1.5mm}$&\hspace{-2mm} $m_\mathrm{wind}^\mathrm{fp}\hspace{-0.5mm}=\hspace{-0.5mm}(1\hspace{-0.5mm}-\hspace{-0.5mm}m_\mathrm{hydro}^\mathrm{fp})\tfrac{\bar{p}^\mathrm{max}_\mathrm{wind}}{\bar{p}^\mathrm{max}_\mathrm{wind}+\bar{p}^\mathrm{max}_\mathrm{pv}}\hspace{-1.5mm}$&\hspace{-1.25mm}$m_\mathrm{wind}^\mathrm{fp}(s)\hspace{-0.5mm}=\hspace{-0.5mm}(1\hspace{-0.5mm}-\hspace{-0.5mm}m_\mathrm{hydro}^\mathrm{fp}(s\hspace{-0.5mm}=\hspace{-0.5mm}0)\hspace{-0.25mm})\tfrac{\bar{p}^\mathrm{max}_\mathrm{wind}}{\bar{p}^\mathrm{max}_\mathrm{wind}+\bar{p}^\mathrm{max}_\mathrm{pv}}\tfrac{1}{\tau_\mathrm{wind}s+1}\hspace{-1.5mm}$&\hspace{-1.25mm}$m_\mathrm{wind}^\mathrm{fp}(s)\hspace{-0.5mm}=\hspace{-0.5mm}(1\hspace{-0.5mm}-\hspace{-0.5mm}m_\mathrm{hydro}^\mathrm{fp}(s\hspace{-0.5mm}=\hspace{-0.5mm}0)\hspace{-0.25mm})\tfrac{{p}^\mathrm{max}_\mathrm{wind}(t)}{{p}^\mathrm{max}_\mathrm{wind}(t)+{p}^\mathrm{max}_\mathrm{pv}(t)}\tfrac{1}{\tau_\mathrm{wind}s+1}\hspace{-3mm}$ \\
         \hspace{-1.75mm}PV$\hspace{-1.5mm}$&\hspace{-2mm} $m_\mathrm{pv}^\mathrm{fp}\hspace{-0.5mm}=\hspace{-0.5mm}(1\hspace{-0.5mm}-\hspace{-0.5mm}m_\mathrm{hydro}^\mathrm{fp})\tfrac{\bar{p}^\mathrm{max}_\mathrm{pv}}{\bar{p}^\mathrm{max}_\mathrm{wind}+\bar{p}^\mathrm{max}_\mathrm{pv}}\hspace{-1.5mm}$&\hspace{-1.25mm}$m_\mathrm{pv}^\mathrm{fp}(s)\hspace{-0.5mm}=\hspace{-0.5mm}(1\hspace{-0.5mm}-\hspace{-0.5mm}m_\mathrm{hydro}^\mathrm{fp}(s\hspace{-0.5mm}=\hspace{-0.5mm}0)\hspace{-0.25mm})\tfrac{\bar{p}^\mathrm{max}_\mathrm{pv}}{\bar{p}^\mathrm{max}_\mathrm{wind}+\bar{p}^\mathrm{max}_\mathrm{pv}}\tfrac{1}{\tau_\mathrm{pv}s+1}\hspace{-1.5mm}$&\hspace{-1.25mm}$m_\mathrm{pv}^\mathrm{fp}(s)\hspace{-0.5mm}=\hspace{-0.5mm}(1\hspace{-0.5mm}-\hspace{-0.5mm}m_\mathrm{hydro}^\mathrm{fp}(s\hspace{-0.5mm}=\hspace{-0.5mm}0)\hspace{-0.25mm})\tfrac{{p}^\mathrm{max}_\mathrm{pv}(t)}{{p}^\mathrm{max}_\mathrm{wind}(t)+{p}^\mathrm{max}_\mathrm{pv}(t)}\tfrac{1}{\tau_\mathrm{pv}s+1}\hspace{-3mm}$ \\
         \hspace{-1.75mm}BESS$\hspace{-1.5mm}$&\hspace{-2mm} $m_\mathrm{bess}^\mathrm{fp}\hspace{-0.5mm}=\hspace{-0.5mm}0\hspace{-1.5mm}$&\hspace{-1.25mm}$m_\mathrm{bess}^\mathrm{fp}(s) \hspace{-0.5mm}=\hspace{-0.5mm} 1\hspace{-0.5mm}-\hspace{-0.5mm}m_\mathrm{pv}^\mathrm{fp}(s)\hspace{-0.5mm}-\hspace{-0.5mm}m_\mathrm{wind}^\mathrm{fp}(s)\hspace{-0.5mm}-\hspace{-0.5mm}m_\mathrm{hydro}^\mathrm{fp}(s)\hspace{-1.5mm}$&\hspace{-1.25mm}$m_\mathrm{bess}^\mathrm{fp}(s) \hspace{-0.5mm}=\hspace{-0.5mm} 1\hspace{-0.5mm}-\hspace{-0.5mm}m_\mathrm{pv}^\mathrm{fp}(s)\hspace{-0.5mm}-\hspace{-0.5mm}m_\mathrm{wind}^\mathrm{fp}(s)\hspace{-0.5mm}-\hspace{-0.5mm}m_\mathrm{hydro}^\mathrm{fp}(s)\hspace{-3mm}$ \\
         \bottomrule
    \end{tabular}
    \caption{Overview of the different types of participation factors for each DVPP device in case study III. To relax the restrictions on the matching control, we additionally apply the relaxation in the high frequency range according to~\cref{eq:sum_part_one_relaxed} for the DPFs and the ADPFs (not shown in this table).}
    \label{tab:different_PFs}
\end{table*}
\renewcommand{\arraystretch}{1} \normalsize

\subsection{Case Study III: Comparison with Existing DVPP Concepts}
After having studied different use cases of our DVPP control strategy in case studies I and II, we now aim to demonstrate the benefits of our method over competing approaches to DVPP control that already exist in literature, namely\cite{joak} and\cite{zhong2021coordinated}. In particular, the works in\cite{joak,zhong2021coordinated} can conceptually be considered as special cases of our method, both with respect to the aggregate DVPP specification $T_\mathrm{des}$, as well as the disaggregation strategy by means of participation factors. On the one hand, both\cite{joak} and\cite{zhong2021coordinated} consider a desired short-term frequency response on an aggregate level, therefore resembling a one-dimensional version of our aggregate MIMO specification in~\cref{eq:MIMO_spec}. On the other hand, the disaggregation strategy in\cite{zhong2021coordinated} is based on static participation factors (SPF) similar to conventional secondary frequency control, while the one in\cite{joak} relies on dynamic participation factors (DPF). In this regard, they represent a (static and) non-adaptive version of the ADPFs in our approach (cf.~\cref{sec:disaggregation}), respectively.

To demonstrate the \textit{conceptual} differences of our proposed DVPP control strategy and the existing methods in\cite{joak,zhong2021coordinated} \textit{with respect to their disaggregation strategies}, we compare the three different types of participation factors, i.e., SPFs, DPFs and ADPFs, for a DVPP with a one-dimensional frequency control specification $T_\mathrm{des}$, given by
\begin{align}\label{eq:T_des_showcase_DVPP}
	\Delta p(s) = T_\mathrm{des}(s) \Delta f(s),\quad T_\mathrm{des}(s)\hspace{-0.5mm}:=\tfrac{-D}{\tau s+1},
\end{align}
where $D$ is the desired droop coefficient, and the denominator with $\tau$ is included to filter out high frequency dynamics. The parameter values are provided in~\cref{tab:showcase_DVPP_parameters}. To get an illustrative comparison of the different characteristics of the participation factors, we combine those types of DVPP devices in case studies I and II with the most distinct heterogeneities into one exemplary ``showcase DVPP'' at bus 1 (\cref{fig:9bus_showcase_DVPP}). In particular, we complement the existing hydro generator, characterized by a very slow short-term frequency response behavior, by a weather-driven wind and PV power plant, and additionally add a BESS to obtain a reliable fast frequency response behavior. The relevant parameter values of the devices are given in~\cref{tab:showcase_DVPP_parameters}. As in case study I, the voltage control at bus 1 is fully provided by the AVR of the hydro plant, therefore not part of the aggregated DVPP control.

\begin{figure}[t!]
	\centering
	\scalebox{0.6}{\includegraphics[]{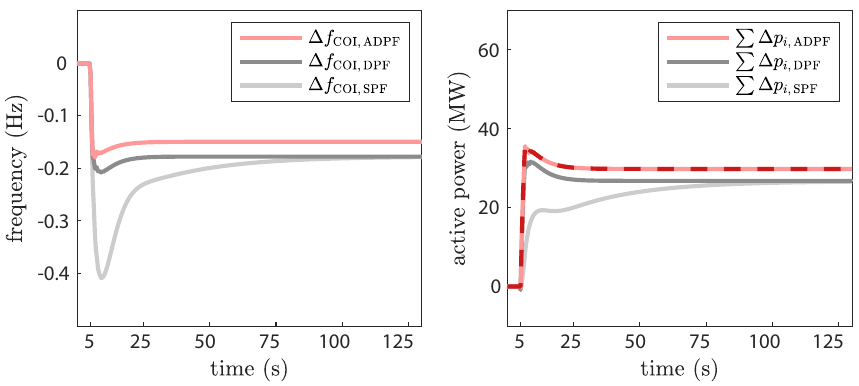}}
	\caption{Frequency and active power response of the DVPP in \cref{fig:9bus_showcase_DVPP} for different participation factors during a loss of PV generation. The dashed line indicates the sum of desired active power injections of the DVPP devices.}
	\label{fig:comparison_DVPP}
\end{figure}

We compare the DVPP behavior for the three different types of participation factors during a step decrease of the PV active power capacity limit $p_\mathrm{pv}^\mathrm{max}(t)$ at $t=5\,\text{s}$, resulting in an overall generation loss of $45.7\,\text{MW}$ at bus 1 (\cref{fig:comparison_DVPP}). To do so, we consider three simulation runs, i.e., one for each disaggregation strategy of $T_\mathrm{des}$. In order to get a fair comparison, we use our previously introduced nonlinear system and device models, and employ the $\mathcal{H}_\infty$ matching control framework for the converter-interfaced generation units in all simulation runs. An overview of the different participation factors for each DVPP device is provided in~\cref{tab:different_PFs}.

\renewcommand{\arraystretch}{1.2}
\begin{table}[t!]\scriptsize
    \centering
        \caption{Showcase DVPP parameters.}
        \vspace{-1mm}
    \begin{tabular}{c||c|c}
     \toprule
         Parameter & Symbol & Value  \\ \hline
         Power rating, DVPP & $S_\mathrm{dvpp,r}$ & $250\,\text{MVA}$ \\
          Power rating, hydro & $S_\mathrm{hydro,r}$ & $250\,\text{MVA}$ \\
         Power rating, wind & $S_\mathrm{wind,r}$ & $38\,\text{MVA}$ \\
         Power rating, PV & $S_\mathrm{pv,r}$ & $70\,\text{MVA}$ \\
         Power rating, BESS & $S_\mathrm{bess,r}$ & $30\,\text{MVA}$ \\\hline
         \hspace{-1mm}Upper/lower capacity limit, wind\hspace{-1mm} & $\ubar{p}^\mathrm{max}_\mathrm{wind},\,\bar{p}^\mathrm{max}_\mathrm{wind}$ & $0\,\text{MW},\,38\,\text{MW}$ \\
         \hspace{-1mm}Upper/lower capacity limit, PV \hspace{-1mm}& $\ubar{p}^\mathrm{max}_\mathrm{pv},\,\bar{p}^\mathrm{max}_\mathrm{pv}$ & $0\,\text{MW},\,70\,\text{MW}$ \\ \hline
         Hydro droop control gains & $R_\mathrm{g},\, R_\mathrm{t}$ & $0.04,\,0.38$ \\
         Hydro time constants & $\tau_\mathrm{g},\,\tau_\mathrm{r},\,\tau_\mathrm{w}$ & $0.2\,\text{s},\,5\,\text{s},\,1\,\text{s}$ \\
         Device time constants & $\tau_\mathrm{wind},\,\tau_\mathrm{pv}$ & $1.5\,\text{s},\,0.6\,\text{s}$ \\
    Device time constants & $\tau_\mathrm{bess},\,\tau_\mathrm{c}$ & $0.2\,\text{s},\,0.081\,\text{s}$ \\\hline
    $T_\mathrm{des}$ parameters & $D,\,\tau$ & $1/0.025,\,0.2\,\text{s}$ \\
    \bottomrule
    \end{tabular}

    \label{tab:showcase_DVPP_parameters}
\end{table}
\renewcommand{\arraystretch}{1} \normalsize

\subsubsection*{Simulation results} 
From the simulation results in~\cref{fig:comparison_DVPP}, we can clearly see how the different characteristics of the participation factors influence the frequency response behavior of the DVPP. In particular, due to the online adaptation of the local device controls, the response obtained by our proposed ADPF-based DVPP control strategy (light red lines) correlates with the actual desired behavior of the DVPP (we have already observed this in case study II when comparing Figures \ref{fig:response_DVPP3_load} and \ref{fig:response_DVPP3_cloud}). In contrast, the SPF-based DVPP control results in a rather poor behavior, most of all because the slow frequency response behavior of the hydro power plant is not addressed by the SPFs of the faster DVPP devices. The use of DPFs, in turn, drastically improves the transient DVPP behavior, since the DPFs are chosen in such a way that different time-scales of local device dynamics are taken into account (cf. \cref{sec:disaggregation}). However, although the DPFs incorporate the response time limitations of the DVPP devices, they are not able to adapt to the changing PV conditions, and hence suffer compensating for the loss of the DVPP control provided by the PV plant, which is in contrast to our superior ADPF-based strategy.

\section{Conclusion}\label{sec:conclusion}
We have proposed a novel multivariable control approach to DVPPs, with the objective to provide desired dynamic ancillary services in the form of fast frequency and voltage control. We employ an optimal adaptive control strategy that takes into account the DVPP internal constraints of the devices, and can additionally handle temporal variability of weather-driven DER in a robust way. Our numerical case study in the IEEE nine-bus system shows the successful performance of our controls, and, in particular demonstrates how our DVPP control strategy can be used to improve the fast frequency response of the initial system, and to facilitate the dynamic ancillary services provision by weather-driven DERs in future power systems. 

Ongoing research includes the incorporation of grid-forming converter controls into our DVPP setup, as well as the extension of the DVPP setup to a geographically distributed scenario, where the DVPP devices are located at different geographical regions within the power system. Moreover, future work should address the design of multivariable and robust specifications for the desired dynamic DVPP behavior.

\bibliographystyle{IEEEtran}
\bibliography{IEEEabrv,mybibliography}

\vspace{-0.8cm}

\begin{IEEEbiography}[{\includegraphics[width=1in,height=1.25in,clip,keepaspectratio]{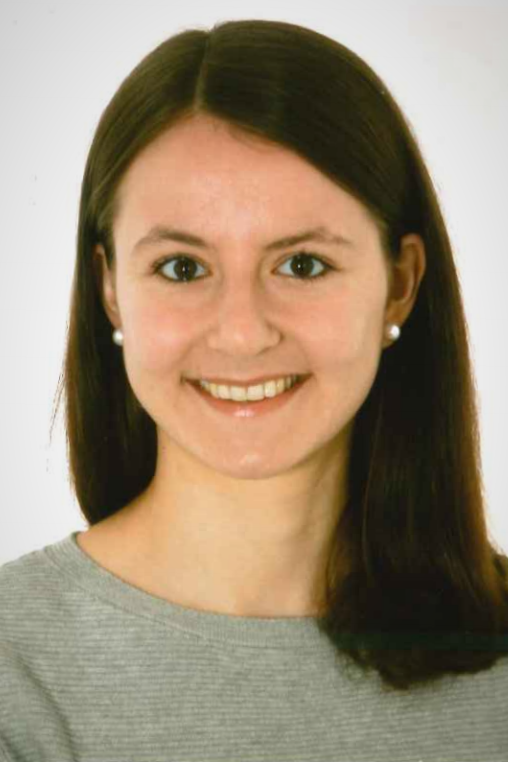}}]{Verena Häberle} is a Ph.D. student with the Automatic Control Laboratory at ETH Zurich, Switzerland, since 2020. She received the B.Sc. and M.Sc. degree in electrical engineering and information technology from ETH Zurich, in 2018 and 2020, respectively. For her outstanding academic achievements during her Master's thesis at the Automatic Control Laboratory, ETH Zurich, under Professor Florian Dörfler, she was honored with the ETH Medal and the SGA Award from the Swiss Society of Automatic Control (SSAC). Her research focuses on the control design of dynamic virtual power plants for future power systems.
\end{IEEEbiography}

\vspace{-1cm}

\begin{IEEEbiography}[{\includegraphics[width=1in,height=1.25in,clip,keepaspectratio]{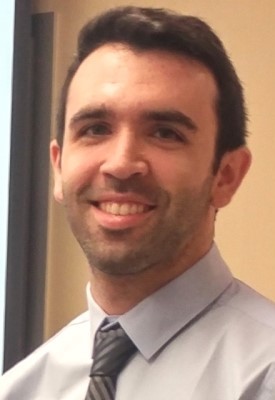}}]{Michael W. Fisher} is a postdoctoral researcher at ETH Zurich, Switzerland, since 2020, affiliated with both the Automatic Control Laboratory and the Power Systems Laboratory. He received his Ph.D. in Electrical Engineering: Systems at the University of Michigan, Ann Arbor in 2020, and a M.Sc. in Mathematics from the same institution in 2017. He received his B.Sc. in Mathematics and Physics from Swarthmore College in 2014. His research interests combine power systems analysis with dynamics, control, and optimization of complex, networked systems. The current focus has been on stability and control of power systems with high penetration of renewable energy.
\end{IEEEbiography}

\vspace{-1cm}

\begin{IEEEbiography}[{\includegraphics[width=1in,height=1.25in,clip,keepaspectratio]{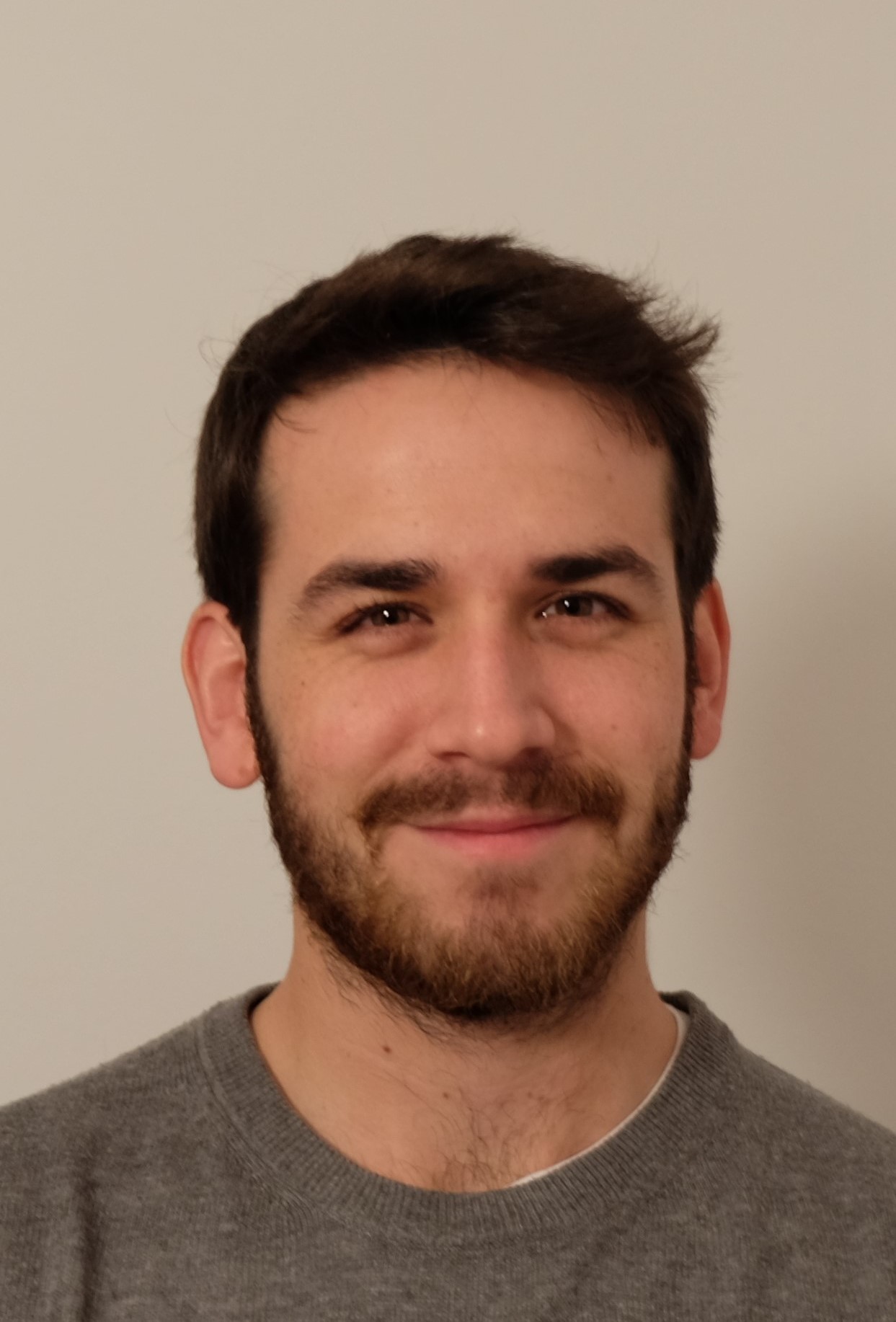}}]{Eduardo Prieto-Araujo} is a Serra Húnter Lecturer with the Electrical Engineering Department at the Technical University of Catalonia (UPC), Barcelona, Spain, where he is part of the CITCEA-UPC research group since 2010. He received the degree in industrial engineering from the School of Industrial Engineering of Barcelona (ETSEIB), UPC, in 2011 and the Ph.D. degree in electrical engineering from UPC in 2016. During 2021, he is a visiting professor at the Automatic Control Laboratory, ETH Zurich, Switzerland. His main interests are renewable generation systems, control of power converters for HVDC applications, interaction analysis between converters, and power electronics dominated power systems.
\end{IEEEbiography}

\vspace{-1cm}

\begin{IEEEbiography}[{\includegraphics[width=1in,height=1.25in,clip,keepaspectratio]{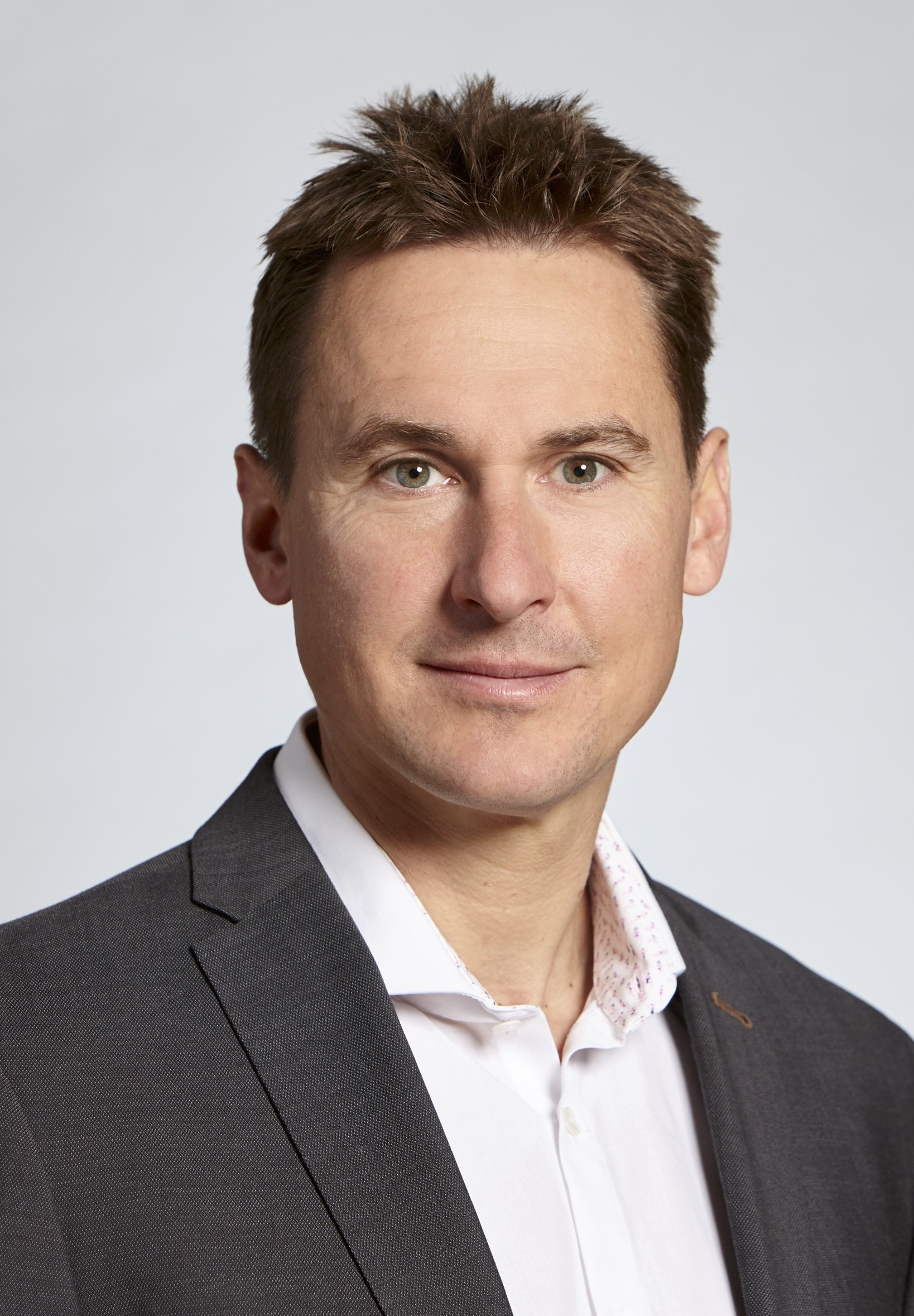}}]{Florian Dörfler} is an Associate Professor at the Automatic Control Laboratory at ETH Zurich, Switzerland, and the Associate Head of the Department of Information Technology and Electrical Engineering. He received his Ph.D. degree in Mechanical Engineering from the University of California at Santa Barbara in 2013, and a Diplom degree in Engineering Cybernetics from the University of Stuttgart, Germany, in 2008. From 2013 to 2014 he was an Assistant Professor at the University of California Los Angeles. His primary research interests are centered around control, optimization, and system theory with applications in network systems, especially electric power grids. He is a recipient of the distinguished young research awards by IFAC (Manfred Thoma Medal 2020) and EUCA (European Control Award 2020). His students were winners or finalists for Best Student Paper awards at the European Control Conference (2013, 2019), the American Control Conference (2016), the Conference on Decision and Control (2020), the PES General Meeting (2020), the PES PowerTech Conference (2017), and the International Conference on Intelligent Transportation Systems (2021). He is furthermore a recipient of the 2010 ACC Student Best Paper Award, the 2011 O. Hugo Schuck Best Paper Award, the 2012-2014 Automatica Best Paper Award, the 2016 IEEE Circuits and Systems Guillemin-Cauer Best Paper Award, and the 2015 UCSB ME Best PhD award.
\end{IEEEbiography}

\end{document}

%% file: Figures/DVPP_ctrl_setup.tex
\tikzstyle{roundnode} =[circle, draw=black!60, fill=black!5, scale=0.85]
\usetikzlibrary{arrows}
\usetikzlibrary{shapes.misc, positioning}
\begin{tikzpicture}[scale=0.32, every node/.style={scale=0.68}]
\draw [rounded corners = 3, dashed,color=black!50, fill=black!5] (-9.4,19.4) rectangle (8.1,-6.75);
\draw  [rounded corners = 3, dashed, color = black!50, fill=black!5](-9.4,-7.75) rectangle (8.1,-9.85);
\draw [rounded corners = 3,-latex, dashed, color = backgroundblue, fill = backgroundblue!10] (-8.4,18.2) rectangle (7.1,10.45);
\draw [rounded corners = 3,-latex, dashed, color = backgroundblue, fill = backgroundblue!10] (-8.4,8.45) rectangle (7.1,-6.35);
\filldraw [-latex, color=black,fill=black!15] (-4.4,-0.35) rectangle (3,-5.15); 
\filldraw [-latex, color=black,fill=black!15] (-4.4,7.25) rectangle (3,2.45);
\draw [rounded corners = 3, fill = backgroundgreen!30] (-2.5,4.45) rectangle (1,2.75); 
\draw [fill=black!40]  (-2.5,6.85) rectangle (1,4.85);

\draw [-latex](1,5.35) -- (1.9,5.35) -- (1.9,3.85) -- (1,3.85);

\draw [-latex](-2.5,3.85) -- (-3.3,3.85) -- (-3.3,5.35) -- (-2.5,5.35);

\draw [-latex, fill = black!40] (-2.5,-0.75) rectangle (1,-2.75);
\draw [rounded corners = 3, -latex,fill=backgroundgreen!30] (-2.5,-3.15) rectangle (1,-4.85);
\draw [-latex](1,-2.25) -- (1.9,-2.25) -- (1.9,-3.65) -- (1,-3.65);

\draw [-latex](-2.5,-3.65) -- (-3.3,-3.65) -- (-3.3,-2.25) -- (-2.5,-2.25);
triangle 60triangle 60
\node [roundnode] at (6.1,9.45) {};

\node (v1) at (6.1,9.45) {};
\draw [-latex](1,6.35) -- (5.9,6.35) ;

\draw [-latex](1,-1.25) -- (6.1,-1.25) --(6.1,6.15) ;

\node [roundnode] at (6.1,6.35) {};

\node [color=black] at (-0.7,1.95) {controllable device $r$};

\node [color=black] at (-0.7,-5.65) {controllable device $n$};

\draw[dotted, thick,color=black] (-0.7,1.45) -- (-0.7,0.75);
\node  at (-10.9,10.65) {$\begin{bmatrix}\Delta f\\ \Delta v\end{bmatrix}$};
\node   [scale = 0.8] at (4.4,5.4) {$\begin{bmatrix}\Delta p_r\\\Delta q_r\end{bmatrix}$};
\node   [scale=0.8] at (4.4,-2.2) {$\begin{bmatrix}\Delta p_n\\\Delta q_n\end{bmatrix}$};
\node   [scale=0.7] at (6.6,8.95) {$+$};
\node   [scale=0.7] at (6.6,9.95) {$+$};

\node   at (-0.7,5.9) {plant $r$};
\node  at (-0.7,-1.7) {plant $n$};
\node  at (-0.7,3.7) {control};
\node  at (-0.7,-3.9) {control};
\node [color=black] at (-0.7,7.75) {$T_r(s)$};
\node [color=black] at (-0.7,0.15) {$T_n(s)$};

%\node [color=black!50] at (-0.7,-7.5) {aggregated DVPP};
\node [color=black!50] at (-0.7,19.9) {DVPP: collection of heterogeneous DERs};
\node  at (9.7,10.65) {$\begin{bmatrix}\Delta p_\mathrm{agg}\\ \Delta q_\mathrm{agg}\end{bmatrix}$};

\draw [-latex](-12.8,9.45)--(-7.4,9.45) node (v2) {} -- (-7.4,-1.25) -- (-2.5,-1.25);

\draw [-latex] (-7.4,6.35) -- (-2.5,6.35);
\fill[black](-7.4,6.35)circle (0.7 mm); 

\draw [-latex](6.35,9.45) -- (11.6,9.45);

\node [scale = 1.5] at (-0.7,-7.3) {$\approx$};
\node [color=black!50] at (-0.7,-10.35) {desired aggregate behavior};
\draw  (-3,-8.05) rectangle (1.5,-9.55);
\node at (-0.7,-8.8) {$T_\mathrm{des}(s)$};
\draw [-latex](1.5,-8.8) -- (11.6,-8.85); 
\draw [-latex](-12.8,-8.8) -- (-3,-8.8);
\node at (-10.9,-7.65) {$\begin{bmatrix}\Delta f\\ \Delta v\end{bmatrix}$};
\node at (9.7,-7.65) {$\begin{bmatrix}\Delta p_\mathrm{des}\\ \Delta q_\mathrm{des}\end{bmatrix}$};
\draw  [color=black,fill=black!15]  (-4.4,17.7) rectangle (3,15.7);
\draw  [color=black,fill=black!15] (-4.4,13.7) rectangle (3,11.7);
\draw [-latex] (-7.4,9.45)--(-7.4,12.7)--(-4.4,12.7);
\fill[black](-7.4,9.45)circle (0.5 mm); 
 -- (-5.5,12.5) node (v3) {} -- (-4.2,12.5); 
\draw [-latex](-7.4,12.7) -- (-7.4,16.7) -- (-4.4,16.7);
\fill[black](-7.4,12.7)circle (0.7 mm); 
\node [scale=0.7] at (5.6,5.85) {$+$};
\node [scale=0.7] at (6.6,5.85) {$+$};

\node [roundnode] at (6.1,12.7) {};

\draw [-latex](3,16.7) -- (6.1,16.7) -- (6.1,13);

\draw [-latex](3,12.7) -- (5.9,12.7);
\draw [-latex](6.1,12.4) -- (6.1,9.75);
\draw [-latex](6.1,6.65) -- (6.1,9.15);
\node [scale=0.7] at (6.6,13.2) {$+$};
\node [scale=0.7] at (5.6,13.2) {$+$};

\draw[dotted, thick,color=black] (-0.7,14.7)  -- (-0.7,14) ;

\node at (-0.7,15.2) {non-controllable device 1};
\node at (-0.85,11.1) {non-controllable device $r\text{-}1$};
\node at (-0.7,16.7) {$T_1(s)$};
\node at (-0.7,12.7) {$T_{r-1}(s)$};
\node [scale = 0.8] at (4.4,11.8) {$\begin{bmatrix}\Delta p_{r\text{-}1}\\\Delta q_{r\text{-}1}\end{bmatrix}$};
\node [scale = 0.8] at (4.4,15.8) {$\begin{bmatrix}\Delta p_1\\\Delta q_1\end{bmatrix}$};

\node [color=backgroundblue] at (-0.7,8.95) {set of controllable devices $\mathcal{C}$};
\node  [color=backgroundblue] at (-0.7,18.7) {set of non-controllable devices $\mathcal{N}$};
\draw [-latex](-3.7,6.35) -- (-3.7,3.15) -- (-2.5,3.15); 
\fill[black](-3.7,6.35)circle (0.5 mm); 
\draw [-latex](-3.7,-1.25) -- (-3.7,-4.35) -- (-2.5,-4.35);
\fill[black](-3.7,-1.25)circle (0.5 mm); 
\draw [-latex](2.3,6.35) -- (2.3,3.15) -- (1,3.15);
\fill[black](2.3,6.35) circle (0.5 mm); 
\draw [-latex](2.3,-1.25) -- (2.3,-4.35)--(1,-4.35);
\fill[black](2.3,-1.25) circle (0.5 mm); 
\end{tikzpicture}

%% file: Figures/communication_strucutre_centralized.tex
\begin{tikzpicture}[scale=0.42, every node/.style={scale=0.6}]
\draw  (-2.5,1) ellipse (2 and 1.25);
\node at (-2.5,1.35) {DVPP};
\node at (-2.5,0.65) {operator};

\draw [-latex,color=black!15,fill=black!15] (-8.9,2.8) ellipse (2 and 1.5);
\draw [fill = black!40] (-9.8,3.9) rectangle (-8,2.9);
\draw [rounded corners = 1.5, fill = backgroundgreen!30] (-9.8,2.5) rectangle (-8,1.7);
\draw [-latex](-8,3.3) -- (-7.4,3.3) -- (-7.4,2.1) -- (-8,2.1); 
\draw [-latex](-9.8,2.1) -- (-10.4,2.1) -- (-10.4,3.3) -- (-9.8,3.3);
\node at (-8.9,3.4) {plant};
\node at (-8.9,2.1) {control};

\draw [-latex,color=black!15,fill=black!15] (-2.7,5.7) ellipse (2 and 1.5);
\draw [fill = black!40]  (-3.6,6.8) rectangle (-1.8,5.8);
\draw [rounded corners = 1.5, fill = backgroundgreen!30] (-3.6,5.4) rectangle (-1.8,4.6);
\draw [-latex](-1.8,6.2) -- (-1.2,6.2) -- (-1.2,5) -- (-1.8,5); 
\draw [-latex](-3.6,5) -- (-4.2,5) -- (-4.2,6.2) -- (-3.6,6.2);
\node at (-2.7,6.3) {plant};
\node at (-2.7,5) {control};

\draw [-latex](-2.4,2.25) -- (-2.4,4.2); 
\draw [-latex](-2.8,4.2) -- (-2.8,2.2);

\draw [-latex](-2.5,-1.4) -- (-2.5,-0.25); 
\draw [-latex](-2.9,-0.25) -- (-2.9,-1.4);

\draw [-latex](-7.1,2.1) -- (-4.2,1.7); 
\draw [-latex](-4.4,1.4) -- (-7.4,1.8);

\draw [-latex](-0.5,1) -- (1.3,0.1); 
\draw [-latex](1.3,-0.3) -- (-0.6,0.6);

\node at (-2.7,-1.9) {$\cdots$};

\node at (-6.3,1.25) {$\theta_i^k(t)$};
\node at (-5.8,2.45) {$y_i^{\mathrm{max},k}(t)$};

\node at (-1.6,3.6) {$\theta_j^k(t)$};
\node at (-3.9,3.6) {$y_j^{\mathrm{max},k}(t)$};

\node at (-8.9,4.6) {device $i$};
\node at (-2.7,7.5) {device $j$};
\node at (2,-0.5) {$\cdots$};

\end{tikzpicture}

%% file: Figures/communication_structure_distributed.tex
\begin{tikzpicture}[scale=0.42, every node/.style={scale=0.6}]

\draw [-latex,color=black!15,fill=black!15] (8,-0.2) ellipse (2 and 1.5);
\draw [fill = black!40]  (7.1,0.9) rectangle (8.9,-0.1);
\draw [rounded corners = 1.5, fill=backgroundgreen!30] (7.1,-0.5) rectangle (8.9,-1.3);
\draw [-latex](8.9,0.3) -- (9.5,0.3) -- (9.5,-0.9) -- (8.9,-0.9); 
\draw [-latex](7.1,-0.9) -- (6.5,-0.9) -- (6.5,0.3) -- (7.1,0.3);
\node at (8,0.4) {plant};
\node at (8,-0.9) {control};

\draw [-latex,color=black!15,fill=black!15] (7.8,5.7) ellipse (2 and 1.5);
\draw [fill = black!40]  (6.9,6.8) rectangle (8.7,5.8);
\draw [rounded corners = 1.5, fill = backgroundgreen!30] (6.9,5.4) rectangle (8.7,4.6);
\draw [-latex](8.7,6.2) -- (9.3,6.2) -- (9.3,5) -- (8.7,5); 
\draw [-latex](6.9,5) -- (6.3,5) -- (6.3,6.2) -- (6.9,6.2);
\node at (7.8,6.3) {plant};
\node at (7.8,5) {control};

\draw [-latex](8.1,1.3) -- (8.1,4.2); 
\draw [-latex](7.7,4.2) -- (7.7,1.3);

\draw [-latex](6.4,0.7)--(4.2,1.9) ; 
\draw [-latex](3.9,1.6)--(6.1,0.4);
\node at (3.5,2) {$\cdots$};

\node at (7,3) {$\theta_i^k(t)$};
\node at (6.6,2.2) {$y_i^{\mathrm{max},k}(t)$};

\node at (8.8,3.7) {$\theta_j^k(t)$};
\node at (9.2,2.9) {$y_j^{\mathrm{max},k}(t)$};

\node at (7.8,7.5) {device $i$};
\node at (8,-2) {device $j$};

\draw [-latex](5.8,6)--(3.4,5.4) ; 
\draw [-latex](3.8,5.1)--(5.8,5.6) ;
\draw [-latex](9.9,0.4) -- (10.5,1.4); 
\draw [-latex](10.7,1) -- (10,-0.2);

\node at (3.1,4.8) {$\cdots$};
\node at (11.1,1.4) {$\cdots$};

\end{tikzpicture}

%% file: Figures/Local_matching_ctrl.tex
\tikzstyle{roundnode} =[circle, draw=black!60, fill=black!5,scale=0.75]
\begin{tikzpicture}[scale=0.5, every node/.style={scale=0.65}]
\draw [fill=black!15](-6.1,2.4) rectangle (6.1,-5.1);
\draw  [fill=black!40](-3.4,1.7) rectangle (3.4,0.4);
\draw [rounded corners = 3,fill=backgroundgreen!30](-4.9,-0.4) rectangle (4.9,-4.8);
\node at (0,1.3) {$\Dot{x}=Ax+Bu+Ew$};
\node at (0,0.8) {$y=Cx+Du+Fw$};
\draw  [fill=backgroundgreen!60](-1,-1) rectangle (1,-2.6);
\draw  [fill=backgroundgreen!60](-3.7,-3.2) rectangle (3.75,-4.55);

\node at (0,2.7) {$T(s)$};
\node at (0,-1.8) {$K(\Theta)$};
\node at (0,-3.65) {$\Dot{x}^{\mathrm{r}}=A^{\mathrm{r}}(\Theta)x^{\mathrm{r}}+E^{\mathrm{r}}(\Theta)w$};

\node at (0,-4.15) {$y^{\mathrm{r}}=C^{\mathrm{r}}(\Theta)x^{\mathrm{r}}+F^{\mathrm{r}}(\Theta)w$};
\node at (0,2) {linearized plant};
\node at (0,-0.15) {matching control};
\node at (0,-0.75) {feedback gain};
\node at (0,-2.9) {LPV reference model $M(s)\cdot T_\mathrm{des}(s)$};
\draw [-latex](-7.3,1.5) -- (-3.4,1.5);
\draw [-latex](-5.6,1.5) -- (-5.6,-3.9) -- (-3.7,-3.9);
\draw [-latex](3.4,0.6) -- (5.2,0.6) -- (5.2,-1.25) -- (1,-1.25); 
\draw [-latex](3.4,1.5) -- (7.3,1.5); 
\draw [-latex](5.6,1.5) -- (5.6,-1.8) -- (4.4,-1.8); 
\draw [-latex](3.75,-3.4) -- (4,-3.4) -- (4,-2.4) -- (1,-2.4);
\node [roundnode] at (4.25,-1.8) {};
\draw [-latex](3.75,-4.3) -- (4.25,-4.3) -- (4.25,-1.95);
\draw [-latex] (2.6,-1.45) rectangle (3.1,-2.15);
\draw [-latex](2.6,-1.8) -- (1,-1.8); 
\draw [-latex](4.1,-1.8) -- (3.1,-1.8);
\node at (2.85,-1.8) {$\int$};
\node at (-6.8,1.75) {$w$};
\node at (6.8,1.75) {$y$};
\draw [-latex](-1,-1.8) -- (-5.2,-1.8) -- (-5.2,0.6) -- (-3.4,0.6);
\node [scale=0.7] at (4.5,-2.4) {$-$};
\node [scale=0.7] at (4.5,-1.55) {$+$};
\node at (3.9,0.85) {$x$};
\node at (2.2,-1.6) {$\sigma$};
\node at (3.8,-1.6) {$\varepsilon$};
\node at (2.2,-2.15) {$x^{\mathrm{r}}$};
\node at (4.6,-4.2) {$y^{\mathrm{r}}$};
\node at (-3.9,0.85) {$u$};
\fill[black](-5.6,1.5)circle (0.5 mm); 
\fill[black](5.6,1.5)circle (0.5 mm); 
\end{tikzpicture}

%% file: Figures/H_inf_matching_ctrl.tex
\tikzstyle{roundnode} =[circle, draw=black!60, fill=black!5, very thick, minimum size=3mm]
\usetikzlibrary{arrows}
\usetikzlibrary{shapes.misc, positioning}

\begin{tikzpicture}[scale=0.5, every node/.style={scale=0.65}]
\draw [rounded corners=8,dashed] (-4.6,-2.9) rectangle (4.6,-6.8);
\draw  [fill=backgroundgreen!60](-1,-5.1) rectangle (1,-6.6);
\draw  [](-3.8,-3.1) rectangle (3.8,-4.8);
\node at (0,-5.85) {$K(\Theta)$};
\node at (0,-3.65) {$\dot{z}=\mathcal{A}(\Theta)z+\mathcal{B}(\Theta)u+{\mathcal{E}}(\Theta)w$};
\node at (0,-4.35) {$\varepsilon=\mathcal{C}(\Theta)z+\mathcal{D}(\Theta)u+{\mathcal{F}}(\Theta)w$};
\draw [-latex](-6.05,-3.5) -- (-3.8,-3.5); 
\draw [-latex](3.8,-3.5) -- (6.05,-3.5); 
\draw [-latex](3.8,-4.5) -- (4.3,-4.5) -- (4.3,-5.85) -- (1,-5.85); 
\draw [-latex](-1,-5.85) -- (-4.3,-5.85) -- (-4.3,-4.5) -- (-3.8,-4.5);
\node at (-5.55,-3.25) {$w$};
\node at (5.55,-3.25) {$\varepsilon$};
\node at (2,-5.6) {$z$};
\node at (-2,-5.6) {$u$};
\node at (0,-2.6) {$N(s)$};
\end{tikzpicture}

%% file: Figures/9bus_system.tex
\usetikzlibrary{arrows}
\begin{tikzpicture}[scale=0.4, every node/.style={scale=0.65}]
	
	\draw(-7.1,19.4) -- (0.1,19.4);
	\draw [ultra thick](-3.4,20.1) -- (-3.4,18.6);
	\draw [ultra thick](-9.5,20.1) -- (-9.5,18.6);
	
	\draw [ultra thick](-6.4,20.1) -- (-6.4,18.6);
	
	\draw [ultra thick](2.7,20.1) -- (2.7,18.6);
	
	\draw [ultra thick](-0.4,20.1) -- (-0.4,18.6);
	\draw[ultra thick] (-5.7,17.9) -- (-4.1,17.9);
	
	\draw [ultra thick](-2.7,17.9) -- (-1.1,17.9);
	\draw [ultra thick](-4.2,16.1) -- (-2.6,16.1);
	\draw [ultra thick](-4.2,13.5) -- (-2.6,13.5);
	\fill[black] (-9.5,19.4)circle (0.7 mm); 
	\fill[black]  (-6.4,19.4)circle (0.7 mm); 
	\fill[black] (-3.4,19.4)circle (0.7 mm); 
	\fill[black]  (-0.4,19.4)circle (0.7 mm); 
	\fill[black] (2.7,19.4) circle (0.7 mm);

	\fill[black] (-1.4,17.9)circle (0.7 mm); 
	\fill[black] (-2.4,17.9)circle (0.7 mm); 
	\fill[black] (-4.4,17.9)circle (0.7 mm); 
	\fill[black] (-5.4,17.9)circle (0.7 mm); 
	\fill[black] (-3.9,16.1)circle (0.7 mm); 
	\fill[black] (-2.9,16.1)circle (0.7 mm); 
	\fill[black] (-3.4,16.1)circle (0.7 mm); 
	\fill[black]  (-3.4,13.5) circle (0.7 mm); 
	
	\node at (-9.5,20.6) {2};
	\node at (-6.4,20.6) {7};
	\node at (-3.4,20.6) {8};
	\node at (-0.4,20.6) {9};
	\node at (2.7,20.6) {3};
	\node at (-6.2,17.9) {5};
	\node at (-0.6,17.9) {6};
	\node at (-2.1,16.1) {4};
	\node at (-2.1,13.5) {1};

	\node (v2) at (-6.4,18.9) {};
	\node at (2.7,18.9) {};
	\draw (-6.4,18.9) -- (-5.4,18.9) -- (-5.4,17.9);
	\fill[black] (-6.4,18.9) circle (0.7 mm); 
	\draw (-4.4,17.9) -- (-4.4,17.2) -- (-3.9,16.8) -- (-3.9,16.1);
	\draw (-2.4,17.9) -- (-2.4,17.2) -- (-2.9,16.8) -- (-2.9,16.1);
	
	\draw (-3.4,16.1) -- (-3.4,15.6) node (v1) {};
	\draw (-0.4,18.9) -- (-1.4,18.9) -- (-1.4,17.9);
	\fill[black] (-0.4,18.9) circle (0.7 mm); 
	\draw [-latex,thick](-3.4,18.9) -- (-2.9,18.9) -- (-2.9,18.1);
	\fill[black] (-3.4,18.9) circle (0.7 mm); 
	\draw [-latex,thick](-4.9,17.9) -- (-4.9,16.7);
	\fill[black](-4.9,17.9)circle (0.7 mm); 
	\draw [-latex, thick](-1.9,17.9) -- (-1.9,16.7);
	\fill[black](-1.9,17.9)circle (0.7 mm);

	%\draw [-latex] (-4.5,4) rectangle (-2.5,2.5);
	\draw[fill=black!20](-10.7,19.4) node (v3) {} circle (7 mm); 
	\draw(-8.3,19.4)  circle (5 mm); 
	\draw(-7.6,19.4)  circle (5 mm); 
	\draw  plot[smooth, tension=.7] coordinates {(-11.2,19.4) (-10.987,19.9) (v3) (-10.46,18.9) (-10.2,19.4)};
	
	%\node at (-3.5,3.25) {DVPP};
	\draw [fill=black!20](3.9,19.4) node (v8) {} circle (7 mm); 
	\draw(-3.4,15.1)  circle (5 mm); 
	\draw(-3.4,14.4)  circle (5 mm); 
	\draw  plot[smooth, tension=.7] coordinates {(3.31,19.4) (3.6,19.9) (v8) (4.2,18.9) (4.47,19.4)};
	\draw(1.3,19.4)  circle (5 mm); 
	\draw(0.6,19.4)  circle (5 mm); 
	\draw (-10,19.4) -- (-8.8,19.4);
	\draw (3.2,19.4) -- (1.8,19.4);
	\draw (-3.4,13) -- (-3.4,13.9);

	\draw[fill=hydro](-3.4,12.3) node {} circle (7 mm); 
	%(-12.7,16.5)
	\draw  plot[smooth, tension=.7] coordinates {(-3.9,12.3) (-3.687,12.8)(-3.4,12.3)(-3.14,11.8) (-2.9,12.3)};
	\node at (-10.7,18.3) {SG 2};
	\node at (-10.7,17.7) {(thermal-based)};
	\node at (3.8,18.3) {SG 3};
	\node at (3.8,17.7) {(thermal-based)};
	\node at (-1.8,12.45) {SG 1};
	\node at (-1.75,11.8) {(hydro)};
	\draw [-latex,thick](2.7,19.8) -- (2.1,19.8) -- (2.1,20.6);
\end{tikzpicture}

%% file: Figures/Conv_model_extended.tex
\usetikzlibrary{circuits.ee.IEC}
\usetikzlibrary{arrows}
\tikzstyle{roundnode} =[circle, draw=backgroundblue!60, fill=backgroundblue!5, scale = 0.5]
\begin{tikzpicture}[circuit ee IEC,scale=0.47, every node/.style={scale=0.65}]

\draw (-2.8,3.2) rectangle (-1.2,2);
\node at (-2,2.6) {$\frac{1}{\tau_\mathrm{dc}s+1}$};

\draw  (0,2.6) ellipse (0.35 and 0.35);
\draw [-latex](0,2.3) node (v2) {} -- (0,2.9) node (v1) {};
\draw(0,2.9)-- (0,3.6) -- (1.5,3.6) node (v3) {} -- (1.5,3.1);
\draw  (1.3,3.1) rectangle (1.7,2.1);
\draw (1.5,2.1) -- (1.5,1.6) node (v4) {} -- (0,1.6) -- (0,2.3); 
\draw (1.5,3.6)-- (2.9,3.6) node (v5) {}--(2.9,2.65);
\draw (2.9,2.55) -- (2.9,1.6) node (v6) {} -- (1.5,1.6) ; 
\draw (2.65,2.65) -- (3.15,2.65);
\draw (2.65,2.55) -- (3.15,2.55); 
\draw (2.9,3.6) -- (4.2,3.6);
\draw (2.9,1.6) -- (4.2,1.6);
\draw  (4.2,3.8) node (v13) {} rectangle (5.9,1.4);

\draw (5.5,3.4) -- (5.5,3.1) -- (5,2.85) node (v3) {};
\draw[-latex](5,2.35) -- (5.5,2.1) ;
\draw (5.5,2.1) -- (5.5,1.6);
\draw (5,3.1) -- (5,2.1);
\draw (4.9,3.1)--(4.9,2.1);
\draw (4.5,2.6) -- (4.9,2.6);

\draw (8.7,2.6) node (v7) {} to [inductor={}] (10.9,2.6) node (v8) {};

\draw  (6.7,2.8) rectangle (7.7,2.4);
\draw (5.9,2.6) -- (6.7,2.6); 
\draw (7.7,2.6) -- (8.7,2.6); 
\draw (10.9,2.6) -- (12.1,2.6) node (v14) {};
\fill[black](11.6,2.6) node (v10) {}circle (0.5 mm); 
\draw [-latex](-1.2,2.6) -- (-0.35,2.6);  
\draw [dashed,-latex,rounded corners = 3, black!50] (1.85,4) rectangle (-3.3,1.2);
\node [orange] at (-3.1,0.4) {$i_\mathrm{dc}^\star$};
\node at (-0.8,3) {$i_\mathrm{dc}$};
\node at (0.8,2.6) {$G_\mathrm{dc}$};
\node at (2.4,2.5) {$C_\mathrm{dc}$};
\node at (3.3,3.2) {$+$};
\node at (3.3,2) {$-$};
\node at (3.65,2.6) {$v_\mathrm{dc}$};
\node at (7.2,3.1) {$R_\mathrm{f}$};
\node at (9.8,3.1) {$L_\mathrm{f}$};

\fill[black](2.9,3.6) circle (0.5 mm); 
\fill[black](2.9,1.6) node (v9) {} circle (0.5 mm); 
\fill[black](1.5,3.6) circle (0.5 mm); 
\fill[black](1.5,1.6) circle (0.5 mm); 
\draw [orange,-latex](2.9,1.6) -- (2.9,0) -- (1.8,0);
\draw [orange, rounded corners = 3,-latex] (-2.7,0.3) rectangle (1.8,-1.2);
\draw [orange,-latex](-2.7,0) -- (-3.7,0) -- (-3.7,2.2)--(-3.7,2.6) -- (-2.8,2.6);
\node [orange] at (-0.4,-0.1) {dc voltage control};
\node [orange] at (2.3,0.4) {$v_\mathrm{dc}$};
\node [black!50] at (-0.8,4.4) {dc energy source model};
\draw [backgroundblue,-latex](5.1,0.6) -- (5.1,1.4); 
\draw [backgroundblue,-latex, rounded corners = 3] (3.4,0.6) rectangle (6.8,-1);
\node [backgroundblue] at (5.1,0.3) {modulation};
\draw [backgroundblue,-latex] (7.9,-0.8) rectangle (9.3,-2);
\draw[backgroundblue] (7.9,-2) -- (9.3,-0.8);
\node [backgroundblue] at (8.4,-1.05) {abc};
\node [backgroundblue] at (8.8,-1.7) {dq};
\draw [backgroundblue,-latex](5.1,-1.8) -- (5.1,-1); 
\draw [backgroundblue, rounded corners = 3, -latex] (10.3,-1) rectangle (15.1,-2.9);
\node [backgroundblue] at (12.7,-1.4) {PLL};
\draw [backgroundblue, -latex](11.6,2.6)  -- (11.6,1);
\node [backgroundblue] at (11,1.6) {$v_\mathrm{abc}$};
\draw [rounded corners = 3, -latex, backgroundblue] (1.9,-3.8) rectangle (8,-6.4);
\node [backgroundblue ] at (5,-4.2) {inner current control loop};
\draw [backgroundblue,-latex](5.1,-3.8) -- (5.1,-3); 
\draw [backgroundblue,-latex](8.6,2.6) -- (8.6,-0.8);
\draw [backgroundblue,-latex] (4.4,-1.8) rectangle (5.8,-3);
\draw [backgroundblue](4.4,-3) -- (5.8,-1.8);
\node [backgroundblue] at (4.9,-2.05) {abc};
\node [backgroundblue] at (5.3,-2.7) {dq};

\draw [backgroundblue,-latex](10.3,-1.4) -- (9.3,-1.4);
\draw [backgroundblue,-latex](9.9,-1.4) -- (9.9,-2.4) -- (5.8,-2.4);

\fill[black](8.6,2.6) circle (0.5 mm); 
\node at (5.1,4.4) {power};
\node at (5.1,4.1) {converter};
\fill[backgroundblue](9.9,-1.4) node (v15) {} circle (0.5 mm); 
\draw [backgroundblue] (8.6,-2) -- (8.6,-2.3);
 \draw[-latex,backgroundblue] (8.6,-2.5) -- (8.6,-4.6) -- (8,-4.6);
\draw [-latex, rounded corners = 3, fill = backgroundgreen!40] (9.6,-8.15) rectangle (12.2,-9.35) node (v11) {};
\node at (10.9,-8.55) {matching};
\node at (10.9,-8.95) {control};
\draw [backgroundblue,-latex](11.6,-2.9) -- (11.6,-6.7);
\node [backgroundblue] at (14.5,-6.1) {$||v_\mathrm{dq}||$};
\node [backgroundblue] at (9.1,1.6) {$i_\mathrm{abc}$};
\draw [backgroundblue,-latex] (1.5,-7.5) rectangle (-2.5,-9.7);

\node [backgroundblue] at (10.3,-0.5) {$\theta_\mathrm{pll}$};
\node [backgroundblue] at (9.1,-4.4) {$i_\mathrm{dq}$};
\node [backgroundblue] at (9.1,-6) {$i_\mathrm{dq}^\star$};

\node [backgroundblue] at (-0.5,-7.9) {power calculation};

\draw [backgroundblue,-latex](-3.3,-8.2) -- (-2.5,-8.2); 
\draw [backgroundblue,-latex ](-3.3,-9.1) -- (-2.5,-9.1); 
\draw [backgroundblue,-latex ](1.5,-8.2) -- (2.2,-8.2); 
\draw [backgroundblue,-latex ](1.5,-9.1) -- (2.2,-9.1); 
\node [backgroundblue] at (1.9,-7.9) {$p$};
\node [backgroundblue] at (1.9,-8.8) {$q$};
\node [backgroundblue] at (-3.1,-7.9) {$v_\mathrm{dq}$};
\node [backgroundblue] at (-3.1,-8.75) {$i_\mathrm{dq}$};
\draw [-latex ](13.6,-8.35) -- (12.2,-8.35); 

\draw [-latex ](13.6,-9.15) -- (12.2,-9.15);

\draw [backgroundblue,-latex ](8.6,-6.7) -- (8.6,-5.7)--(8,-5.7);
\node [backgroundblue] at (5.8,-3.4) {$v_\mathrm{c,dq}^\star$};
\node [backgroundblue] at (6.1,-1.4) {$v_\mathrm{c,abc}^\star$};
\node at (6.6,2) {$v_\mathrm{c,abc}$};
\draw [backgroundblue,-latex](13.8,-2.9) -- (13.8,-6.7) node (v12) {};

\draw [-latex](9.6,-8.35) -- (8.4,-8.35);
\draw [-latex](8.4,-9.15) -- (9.6,-9.15);

\draw [backgroundblue,rounded corners=3,-latex] (7.1,-6.7) rectangle (15,-9.7);
\node [backgroundblue] at (11.1,-6.1) {$f_\mathrm{pll}$};
\draw [backgroundblue,-latex](4.9,-7.5) -- (7.1,-7.5); 

\draw [backgroundblue,-latex](4.9,-8.9) -- (7.1,-8.9); 
\node [backgroundblue] at (5.9,-7.1) {$i_\mathrm{dq}, x_\mathrm{i,dq}$};
\node [backgroundblue] at (11,-7.2) {outer control loop including};
\node [backgroundblue] at (5.9,-8.6) {$p, q$};
\draw [backgroundblue,-latex](0.9,-5) -- (1.9,-5);
\node [backgroundblue] at (1.3,-4.6) {$v_\mathrm{dq}$};
\draw [-stealth](3.1,3.8) -- (v13);
\node at (3.5,4.2) {$i_\mathrm{x}$};
\node [backgroundblue] at (11,-7.65) {$\mathcal{H}_\infty$ matching control};
\node at (14.4,-8.65) {$\cdots$};
\node at (7.8,-8.65) {$\cdots$};
\node [backgroundblue] at (5.8,0.9) {$m_\mathrm{abc}$};
\draw  (12.45,2.6) ellipse (0.35 and 0.35);
\draw (12.85,2.6) ellipse (0.35 and 0.35); 
\draw [dashed](13.15,2.6) -- (15,2.6);
\node at (12.65,3.3) {LV/MV};

\node [orange] at (-0.4,-0.7) {$i_\mathrm{dc}^\star=k_\mathrm{dc}(v_\mathrm{dc}^\star-\,v_\mathrm{dc})$};
\node [backgroundblue] at (11.3,-1.8) {$\dot{x}_\mathrm{pll}=v_\mathrm{q}$};
\node [backgroundblue] at (12.7,-2.4) {$\dot{\theta}_\mathrm{pll}=k_\mathrm{p}^\mathrm{pll}v_\mathrm{q}+k_\mathrm{i}^\mathrm{pll}x_\mathrm{pll}$};
\node [backgroundblue] at (5.1,-0.4) {$m_\mathrm{abc}=\tfrac{2\,v_\mathrm{c,abc}^\star}{v_\mathrm{dc}^\star}$};
\node [backgroundblue] at (3.77,-4.75) {$\dot{x}_\mathrm{i,dq}=i^\star_\mathrm{dq}-\,i_\mathrm{dq}$};

\node [backgroundblue] at (4.2,-5.35) {$v^\star_\mathrm{c,dq}=v_\mathrm{dq}+\mathcal{Z}_\mathrm{f}i_\mathrm{dq}+$};
\node [backgroundblue] at (5.4,-5.95) {$+k_\mathrm{p}^\mathrm{i}(i^\star_\mathrm{dq}-\,i_\mathrm{dq})+k_\mathrm{i}^\mathrm{i}x_\mathrm{i,dq}$};
\node [backgroundblue] at (-0.5,-8.5) {$p=v_\mathrm{d}i_\mathrm{d}+v_\mathrm{q}i_\mathrm{q}$};
\node [backgroundblue] at (-0.5,-9.2) {$q=v_\mathrm{q}i_\mathrm{d}-v_\mathrm{d}i_\mathrm{q}$};
\node at (14.1,2.1) {to DVPP};
\node at (13.95,1.65) {connection};
\node at (14.5,1.2) {point};
\draw  [backgroundblue](10.9,1) rectangle (12.2,-0.2);
\draw [backgroundblue](10.9,-0.2) -- (12.2,1);
\node [backgroundblue] at (11.8,0.1) {dq};
\node [backgroundblue] at (11.4,0.75) {abc};
\draw [backgroundblue,-latex](11.6,-0.2) -- (11.6,-1); 
\draw [backgroundblue,-latex](9.9,-1.4)
 -- (9.9,0.4) -- (10.9,0.4);
\node [backgroundblue]at (9.7,-2.7) {$\theta_\mathrm{pll}$};
\end{tikzpicture}

%% file: Figures/9bus_system_C1.tex
\tikzstyle{roundnode} =[circle, draw=black!60, fill=black!5, very thick, minimum size=3mm]
\usetikzlibrary{arrows}
\usetikzlibrary{shapes.misc, positioning}
\begin{tikzpicture}[scale=0.4, every node/.style={scale=0.65}]
\draw  [rounded corners=8,dashed,black!50,fill=black!5](-6.4,13.2) rectangle (0,10.6);

\draw(-3.6,19) node (v2) {} -- (-0.6,19);
\draw [ultra thick](-3.6,19.7) -- (-3.6,18.2);

\draw [ultra thick](1.8,19.7) -- (1.8,18.2);

\draw [ultra thick](-1,19.7) -- (-1,18.2);

\draw [ultra thick](-2.9,17.5) -- (-1.3,17.5);
\draw [ultra thick](-4.2,16.1) -- (-2.6,16.1);
\draw [ultra thick](-4.2,13.5) -- (-2.6,13.5);

\fill[black] (-3.6,19) node (v8) {}circle (0.7 mm); 
\fill[black]  (-1,19)circle (0.7 mm); 
\fill[black] (1.8,19) circle (0.7 mm);

\fill[black] (-1.6,17.5)circle (0.7 mm); 
\fill[black] (-2.6,17.5)circle (0.7 mm); 
\draw (-3.9,16.6) -- (-3.9,16.1);
\draw [dotted] (-4.2,17.4) -- (-4.2,17)--(-3.9,16.6);

\fill[black] (-3.9,16.1)circle (0.7 mm); 
\fill[black] (-2.9,16.1)circle (0.7 mm); 
\fill[black] (-3.4,16.1)circle (0.7 mm); 
\fill[black]  (-3.4,13.5) circle (0.7 mm);

\node at (-3.6,20.2) {8};
\node at (-1,20.2) {9};
\node at (1.8,20.2) {3};

\node at (-0.9,17.5) {6};
\node at (-2.1,16.1) {4};
\node at (-2.1,13.5) {1};
\node at (1.8,18.5) {};

\draw (-2.6,17.5) -- (-2.6,17) -- (-2.9,16.6) -- (-2.9,16.1);

\draw (-3.4,16.1) -- (-3.4,15.6) node (v1) {};
\draw (-1,18.5) -- (-1.6,18.5) -- (-1.6,17.5);
\fill[black] (-1,18.5) circle (0.7 mm); 
\draw [-latex,thick](-3.6,18.5) -- (-3.1,18.5) -- (-3.1,17.7);
\fill[black] (-3.6,18.5) circle (0.7 mm);

\draw [-latex,thick](-2.1,17.5) -- (-2.1,16.6);
\fill[black](-2.1,17.5)circle (0.7 mm);

%\draw [-latex] (-4.5,4) rectangle (-2.5,2.5);

%\node at (-3.5,3.25) {DVPP};
\draw(-3.4,15.1)  circle (5 mm); 
\draw(-3.4,14.4)  circle (5 mm); 
\draw(0.6,19)  circle (5 mm); 
\draw(-0.1,19)  circle (5 mm); 

\draw (2.4,19) -- (1.1,19);
\draw (-3.4,13) node (v4) {} -- (-3.4,13.9);

\draw[fill=hydro](-5.4,12) node {} circle (7 mm); 
%(-12.7,16.5)
\draw  plot[smooth, tension=.7] coordinates {(-5.9,12) (-5.687,12.5)(-5.4,12)(-5.14,11.5) (-4.9,12)};
\node at (-5.4,10.95) {hydro};
\draw [fill=battery!60] (-4.3,12.5) rectangle (-2.5,11.6);
\node at (-3.4,12.05) {BESS};
\draw  [fill=SC!80](-2.1,12.5) rectangle (-0.3,11.6);
\node  [scale=0.8] at (-1.2,12.15) {super-};
\node [scale=0.8] at (-1.2,11.85) {capacitor};
\draw  (-3.4,13) node (v5) {} -- (-5.4,13) -- (-5.4,12.7);

\draw  (-3.4,13) node (v6) {}-- (-3.4,12.5);
\draw  (-3.4,13) -- (-1.2,13) -- (-1.2,12.5);

\draw [fill=black!20](3.1,19)node(v10) {} circle (7 mm); 
	\draw  plot[smooth, tension=.7] coordinates {(2.51,19) (2.8,19.5) (v10) (3.4,18.5) (3.67,19)};
\draw [dotted](v8) -- (-4.6,19);
	\node at (3,17.9) {SG 3};
% 	\node at (3,17.3) {(thermal-based)};
	\node at (1.2,11.9) {DVPP 1};
		\draw [-latex,thick](1.8,19.4) -- (1.2,19.4) -- (1.2,20.2);
\end{tikzpicture}

    

%% file: Figures/Conv_matching_DVPP1.tex
\tikzstyle{roundnode} =[circle, draw=black!60, fill=black!5, scale=0.75]
\begin{tikzpicture}[scale=0.55, every node/.style={scale=0.65}]
\draw [rounded corners = 3,fill=backgroundgreen!30] (-2.9,7.2) rectangle (7,3);
\draw  [fill=backgroundgreen!60](1.3,6.6) rectangle (2.9,5.2);
\node at (2.1,5.9) {$K_i$};
\draw  [fill=backgroundgreen!60](-1,4.6) rectangle (5.2,3.3);
\node at (2.2,4.2) {$\dot{x}^\mathrm{r}=A^\mathrm{r}_ix^\mathrm{r}+E^\mathrm{r}_i\Delta f_\mathrm{pll}$};

\node at (2.06,3.7) {$\Delta p^\mathrm{r}=C^\mathrm{r}_ix^\mathrm{r}+F^\mathrm{r}_i\Delta f_\mathrm{pll}$};

\draw [-latex](5.2,4.3) -- (5.5,4.3) -- (5.5,5.4) -- (2.9,5.4); 
\draw [-latex](7.15,6.4) -- (2.9,6.4); 
\draw [-latex](5.2,3.8) -- (5.8,3.8) -- (5.8,5.8); 

\draw [-latex] (4.3,6.2) rectangle (4.7,5.6);

\node [roundnode] at (5.8,5.9) {}; 
\draw [-latex](5.65,5.9) -- (4.7,5.9); 
\draw [-latex](4.3,5.9) -- (2.9,5.9);

\node [roundnode] at (7.3,6.4) {};
\draw [backgroundblue,-latex](9.3,6.4) -- (7.45,6.4); 
\draw [-latex](7.3,7.2) -- (7.3,6.5);

\node [roundnode] at (7.3,5.9) {};
\draw [backgroundblue,-latex](9.3,5.9) -- (7.45,5.9); 
\draw (7.15,5.9) -- (6.45,5.9); 
\draw [-latex](7.15,5.9)-- (5.9,5.9);

\draw [-latex](1.3,6) -- (-3.5,6);

\node [roundnode] at (-3.6,6) {};

\draw [-latex](-3.6,6.6) -- (-3.6,6.1); 

\draw [backgroundblue,-latex](-3.75,6) -- (-5.2,6);

\draw [-latex](-3.45,4) -- (-1,4);

\node [roundnode] at (-3.6,4) {};

\draw [-latex](-3.6,4.6) -- (-3.6,4.1); 

\draw [backgroundblue,-latex](-5.2,4) -- (-3.7,4); 

\draw [backgroundblue,rounded corners =3,-latex] (-4.3,8.5) rectangle (8.4,1);
\node at (2.1,7.4) {matching control};
\node at (2.1,6.85) {feedback gain};
\node at (2.2,4.85) {reference model $M_i(s)\cdot T_{\mathrm{des}}(s)$};
\node [backgroundblue] at (2.1,8.2) {outer control loop including $\mathcal{H}_\infty$ matching control (DVPP 1)};
\node at (-1.7,4.3) {$\Delta f_\mathrm{pll}$};

\node at (-3.9,4.6) {$f^\star$};
\node [backgroundblue] at (-4.9,4.3) {$f_\mathrm{pll}$};

\node [backgroundblue] at (-4.9,6.3) {$i_\mathrm{d}^\star$};
\node [backgroundblue] at (-4.9,2.3) {$i_\mathrm{q}^\star$};
\node at (-3.95,6.6) {$i_\mathrm{d,0}^\star$};

\node at (0.4,6.3) {$\Delta i_\mathrm{d}^\star$};

\draw [-latex](7.3,5.1) -- (7.3,5.78);
\node at (4.5,5.9) {$\int$};
\node at (3.8,6.1) {$\sigma_\mathrm{p}$};
\node at (5.6,3.55) {$\Delta p^\mathrm{r}$};

\node at (4.5,6.65) {$\Delta i_\mathrm{d},\Delta x_\mathrm{i,d}$};
\node at (5.45,4.1) {$x^\mathrm{r}$};
\node at (7.85,7.75) {$i_\mathrm{d,0}$};
\node at (7.85,7.4) {$x_\mathrm{i,d,0}$};
\node at (7.6,5.3) {$p_0$};

\node [backgroundblue] at (9.2,6.65) {$i_\mathrm{d}, x_\mathrm{i,d}$};
\node [backgroundblue] at (8.65,2.2) {$q$};
\node [backgroundblue] at (8.65,6.1) {$p$};
\node [scale=0.7] at (-3.4,6.3) {+};

\node [scale=0.7] at (-3.2,6.2) {+};

\node [scale=0.7] at (-3.8,4.3) {-};
\node [scale=0.7] at (-4,4.2) {+};

\node [scale=0.7] at (6.1,5.7) {+};
\node [scale=0.7] at (6,5.5) {-};

\node [scale=0.7] at (7.5,6.7) {-};
\node [scale=0.7] at (7.7,6.6) {+};

\node [scale=0.7] at (7.5,5.6) {-};
\node [scale=0.7] at (7.7,5.7) {+};

\node at (6.6,5.65) {$\Delta p$};

\draw  [rounded corners=3](-0.2,2.8) rectangle (4.3,1.2);
\node at (1,2) {$\dot{x}_\mathrm{q}=q-q^\star$};
\draw [-latex,backgroundblue](9.3,2) -- (4.3,2);
\draw [-latex,backgroundblue](-0.2,2) -- (-5.2,2);
\node at (2.1,1.5) {$i_\mathrm{q}^\star=k_\mathrm{p}^\mathrm{q}(q-q^\star)+k_\mathrm{i}^\mathrm{q}x_\mathrm{q}$};
\node at (2.1,2.5) {reactive power control};

\end{tikzpicture}

%% file: Figures/9bus_system_C2.tex
\tikzstyle{roundnode} =[circle, draw=black!60, fill=black!5, very thick, minimum size=3mm]
\usetikzlibrary{arrows}
\usetikzlibrary{shapes.misc, positioning}
\begin{tikzpicture}[scale=0.4, every node/.style={scale=0.65}]
\draw  [rounded corners=8,dashed,black!50,fill=black!5](-6.4,13.2) rectangle (0,10.6);
\draw  [rounded corners =8,dashed,black!50,fill=black!5](2.2,20.9) rectangle (4.9,17.1);
\draw(-3.6,19) node (v2) {} -- (-0.6,19);
\draw [ultra thick](-3.6,19.7) -- (-3.6,18.2);

\draw [ultra thick](1.8,19.7) -- (1.8,18.2);

\draw [ultra thick](-1,19.7) -- (-1,18.2);

\draw [ultra thick](-2.9,17.5) -- (-1.3,17.5);
\draw [ultra thick](-4.2,16.1) -- (-2.6,16.1);
\draw [ultra thick](-4.2,13.5) -- (-2.6,13.5);

\fill[black] (-3.6,19) node (v8) {}circle (0.7 mm); 
\fill[black]  (-1,19)circle (0.7 mm); 
\fill[black] (1.8,19) circle (0.7 mm);

\fill[black] (-1.6,17.5)circle (0.7 mm); 
\fill[black] (-2.6,17.5)circle (0.7 mm); 
\draw (-3.9,16.6) -- (-3.9,16.1);
\draw [dotted] (-4.2,17.4) -- (-4.2,17)--(-3.9,16.6);

\fill[black] (-3.9,16.1)circle (0.7 mm); 
\fill[black] (-2.9,16.1)circle (0.7 mm); 
\fill[black] (-3.4,16.1)circle (0.7 mm); 
\fill[black]  (-3.4,13.5) circle (0.7 mm);

\node at (-3.6,20.2) {8};
\node at (-1,20.2) {9};
\node at (1.8,20.2) {3};

\node at (-0.9,17.5) {6};
\node at (-2.1,16.1) {4};
\node at (-2.1,13.5) {1};
\node at (1.8,18.5) {};

\draw (-2.6,17.5) -- (-2.6,17) -- (-2.9,16.6) -- (-2.9,16.1);

\draw (-3.4,16.1) -- (-3.4,15.6) node (v1) {};
\draw (-1,18.5) -- (-1.6,18.5) -- (-1.6,17.5);
\fill[black] (-1,18.5) circle (0.7 mm); 
\draw [-latex,thick](-3.6,18.5) -- (-3.1,18.5) -- (-3.1,17.7);
\fill[black] (-3.6,18.5) circle (0.7 mm);

\draw [-latex,thick](-2.1,17.5) -- (-2.1,16.5);
\fill[black](-2.1,17.5)circle (0.7 mm);

%\draw [-latex] (-4.5,4) rectangle (-2.5,2.5);

%\node at (-3.5,3.25) {DVPP};
\draw(-3.4,15.1)  circle (5 mm); 
\draw(-3.4,14.4)  circle (5 mm); 
\draw(0.6,19)  circle (5 mm); 
\draw(-0.1,19)  circle (5 mm); 

\draw (2.4,19) -- (1.1,19);
\draw (-3.4,13) node (v4) {} -- (-3.4,13.9);

\draw[fill=hydro](-5.4,12) node {} circle (7 mm); 
%(-12.7,16.5)
\draw  plot[smooth, tension=.7] coordinates {(-5.9,12) (-5.687,12.5)(-5.4,12)(-5.14,11.5) (-4.9,12)};
\node at (-5.4,10.95) {hydro};
\draw [fill=battery!60] (-4.3,12.5) rectangle (-2.5,11.6);
\node at (-3.4,12.05) {BESS};
\draw  [fill=SC!80](-2.1,12.5) rectangle (-0.3,11.6);
\node  [scale=0.8] at (-1.2,12.15) {super-};
\node [scale=0.8] at (-1.2,11.85) {capacitor};
\draw  (-3.4,13) node (v5) {} -- (-5.4,13) -- (-5.4,12.7);

\draw  (-3.4,13) node (v6) {}-- (-3.4,12.5);
\draw  (-3.4,13) -- (-1.2,13) -- (-1.2,12.5);

\node at (1.2,11.9) {DVPP 1};
\draw [fill=wind!60] (2.7,20.5) rectangle (4.5,19.6);
\draw [fill=PV] (2.7,19.4) rectangle (4.5,18.5);
\draw [fill=STATCOM] (2.7,18.3) rectangle (4.5,17.4);
\node at (3.6,20.05) {wind};
\node at (3.6,18.95) {PV};
\node  [scale=0.65] at (3.6,18.02) {STATCOM};
\node  [scale=0.65] at (3.6,17.68) {\& battery};
\draw (2.4,19) node (v7) {} -- (2.4,20.1) -- (2.7,20.1);
\draw (2.4,19) -- (2.7,19) --(2.4,19)  -- (2.4,17.8) -- (2.7,17.8);
\node at (3.6,16.6) {DVPP 3};

\draw [dotted](v8) -- (-4.6,19);
		\draw [-latex,thick](1.8,19.4) -- (1.2,19.4) -- (1.2,20.2);
\end{tikzpicture}

    

%% file: Figures/Conv_matching_DVPP3.tex
\tikzstyle{roundnode} =[circle, draw=black!60, fill=black!5, scale=0.75]
\begin{tikzpicture}[scale=0.55, every node/.style={scale=0.65}]
\draw [fill=backgroundgreen!30,rounded corners = 3] (-3,-0.5) rectangle (7.1,-6.2);
\draw  [fill=backgroundgreen!60](1.3,-1) rectangle (2.9,-3.2);
\node at (2.1,-2.1) {$K_i(\Theta)$};
\draw  [fill=backgroundgreen!60](-1.4,-3.8) rectangle (5.4,-6.05);
\node at (2.38,-4.4) {$\dot{x}^\mathrm{r}=A_i^\mathrm{r}(\Theta)x^\mathrm{r}+E_i^\mathrm{r}(\Theta)\begin{bmatrix}\Delta f_\mathrm{pll}\\||\Delta v_\mathrm{dq}||\end{bmatrix}$};

\node at (2,-5.45) {$\begin{bmatrix}\Delta p^\mathrm{r}\\\Delta q^\mathrm{r}\end{bmatrix}=C_i^\mathrm{r}(\Theta)x^\mathrm{r}+F_i^\mathrm{r}(\Theta)\begin{bmatrix}\Delta f_\mathrm{pll}\\||\Delta v_\mathrm{dq}||\end{bmatrix}$};

\draw [-latex](5.4,-4.3) -- (5.6,-4.3) -- (5.6,-3) -- (2.9,-3); 
\draw [-latex](7.25,-1.2) -- (2.9,-1.2); 
\draw [-latex](5.4,-4.9) -- (5.9,-4.9) -- (5.9,-2.7); 
\draw [-latex](5.4,-5.5) -- (6.4,-5.5) -- (6.4,-1.7);
\draw [-latex] (4.3,-2.3) rectangle (4.7,-2.9);
\draw [-latex] (4.3,-1.3) rectangle (4.7,-1.9);
\node [roundnode] at (5.9,-2.6) {}; 
\draw [-latex](5.75,-2.6) -- (4.7,-2.6); 
\draw [-latex](4.3,-2.6) -- (2.9,-2.6);
\node [roundnode] at (6.4,-1.6) {};
\draw [-latex](6.25,-1.6) -- (4.7,-1.6);
\draw [-latex](4.3,-1.6) -- (2.9,-1.6);
\node [roundnode] at (7.4,-1.2) {};
\draw [backgroundblue,-latex](9.4,-1.2) -- (7.55,-1.2); 
\draw [-latex](7.4,-0.6) -- (7.4,-1.1);
\draw [-latex](7.75,-1.6) -- (6.5,-1.6);
\node [roundnode] at (7.9,-1.6) {};
\draw [backgroundblue,-latex](9.4,-1.6) -- (8.05,-1.6);
\node [roundnode] at (7.4,-2.6) {};
\draw [backgroundblue,-latex](9.4,-2.6) -- (7.55,-2.6); 
\draw (7.25,-2.6) -- (6.45,-2.6); 
\draw [-latex](6.36,-2.6) -- (6,-2.6);
\draw [-latex](7.4,-3.3) -- (7.4,-2.75);
\draw [-latex](7.9,-2.3) -- (7.9,-1.75);
\draw [-latex](1.3,-1.6) -- (-3.5,-1.6);
\draw [-latex](1.3,-2.6) -- (-3.5,-2.6);
\node [roundnode] at (-3.6,-1.6) {};
\node [roundnode] at (-3.6,-2.6) {};
\draw [-latex](-3.6,-1) -- (-3.6,-1.5); 
\draw [-latex](-3.6,-3.2) -- (-3.6,-2.7);
\draw [backgroundblue,-latex](-3.75,-1.6) -- (-5.2,-1.6);
\draw [backgroundblue,-latex](-3.75,-2.6) -- (-5.2,-2.6);

\draw [-latex](-3.45,-4.4) -- (-1.4,-4.4);
\draw[-latex](-3.45,-5.4) -- (-1.4,-5.4);
\node [roundnode] at (-3.6,-4.4) {};
\node [roundnode] at (-3.6,-5.4) {}; 
\draw [-latex](-3.6,-3.8) -- (-3.6,-4.3); 
\draw [-latex](-3.6,-6) -- (-3.6,-5.5); 
\draw [backgroundblue,-latex](-5.2,-4.4) -- (-3.7,-4.4); 
\draw [backgroundblue,-latex](-5.2,-5.4) -- (-3.7,-5.4);
\draw [backgroundblue,rounded corners =3,-latex] (-4.3,0.8) rectangle (8.5,-6.45);
\node at (2.1,-0.3) {matching control};
\node at (2.1,-0.8) {controller};
\node at (2.2,-3.55) {reference model $M_i(s)\cdot T_{\mathrm{des}}(s)$};
\node [backgroundblue] at (2.1,0.5) {outer control loop including $\mathcal{H}_\infty$ matching control (DVPP 3)};
\node at (-2.1,-4.1) {$\Delta f_\mathrm{pll}$};
\node at (-2.2,-5.1) {$||\Delta v_\mathrm{dq}||$};
\node at (-3.9,-6) {$v^\star$};
\node at (-3.9,-3.8) {$f^\star$};
\node [backgroundblue] at (-4.9,-4.1) {$f_\mathrm{pll}$};
\node [backgroundblue] at (-4.9,-5.1) {$||v_\mathrm{dq}||$};
\node [backgroundblue] at (-4.9,-1.3) {$i_\mathrm{d}^\star$};
\node [backgroundblue] at (-4.9,-2.3) {$i_\mathrm{q}^\star$};
\node at (-3.95,-1) {$i_\mathrm{d,0}^\star$};
\node at (-3.95,-3.1) {$i_\mathrm{q,0}^\star$};
\node at (0.4,-1.3) {$\Delta i_\mathrm{d}^\star$};
\node at (0.4,-2.3) {$\Delta i_\mathrm{q}^\star$};
\node at (4.5,-1.6) {$\int$};
\node at (3.9,-1.4) {$\sigma_\mathrm{q}$};
\node at (4.5,-2.6) {$\int$};
\node at (3.9,-2.4) {$\sigma_\mathrm{p}$};

\node at (5.8,-5.2) {$\Delta p^\mathrm{r}$};
\node at (5.8,-5.8) {$\Delta q^\mathrm{r}$};
\node at (4.4,-0.95) {$\Delta i_\mathrm{dq},\Delta x_\mathrm{i,dq}$};
\node at (5.65,-4.55) {$x^\mathrm{r}$};
\node at (7.95,-0.15) {$i_\mathrm{dq,0}$};
\node at (7.95,-0.5) {$x_\mathrm{i,dq,0}$};
\node at (7.7,-3.2) {$p_0$};
\node at (8.2,-2.2) {$q_0$};
\node [backgroundblue] at (9.3,-0.95) {$i_\mathrm{dq}, x_\mathrm{i,dq}$};
\node [backgroundblue] at (8.75,-1.4) {$q$};
\node [backgroundblue] at (8.75,-2.4) {$p$};
\node [scale=0.7] at (-3.4,-1.3) {+};

\node [scale=0.7] at (-3.2,-1.4) {+};
\node [scale=0.7] at (-3.2,-2.8) {+};
\node [scale=0.7] at (-3.4,-2.9) {+};
\node [scale=0.7] at (-3.8,-4.1) {-};
\node [scale=0.7] at (-4,-4.2) {+};
\node [scale=0.7] at (-4,-5.6) {+};
\node [scale=0.7] at (-3.8,-5.7) {-};
\node [scale=0.7] at (6.2,-2.8) {+};
\node [scale=0.7] at (6.1,-3) {-};
\node [scale=0.7] at (6.8,-1.8) {+};
\node [scale=0.7] at (6.6,-2) {-};
\node [scale=0.7] at (7.6,-0.9) {-};
\node [scale=0.7] at (7.8,-1) {+};
\node [scale=0.7] at (8.3,-1.8) {+};
\node [scale=0.7] at (8.1,-1.9) {-};
\node [scale=0.7] at (7.6,-2.9) {-};
\node [scale=0.7] at (7.8,-2.8) {+};
\node at (7.45,-1.85) {$\Delta q$};
\node at (6.75,-2.85) {$\Delta p$};
\end{tikzpicture}

%% file: Figures/9bus_system_C3.tex
\usetikzlibrary{arrows}
\begin{tikzpicture}[scale=0.4, every node/.style={scale=0.65}]
	\draw [rounded corners=8,dashed,black!50,fill=black!5] (-7.5,13.2) rectangle (1.1,10.4);
	\draw(-7.1,19.4) -- (0.1,19.4);
	\draw [ultra thick](-3.4,20.1) -- (-3.4,18.6);
	\draw [ultra thick](-9.5,20.1) -- (-9.5,18.6);
	
	\draw [ultra thick](-6.4,20.1) -- (-6.4,18.6);
	
	\draw [ultra thick](2.7,20.1) -- (2.7,18.6);
	
	\draw [ultra thick](-0.4,20.1) -- (-0.4,18.6);
	\draw[ultra thick] (-5.7,17.9) -- (-4.1,17.9);
	
	\draw [ultra thick](-2.7,17.9) -- (-1.1,17.9);
	\draw [ultra thick](-4.2,16.1) -- (-2.6,16.1);
	\draw [ultra thick](-4.2,13.5) -- (-2.6,13.5);
	\fill[black] (-9.5,19.4)circle (0.7 mm); 
	\fill[black]  (-6.4,19.4)circle (0.7 mm); 
	\fill[black] (-3.4,19.4)circle (0.7 mm); 
	\fill[black]  (-0.4,19.4)circle (0.7 mm); 
	\fill[black] (2.7,19.4) circle (0.7 mm);

	\fill[black] (-1.4,17.9)circle (0.7 mm); 
	\fill[black] (-2.4,17.9)circle (0.7 mm); 
	\fill[black] (-4.4,17.9)circle (0.7 mm); 
	\fill[black] (-5.4,17.9)circle (0.7 mm); 
	\fill[black] (-3.9,16.1)circle (0.7 mm); 
	\fill[black] (-2.9,16.1)circle (0.7 mm); 
	\fill[black] (-3.4,16.1)circle (0.7 mm); 
	\fill[black]  (-3.4,13.5) circle (0.7 mm); 
	
	\node at (-9.5,20.6) {2};
	\node at (-6.4,20.6) {7};
	\node at (-3.4,20.6) {8};
	\node at (-0.4,20.6) {9};
	\node at (2.7,20.6) {3};
	\node at (-6.2,17.9) {5};
	\node at (-0.6,17.9) {6};
	\node at (-2.1,16.1) {4};
	\node at (-2.1,13.5) {1};

	\node (v2) at (-6.4,18.9) {};
	\node at (2.7,18.9) {};
	\draw (-6.4,18.9) -- (-5.4,18.9) -- (-5.4,17.9);
	\fill[black] (-6.4,18.9) circle (0.7 mm); 
	\draw (-4.4,17.9) -- (-4.4,17.2) -- (-3.9,16.8) -- (-3.9,16.1);
	\draw (-2.4,17.9) -- (-2.4,17.2) -- (-2.9,16.8) -- (-2.9,16.1);
	
	\draw (-3.4,16.1) -- (-3.4,15.6) node (v1) {};
	\draw (-0.4,18.9) -- (-1.4,18.9) -- (-1.4,17.9);
	\fill[black] (-0.4,18.9) circle (0.7 mm); 
	\draw [-latex,thick](-3.4,18.9) -- (-2.9,18.9) -- (-2.9,18.1);
	\fill[black] (-3.4,18.9) circle (0.7 mm); 
	\draw [-latex,thick](-4.9,17.9) -- (-4.9,16.7);
	\fill[black](-4.9,17.9)circle (0.7 mm); 
	\draw [-latex, thick](-1.9,17.9) -- (-1.9,16.7);
	\fill[black](-1.9,17.9)circle (0.7 mm);

	%\draw [-latex] (-4.5,4) rectangle (-2.5,2.5);
	\draw[fill=black!20](-10.7,19.4) node (v3) {} circle (7 mm); 
	\draw(-8.3,19.4)  circle (5 mm); 
	\draw(-7.6,19.4)  circle (5 mm); 
	\draw  plot[smooth, tension=.7] coordinates {(-11.2,19.4) (-10.987,19.9) (v3) (-10.46,18.9) (-10.2,19.4)};
	
	%\node at (-3.5,3.25) {DVPP};
	\draw [fill=black!20](3.9,19.4) node (v8) {} circle (7 mm); 
	\draw(-3.4,15.1)  circle (5 mm); 
	\draw(-3.4,14.4)  circle (5 mm); 
	\draw  plot[smooth, tension=.7] coordinates {(3.31,19.4) (3.6,19.9) (v8) (4.2,18.9) (4.47,19.4)};
	\draw(1.3,19.4)  circle (5 mm); 
	\draw(0.6,19.4)  circle (5 mm); 
	\draw (-10,19.4) -- (-8.8,19.4);
	\draw (3.2,19.4) -- (1.8,19.4);
	\draw (-3.4,13.5) node (v4) {} -- (-3.4,13.9);

	\draw[fill=hydro](-6.5,11.9) node {} circle (7 mm); 
	%(-12.7,16.5)
	\draw  plot[smooth, tension=.7] coordinates {(-7,11.9) (-6.787,12.4)(-6.5,11.9)(-6.24,11.4) (-6,11.9)};
	\node at (-10.7,18.3) {SG 2};

	\node at (3.8,18.3) {SG 3};

	\node at (-6.5,10.8) {hydro};
	\draw [-latex,thick](2.7,19.8) -- (2.1,19.8) -- (2.1,20.6);
	
	\draw [fill=PV] (-3.2,12.3) rectangle (-1.4,11.4);
\node at (-2.3,11.85) {PV};
\draw  [fill=battery!60](-1,12.3) rectangle (0.8,11.4);
\node  [scale=0.8] at (-0.1,11.85) {BESS};
	
		\draw [fill=wind!60] (-5.4,12.3) rectangle (-3.6,11.4);
\node at (-4.5,11.85) {wind};
\draw (-3.4,13.5) -- (-3.4,12.9) node (v5) {};
\draw  (-3.4,12.9) -- (-6.5,12.9) -- (-6.5,12.6);
\draw(-4.5,12.9) -- (-4.5,12.3); 
\draw  (-3.4,12.9)  -- (-0.1,12.9) -- (-0.1,12.3); 
\draw (-2.3,12.9) -- (-2.3,12.3); 

\node at (2.1,11.8) {DVPP};
\end{tikzpicture}